\documentclass[11pt,a4paper]{article}

\usepackage[utf8x]{inputenc}
\usepackage{graphicx,a4wide,cite,hyperref}
\usepackage[title,titletoc,toc]{appendix}

\def\be{\begin{equation}}
\def\ee{\end{equation}}
\def\bea{\begin{eqnarray}}
\def\eea{\end{eqnarray}}

\def\reff#1{\ref{#1}}
\def\labell#1{\label{#1}}

\def\app#1{Appendix~\reff{#1}}

\def\eq#1{Eq.~(\reff{#1})}
\def\eqs#1#2{Eqs.~(\reff{#1})--(\reff{#2})}

\def\fig#1{Fig.~\reff{#1}}
\def\figs#1#2{Figs.~\reff{#1}--\reff{#2}}
\def\sec#1{Sec.~\reff{#1}}
\def\tab#1{Table~\reff{#1}}

\newcommand{\lsim}{ {\
\lower-1.2pt\vbox{\hbox{\rlap{$<$}\lower5pt\vbox{\hbox{$\sim$}}}}\ } }
\newcommand{\gsim}{ {\
\lower-1.2pt\vbox{\hbox{\rlap{$>$}\lower5pt\vbox{\hbox{$\sim$}}}}\ } }

\def\er#1#2{\relax\ifmmode{}^{+#1}_{-#2}\else$^{+#1}_{-#2}$\fi}
\def\erparen#1#2{\relax\ifmmode{}(^{#1}_{#2})\else$(^{#1}_{#2})$\fi}

\def~{\ifmmode\phantom{0}\else\penalty10000\ \fi}

\def\fm{\mathrm{fm}}
\def\ev{\mathrm{e\kern-0.1em V}}
\def\kev{\mathrm{ke\kern-0.1em V}}
\def\mev{\mathrm{Me\kern-0.1em V}}
\def\gev{\mathrm{Ge\kern-0.1em V}}
\def\tev{\mathrm{Te\kern-0.1em V}}

\def\n#1e#2n{{#1}\times 10^{#2}}

\def\nn{\nonumber}

\def\ods2{\mathcal{O}_{\Delta S=2}}
\def\zds2{Z_{\Delta S=2}}

\def\msbar{{\overline{\mathrm{MS}}}}

\def\RGI{\mathrm{RGI}}

\def\NLO{\mathrm{NLO}}
\def\NNLO{\mathrm{NNLO}}

\def\lqcd{\Lambda_\mathrm{QCD}}

\def\mudrgimax{(m_{ud}^\RGI)^\max}

\makeatletter
\def\slash#1{{\mathpalette\c@ncel{#1}}} 
\def\big#1{{\hbox{$\left#1\vbox to1.012\ht\strutbox{}\right.\n@space$}}}
\def\Big#1{{\hbox{$\left#1\vbox to1.369\ht\strutbox{}\right.\n@space$}}}
\def\bigg#1{{\hbox{$\left#1\vbox to1.726\ht\strutbox{}\right.\n@space$}}}
\def\Bigg#1{{\hbox{$\left#1\vbox
to2.083\ht\strutbox{}\right.\n@space$}}}
\makeatother

\makeatletter
\def\old@comma{,}
\catcode`\,=13
\def,{%
  \ifmmode%
    \old@comma\discretionary{}{}{}%
  \else%
    \old@comma%
  \fi%
}
\makeatother

\makeatletter
\def\old@semicolon{;}
\catcode`\;=13
\def;{%
  \ifmmode%
    \old@semicolon\discretionary{}{}{}%
  \else%
    \old@semicolon%
  \fi%
}
\makeatother

\def\Mss{M_{s\bar{s}}}

\def\max{\mathrm{max}}
\def\min{\mathrm{min}}

\def\phys{\mathrm{ph}}
\def\awi{\mathrm{AWI}}
\def\vwi{\mathrm{VWI}}
\def\pcac{\mathrm{PCAC}}
\def\imp{\mathrm{imp}}
\def\bare{\mathrm{bare}}
\def\qcd{\mathrm{QCD}}

\usepackage{xcolor}

\title{\begin{minipage}{\textwidth}
\vspace{-1.5 cm}
\hspace{10 cm}\begin{small}CPT-P005-2013, WUB/13-14\end{small} \newline 
 \begin{center}
 Lattice QCD at the physical point meets $SU(2)$ chiral perturbation theory\end{center}                
\end{minipage}} 

\author{\begin{minipage}{\textwidth}\begin{center}
{\bf Budapest-Marseille-Wuppertal collaboration}\\[0.2cm]
Stephan D\"urr$^{1,2}$, Zoltán Fodor$^{1,2,3}$, Christian
Hoelbling$^1$, Stefan Krieg$^{1,2}$,
Thorsten Kurth$^1$,
Laurent Lellouch$^{4,5}$, Thomas Lippert$^{2}$,  Rehan Malak$^{4,5,7}$, Thibaut Métivet$^{4,5,8}$, Antonin
Portelli$^{4,5,6}$, Alfonso Sastre$^{4,5}$, Kálmán Szabó$^1$\\[0.4cm]
{\it $^1$ Department of Physics, Wuppertal University,
Gaussstr. 20, 
D-42119 Wuppertal, Germany\\
$^2$ IAS/JSC,
Forschungszentrum J\"ulich, D-52425 J\"ulich, Germany\\
$^3$ Inst.\ for Theor.\ Physics, E\"otv\"os University,
P\'azm\'any P. s\'et.\ 1/A, 
H-1117 Budapest, Hungary\\
$^4$ Aix Marseille Université, CNRS, CPT, UMR 7332, F-13288 Marseille,
France\\
$^5$ Université de Toulon, CNRS, CPT, UMR 7332, F-83957 La Garde,
France\\
$^6$ School of Physics \&\ Astronomy, University of Southampton, SO17
1BJ, UK\\
$^7$ 
CEA/CNRS Maison de la Simulation, USR 3441, CEA-CNRS-INRIA-UPSud-UVSQ,
F-91191 Gif-sur-Yvette Cedex, France\\
$^8$ CEA/IRFU,
CEA-Orme des Merisiers, Bât. 703,
F-91191 Gif-sur-Yvette Cedex, France}
\end{center}
\end{minipage}
}

\date{}

\begin{document}

\maketitle

\begin{abstract}

We perform a detailed, fully-correlated study of the chiral behavior
of the pion mass and decay constant, based on 2+1 flavor lattice QCD
simulations. These calculations are implemented using tree-level,
$O(a)$-improved Wilson fermions, at four values of the lattice spacing
down to 0.054~fm and all the way down to below the physical value of
the pion mass. They allow a sharp comparison with the predictions of
$SU(2)$ chiral perturbation theory ($\chi$PT) and a determination of
some of its low energy constants. In particular, we systematically
explore the range of applicability of NLO $SU(2)$ $\chi$PT in two
different expansions: the first in quark mass ($x$-expansion), and the
second in pion mass ($\xi$-expansion). We find that these
  expansions begin showing signs of failure around $M_\pi=300\,\mev$
  for the typical percent-level precision of our $N_f=2+1$ lattice
  results. We further determine the LO low energy constants (LECs),
  $F=88.0\pm 1.3\pm 0.3$ and $B^\msbar(2\,\gev)=2.58 \pm 0.07 \pm 0.02\,\gev$,
  and the related quark condensate,
  $\Sigma^\msbar(2\,\gev)=(271\pm 4\pm 1\,\mev)^3$, as well as the NLO ones, $\bar
  \ell_3=2.5 \pm 0.5 \pm 0.4$ and $\bar\ell_4=3.8 \pm 0.4 \pm 0.2$, with fully
  controlled uncertainties. Our results are summarized in
  \tab{tab:final-lec-results}. We also explore the NNLO expansions
and the values of NNLO LECs. In addition, we show that the lattice
results favor the presence of chiral logarithms. We further
demonstrate how the absence of lattice results with pion masses below
200~MeV can lead to misleading results and conclusions. Our
calculations allow a fully controlled, ab initio determination of the
pion decay constant with a total 1\% error, which is in excellent
agreement with experiment.

\end{abstract}

\tableofcontents

\section{Introduction}
\labell{sec:intro}
The study of the strong interaction at low energy is hampered by the
highly nonlinear nature of quantum chromodynamics (QCD). Thus, large
scale numerical simulations in lattice QCD have become an essential
tool for investigating, from first principles, the nonperturbative
dynamics of the theory in that domain. In order to account for all of
the relevant physics at the few percent level in low-energy
observables, one must include the vacuum fluctuations of the up, down
and strange quarks. The heavier quarks contribute corrections in
inverse powers of the quark mass squared and of the number of colors,
which can be neglected at that level of precision. Moreover, for most
QCD observables, isospin breaking effects, which are proportional to
powers of the small up-down mass difference, $(m_d-m_u)$, and of the
fine structure constant, $\alpha$, can also be neglected. Thus,
today's state-of-the-art calculations are performed with $N_f\ge 2+1$
flavors of sea quarks, where the 2 stands for mass-degenerate $u$ and
$d$ quarks with $m_u=m_d=m_{ud}\equiv(m_u+m_d)/2 $ and the 1 for a
more massive $s$ quark with mass $m_s$.

One of the main challenges has been to mitigate the fast rising cost
of these calculations as the average mass of the simulated up and down
quarks is lowered towards its very small physical value,
corresponding to a pion mass $M_\pi\simeq 135\,\mev$. Up until fairly
recently, the values of $m_{ud}$ reached were too large to allow a
controlled extrapolation of the results to the physical mass
point. However, in the last few years, a handful of groups have been
able to enter the small mass region, $M_\pi\lsim 200\,\mev$, with
$N_f\ge
2+1$~\cite{Durr:2008zz,Aoki:2008sm,Aoki:2009ix,Durr:2010vn,Durr:2010aw,
Borsanyi:2012zv,Arthur:2012opa,Bazavov:2012xda}. In
particular, we recently performed $N_f=2+1$ simulations which reach
down to $M_\pi\simeq 120\,\mev$ (i.e. even below the physical point)
on lattices with sizes $L$ up to 6~fm and lattice spacings down to
$a\simeq 0.054\,\fm$~\cite{Durr:2010vn,Durr:2010aw}. This puts us in a
very favorable position to probe the low-energy and low-mass domain of
QCD, known as the chiral regime.

In this paper we investigate $SU(2)$ chiral perturbation theory
($\chi$PT), which is a systematic expansion around the $m_u=m_d=0$
chiral limit, at fixed $m_s$ (and possibly $m_c$,
\ldots)~\cite{Weinberg:1978kz,Gasser:1983yg}. In the corresponding chiral effective
Lagrangian there are two low-energy constants (LECs) at leading $O(p^2)$:
\be
\labell{eq:FBdef}  F\equiv F_\pi\,\rule[-0.3cm]{0.01cm}{0.7cm}_{\;m_u,m_d\rightarrow
  0} \; , \qquad B\equiv -\frac{\langle 0\vert \bar u u
  \vert 0\rangle}{F_\pi^2}\,\rule[-0.3cm]{0.01cm}{0.9cm}_{\;m_u,m_d\rightarrow
  0} \; , 
\ee
where $F_\pi$ is the pion, leptonic decay constant, and there are 7
more at next-to-leading $O(p^4)$, denoted by $\ell_i(\mu)$,
$i=1, \ldots, 7$~\cite{Gasser:1983yg}. By definition the LECs are
independent of the $u$ and $d$ quark masses, but do depend on the
masses of the other four quarks. They also acquire a scale dependence,
after renormalization. It is conventional to define them at the
renormalization scale $\mu=\hat M_{\pi^+}=134.8(3)\,\mev$, where $\hat
M_{\pi^+}$ is the $\pi^+$ meson mass, corrected for electromagnetic
effects~\cite{Colangelo:2010et}. Up to negligible corrections, it is
also equal to $\bar M_{\pi}$, the pion mass in the isospin limit
($m_u-m_d\rightarrow0$ at fixed $m_{ud}$) \cite{Colangelo:2010et}, in
which our $N_f=2+1$ lattice calculations are performed.

The observables which we consider here are $M_\pi^2$ and
$F_\pi$. Their expansions in powers of the quark mass are known to
next-to-next-to-leading order (NNLO) in the $SU(2)$ chiral effective
theory. In the isospin limit, the explicit expressions may be written
in the form\footnote{Here and in the following, we work in the
normalization $F_\pi\equiv f_\pi/\sqrt{2}=92.2\mev$.}, 
$m_u=m_d=m_{ud}$ \cite{Colangelo:2001df}
\bea
\labell{eq:MF}
M_\pi^2 & = & M^2\left\{1-\frac{1}{2}x\ln\frac{\Lambda_3^2}{M^2}
  +\frac{17}{8}x^2 \left(\ln\frac{\Lambda_M^2}{M^2}  \right)^2 +x^2 k_M
  +O(x^3)             \right\},
\\
F_\pi & = & F\left\{1+x\ln\frac{\Lambda_4^2}{M^2} -\frac{5}{4}x^2
  \left(\ln\frac{\Lambda_F^2}{M^2}  \right)^2 +x^2k_F   +O(x^3)
\right\}.
 \nonumber
\eea
The expansion parameter is given by 
\be
\labell{eq:xM2}
 x=\frac{M^2}{(4\pi
  F)^2},\;\;\;\;\;\;\;\;\;\;M^2=2Bm_{ud}=\frac{2m_{ud} \Sigma }{F^2}.
\ee
The $O(p^6)$ LECs, $k_M$ and $k_F$, in \eq{eq:MF} are also independent of
the $u$ and $d$ quark masses. The scales in the quadratic logarithms
can be written in terms of $O(p^4)$ LECs through:
\bea
\labell{eq:LMLF}
  \ln\frac{\Lambda_M^2}{M^2} & = &
  \frac{1}{51}\left(60\ln\frac{\Lambda_{12}^2}{M^2}
    -9 \ln\frac{\Lambda_3^2}{M^2}+49
  \right),    \\  
  \ln\frac{\Lambda_F^2}{M^2} & = &
  \frac{1}{30}\left(30\ln\frac{\Lambda_{12}^2}{M^2}
     +6 \ln\frac{\Lambda_3^2}{M^2} 
    - 6 \ln\frac{\Lambda_4^2}{M^2}      +23  \right)\ ,   \nonumber
\eea
where we have defined $\ln \Lambda_{12}^2 = (7\ln \Lambda_1^2 +
8\ln \Lambda_2^2)/15$. The logarithmic scales $\Lambda_n$ in
\eqs{eq:MF}{eq:LMLF} are related to the effective
coupling constants $\bar\ell_3,\bar\ell_4$ of the chiral Lagrangian at
running scale $\hat M_{\pi^+}$ through:
\be
\labell{eq:barelldef}
\bar\ell_n=\ln\frac{\Lambda_n^2}{\hat M_{\pi^+}^2},\quad
n=1,...,7\mbox{ \&}12
\ ,
\ee
where we have generalized the definition to also include
$\Lambda_{12}$ and $\bar\ell_{12}$.

It is interesting to note that once we fix $\Lambda_3$ and
$\Lambda_4$, which appear already at NLO in the expansions of $F_\pi$
and $M_\pi$, the new logarithmic scales $\Lambda_M$ and $\Lambda_F$
are linearly related. This reduces from 8 to 7 the number of parameters in
a combined fit of the dependence of $M_\pi^2$ and $F_\pi$ on
$m_{ud}$. In particular this means that with precise enough lattice
results for the pair $(M_\pi^2,F_\pi)$, at four or more values of
$m_{ud}$, one can in principle determine the 7 independent LECs which
appear in the expansions of \eq{eq:MF} as well as test the
compatibility of the lattice results with NNLO $\chi$PT. Such an
NNLO analysis is still very demanding by today's standards.

The situation is significantly more simple if the expressions of
\eq{eq:MF} are truncated at NLO. Then, only 4 LECs appear, $B$ and
$F$ at $O(p^2)$, and $\bar\ell_3$ and $\bar\ell_4$ at $O(p^4)$. This
is the expansion considered in previous $N_f\ge 2+1$
work 
\cite{Allton:2008pn,Aoki:2008sm,Bazavov:2009fk,Aoki:2010dy,Bazavov:2010hj,
Bazavov:2010yq,Beane:2011zm,Borsanyi:2012zv,Arthur:2012opa,
Baron:2010bv,Baron:2011sf}. Of those, the only calculation whose
simulations reach all the way down to the physical up-down quark mass
is \cite{Borsanyi:2012zv}. In that work, NNLO effects are also
investigated. 

Work on the $x$-expansion has also been performed using $N_f=2$
lattice QCD simulations
in \cite{DelDebbio:2006cn,Noaki:2008iy,Frezzotti:2008dr,JLQCD:2009qn,Baron:2009wt,Chiu:2011bm,Bernardoni:2011kd,Horsley:2013ayv,Brandt:2013dua}. Such
work has provided interesting information about $SU(2)$
$\chi$PT. However, because the effects of the omitted strange, sea
quark in these calculations cannot be quantified {\it a priori}, the
conclusions which are drawn from such studies will differ
qualitatively and quantitatively from ours by an unknown amount. Thus,
we do not consider them further here and refer the interested reader
to \cite{Colangelo:2010et} and the original papers for further
information. 

As with any expansion, the chiral expressions can be reorganized in
terms of any other parameter which is related to $x$ of
\eq{eq:xM2}, through a power series in $x$. In particular, one can
invert \eq{eq:MF}, and express $M$ and $F$ as an expansion in 
\be
\labell{eq:xi}
\xi \equiv \frac{M_\pi^2}{(4 \pi F_\pi)^2} \; \; .
\ee
The corresponding expressions read~\cite{Colangelo:2010et}
\bea
\labell{eq:MpiFpi} 
M^2&=& M_\pi^2\,\left\{ 
1+\frac{1}{2}\,\xi\,\ln\frac{\Lambda_3^2}{M_\pi^2}-
\frac{5}{8}\,\xi^2 \left(\!\ln\frac{\Omega_M^2}{M_\pi^2}\!\right)^2+
\xi^2 c_M+O(\xi^3)\right\} \ , \\
F&=& F_\pi\,\left\{1-\xi\,\ln\frac{\Lambda_4^2}{M_\pi^2}-\frac{1}{4}\,\xi^2
\left(\!\ln\frac{\Omega_F^2}{M_\pi^2}\!\right)^2
+\xi^2 c_F+O(\xi^3)\right\} \ . \nn
\eea
This expansion has the advantage that its parameter $\xi$ is given in
terms of the physical mass and decay constant of the particle which is
actually contributing to the process. Thus, it resums a number of
higher-order contributions which are known to be present, and
therefore might exhibit better convergence. In \eq{eq:MpiFpi}, the
scales of the quadratic logarithms are determined by
$\Lambda_1,\ldots,\Lambda_4$~\cite{Colangelo:2010et}:
\bea
\labell{eq:OmegaM}
\ln\frac{\Omega_M^2}{M_\pi^2}&=&\frac{1}{15}\left(60\ln\frac{\Lambda_{12}^2}{M_\pi^2}
-33\ln\frac{\Lambda_3^2}{M_\pi^2}-12\ln\frac{\Lambda_4^2}{M_\pi^2}
+52\right) \ , \\
\labell{eq:OmegaF}
\ln\frac{\Omega_F^2}{M_\pi^2}&=&\frac{1}{3}\,\left(-15\ln\frac{\Lambda_{12}^2}{M_\pi^2}+
18\ln\frac{\Lambda_4^2}{M_\pi^2}- \frac{29}{2}\right)\nonumber \ .
\eea

Here we study $SU(2)$ $\chi$PT in both the $x$ and
$\xi$-expansions. While most of the work concerns the NLO expansions,
we also investigate the NNLO expansions, in particular in regards to
its range of applicability. 

The remainder of the paper is organized as
follows. In \sec{sec:latticedetails} we detail the lattice ensembles
used in the present study and the various steps in required to
determine the chiral observables $M_\pi$, $F_\pi$ and the quark masses
from our correlation functions.  We also discuss how we perform the
necessary renormalizations and how we account for the various sources
of lattice systematic errors in our analyses.
In \sec{sec:su2validity}, we systematically explore the range of
applicability, in pion or light-quark mass, of the various $SU(2)$
$\chi$PT expressions for $M_\pi^2/2m_{ud}$ and $F_\pi$. In particular,
we assume that $SU(2)$ $\chi$PT is valid around $M_\pi^\phys$, where
here and below the superscript ``$\phys$'' stands for ``physical'' or
from experiment, and explore how far up one can go in pion or
light-quark mass, while still maintaining an acceptable description of
the lattice results. Then, having established the range of
applicability of $SU(2)$ $\chi$PT for $M_\pi^2/2m_{ud}$ and $F_\pi$,
we devote \sec{sec:lecresults} to a determination of the corresponding
LO, NLO and NNLO LEC's, as well as of $F_\pi$ and the quark
condensate. In particular, we perform a complete systematic error
analysis for these quantities. Our main results are summarized
in \tab{tab:final-lec-results}. In \sec{sec:logsandmisuse} we show
that the lattice results favor the presence of chiral logarithms. We
also show how the absence of lattice results with $M_\pi\le 200\,\mev$
can lead to misleading results and conclusions. In the paper's final
section, \sec{sec:conclusion} we present our conclusions. We also
provide an appendix in which we discuss our implementation of
the $\xi$-expansion and the ensuing constraints on the LECs.

\clearpage

\section{Determination of lattice quantities and
  associated systematic errors}
\labell{sec:latticedetails}
In this section, we describe how we compute the values of $M_\pi$,
$F_\pi$ and $m_{ud}$ required for the $\chi$PT studies described
below. We do so for a range of $m_{ud}$ around and above its physical
value to explore the range of applicability of $SU(2)$ $\chi$PT. We
also do so for a large variety of lattice parameters to be able to
control all sources of systematic uncertainties.

As first proposed in \cite{Durr:2008zz}, we determine the central values,
statistical and systematic uncertainties of our results from
histograms obtained by combining the results form a variety of
different analyses. Indeed, for each step of the analysis, we consider
a wide range of possible procedures whose effects we propagate to the
end of the calculation. Thus, our analyses form a tree where each path
corresponds to one of the many different possible ways in which to
compute a given observable.

The trunk of the tree corresponds to the primary observables. In the
present study, they are the hadron correlators. Thus, the first level
of branching occurs in choosing the time interval over which these
correlators are fitted to obtain the bare masses and decays constants
in lattice units. The next level of branching is a result of the
different ways which we have to set the lattice spacing.  Note
that at each level, these same twigs are sprouted from every
branch. For quantities which require renormalization, an additional
level of branching arises, corresponding to the different ways which
we have to compute the renormalization constants. Note that the
renormalization constants are themselves the result of a tree, as
described below.

At that stage in the analysis, we have obtained, in all possible ways,
the renormalized results in physical units for each simulation, which
we will need to study $SU(2)$ $\chi$PT. Note that throughout our
analysis we fully take into account statistical correlations as well
as correlations induced by quantities such as the lattice spacing or
the renormalization constants, which are shared by all ensembles
at a given $\beta$.

In the remainder of the section, we detail the ingredients of the
analysis briefly described here, including the procedure used to determine
the associated systematic uncertainties.

\subsection{Simulation details}

The study presented here is based on the  forty-seven $N_f=2+1$
ensembles that we produced for determining the light quark
masses~\cite{Durr:2010vn,Durr:2010aw}. They were generated using a
tree-level $O(a^2)$-improved Symanzik gauge
action~\cite{Weisz:1982zw,Weisz:1983bn,Luscher:1984xn,Luscher:1985zq},
together with tree-level clover-improved Wilson
fermions~\cite{Sheikholeslami:1985ij}, coupled to links which have
undergone two levels of HEX
smearing~\cite{Hasenfratz:2001hp,Morningstar:2003gk,Capitani:2006ni}. Details
of the action and simulations are given in \cite{Durr:2010aw}. Here we
mention that we use the 26 large-volume ensembles that were generated
at 4 values of the lattice spacing spanning the range $0.054\,\fm\lsim
a\lsim 0.093\,\fm$. We found that the low momentum cutoff of the
coarsest lattice in \cite{Durr:2010aw}, with $a=0.116\,\fm$, does not
allow a precise determination of the renormalization constant of the
axial current, $Z_A$, required for the computation of $F_\pi$. The
uncertainty associated with its determination, of order 1.5\%, is
sufficiently large that it negates any improvement the inclusion of
the results at that lattice spacing could bring to the final
results. Thus, as in \cite{Durr:2011ap}, we have chosen not to
incorporate the results of this simulation in our analysis.

The
strange quark mass in these simulations is varied around the physical
value to allow for a precise interpolation to that value. For the 2
lattice spacings $a\approx 0.077,\,0.093\,\fm$, simulations were
performed all the way down to the physical value of $m_{ud}$ and even
below. For the remaining 2 lattice spacings ($a\approx
0.065,\,0.054\,\fm$), the pion masses reached are 180 and 220 MeV,
respectively. Thus, our simulations allows us to replace the usual
extrapolations to physical $m_{ud}$ by an interpolation, but also to
systematically probe the $SU(2)$ chiral regime. 

The parameters of the
simulations used in this work are summarized in Tables \reff{tab:aZSZA}
and \reff{tab:data}, together with illustrative results for the
lattice spacing, renormalization constants and observables that are
discussed below. 

\subsection{Strategy for determining masses and decay constants}
\labell{sec:mass-and-decay}

We determine $aM_\pi$ and $aF_\pi/Z_A$ for each simulation point
by performing a combined correlated fit of the asymptotic time
behavior of the two, zero-momentum correlators, $\sum_{\vec{x}}\langle
A_0^L({\vec{x}},x_0) P^{G\dagger}(0)\rangle$ and
$\sum_{\vec{x}}\langle P^G({\vec{x}},x_0) P^{G\dagger}(0)\rangle$, to
the appropriate asymptotic forms. Here $A_0$ is the time-component of
the axial-vector current and $P$ is the corresponding pseudoscalar
density. Both are appropriately tree-level
$O(a)$-improved \cite{Sheikholeslami:1985ij,Heatlie:1990kg}. These operators have the flavor
quantum numbers appropriate for annihilating a $\pi^+$. The
superscript $L$ stands for ``local'' (i.e. all quark fields are at
the same spacetime point) and $G$ for ``Gaussian''. Indeed, to reduce the
relative weight of excited states in the correlation functions,
Gaussian sources and sinks are used (except for the axial current, of
course), with a radius of about $0.32\fm$, which was found to be a
good choice~\cite{Durr:2008zz}. The kaon masses, $aM_K$, are obtained
from a correlated fit to the corresponding, two-point, pseudoscalar
density correlators.

To study the $x$-expansion discussed above, we need to determine the
quark masses $m_{ud}$ and $m_s$ for each simulation point. Here we
follow the $O(a)$-improved ratio-difference method put forward
in~\cite{Durr:2010vn,Durr:2010aw}. Thus, for each simulation point we
determine the bare axial-Ward-identity mass combinations
$2m_{ud}^\pcac(g_0)=(m_u+m_d)^\pcac(g_0)$ and
$(m_s+m_{ud})^\pcac(g_0)$ from the relevant ratio of two-point
functions, $\partial_0\sum_{\vec{x}}\langle A_0^L({\vec{x}},x_0)
P^{G\dagger}(0)\rangle$ / $\sum_{\vec{x}}\langle P^G({\vec{x}},x_0)
P^{G\dagger}(0)\rangle$, where $\partial_\mu$ is the symmetric
derivative. The
operators are appropriately tree-level $O(a)$-improved. From this we
obtain the ratio of renormalized, improved quark masses, $r^\imp\equiv
m_s^\awi(\mu)/m_{ud}^\awi(\mu)$, through $r^\imp=
m_s^\pcac(g_0)/m_{ud}^\pcac(g_0)[1+O(a)]$, where the $O(a)$
improvement terms are given in \cite{Durr:2010aw} and $\mu$ is a
renormalization scale. Because the numerator and denominator in this
ratio renormalize identically, all scale and scheme dependence
cancels. This ratio is then combined with the difference of
renormalized, improved vector-Ward-identity masses,
$(m_s-m_{ud})^\vwi(\mu)=d^\imp(g_0)/(aZ_S(a\mu,g_0))$, to obtain the
renormalized values $m_{ud}(\mu)$ and $m_s(\mu)$ of the quark masses
for a given simulation. Here
$d^\imp(g_0)=(am_s^\bare-am_{ud}^\bare)(g_0)[1+O(a)]$, where the $O(a)$
improvement terms are also given in \cite{Durr:2010aw} and where
$am_{ud,s}^\bare(g_0)$ are the bare lagrangian masses used at bare coupling
$g_0$. In the definition of the mass difference, $Z_S(a\mu,g_0)$ is
the renormalization constant of the non-singlet scalar density in any
chosen scheme at scale $\mu$~\cite{Durr:2010vn,Durr:2010aw}. Here we
will mainly use its renormalization group invariant (RGI) value, which
is regularization scheme and renormalization scale independent.

\subsection{Excited state contributions}
\labell{sec:exci}
A source of uncertainty, which often proves important, is the
contamination by excited states of the desired ground state in
two-point correlators. As described above, this contamination is
reduced by working with extended sources and sinks. Moreover, we
tested 1-state and 2-state fits, and found complete agreement if the
1-state fits start at $t_\mathrm{min}{\simeq}0.7\,\fm$ for the
pseudoscalar meson channels and from $t_\mathrm{min}{\simeq}0.8\,\fm$
for the $\Omega$.  In lattice units this amounts to
$at_\mathrm{min}\!=\!\{8,9,11,13\}$ for
$\beta{=}\{3.5,3.61,3.7,3.8\}$ (and $\sim\!20\%$ later for
baryons).  In order to estimate any remaining excited state effects,
we repeat our analysis with an even more conservative fit range,
starting at $at_\mathrm{min}{=}\{9,11,13,15\}$ for mesons and $\sim
20\%$ later for baryons.  The end of the fit interval is always chosen
to be $at_\mathrm{max}=2.7\!\times\!at_\mathrm{min}$ or $T/2-1$ for
lattices with a time extent shorter than
$5.4\!\times\!at_\mathrm{min}$. In total, this yields
2 combined, time-fit intervals.

\subsection{Lattice spacing}
\labell{sec:latspac}

To set the lattice spacing, we follow \cite{Durr:2008zz} and use the
$\Omega$ baryon mass. Thus, we perform a combined interpolation to the
physical mass point of our results for $aM_\Omega$ at all four values
of $\beta$, with the following functional form:
\begin{eqnarray}
\labell{eq:aMOmegainterp}
aM_\Omega=aM_\Omega^\phys(\beta)
&&\left\{1+c_s\left[\left(\frac{aM_{s\bar{s}}}{aM_\Omega}\right)^2-\left(\frac{M_{s\bar{s}}}{M_\Omega}\right)^2_\phys\right]\right.\\
&&+
\left.c_{ud}\left[\left(\frac{aM_\pi}{aM_\Omega}\right)^2-\left(\frac{\hat
  M_{\pi^+}}{M_\Omega}\right)^2_\phys\right]\right\}\nonumber
\ ,\end{eqnarray}
where $(M_{s\bar{s}})^2=2M_K^2-M_\pi^2$. In (\ref{eq:aMOmegainterp}),
there is of course one parameter $aM_\Omega^\phys$ per lattice
spacing, but we find that our fits do not require the parameters
$c_{s,ud}$ to be $\beta$ dependent. Moreover, for the range of quark
masses considered, we find that we do not need higher order terms in
the mass expansion. Thus, these fits have a total of 7 parameters.

To estimate the systematic uncertainties in our final results
associated with the determination of the lattice spacing, we consider
$2\times 2 = 4$ different procedures for its computation, which we
propagate throughout our analysis. In particular, we consider 2
different time-fitting ranges for the extraction of $aM_\Omega$ in
each simulation ($at_\mathrm{min}\!=\!\{10, 11, 13, 16\}$ or
$at_\mathrm{min}{=}\{11, 13, 16, 18\}$ for
$\beta{=}\{3.5,3.61,3.7,3.8\}$) to estimate the possible effects
of excited state contributions to the two-point functions and 2 pion
cuts in the mass interpolation fits described above (380 or 480~MeV),
to estimate the uncertainties associated with the interpolation of
$aM_\Omega$ to the physical mass point. This gives us a total of 4
values of the lattice spacing for each $\beta$. While each of these
procedures enters individually in our determination of systematic
uncertainties for all quantities which depend on the lattice spacing,
we give in \tab{tab:aZSZA} illustrative
numbers, whose central values are the fit-quality weighted averages of
the results from the different procedures and whose statistical errors
are the variance of these central values over 2000 bootstrap samples. The
systematic errors are obtained from the variance over the
procedures. 

\begin{table}
 \centering
 \begin{tabular}{cccc} 
 \hline
 \hline
  $\beta$ & $a$ [fm] & $1/Z_S^\RGI$ & $Z_A$ \\
 \hline
  3.5 &  0.0904(10)(2) & 1.47(2)(3) & 0.9468(5)(56) \\ \hline 
3.61 & 0.0755(11)(3) & 1.50(3)(2) & 0.9632(4)(53) \\ \hline 
3.7 & 0.0647(11)(3) & 1.54(3)(3) & 0.9707(3)(35) \\ \hline 
3.8 & 0.0552(8)(1) & 1.58(1)(1) & 0.9756(1)(15) \\ 
 \hline
 \hline
 \end{tabular}
\caption{Illustrative results for the lattice spacing and the 
renormalization constants at our four
 values of $\beta$. $1/Z_S^\RGI$ is required to convert bare quark
 masses to masses renormalized in the $N_f=3$ RGI scheme. To convert
 results to the $\msbar$ scheme at scale 2~GeV, the numbers in the
 third column of the table must be multiplied by
 0.750 \cite{Durr:2010aw}. $Z_A$ is used to correctly normalize
 $F_\pi$. In the results above, the first error is statistical and the
 second is systematic. The main text explains how these errors are obtained as
 well as why the results cannot be used to reproduce the extensive
 analyses performed in this paper.}
\labell{tab:aZSZA}
\end{table}

\subsection{Renormalization}
\labell{sec:renorm}

To determine the renormalization constants we use the nonperturbative
renormalization and running techniques developed
in \cite{Durr:2010vn,Durr:2010aw,Arthur:2010ht}, which are based on
the RI/MOM methods {\em \`a la
Rome-Southampton} \cite{Martinelli:1994ty}. For $Z_S$, we
follow \cite{Durr:2010aw} and $Z_A$ is determined as
in \cite{Durr:2011ap}. As described in \cite{Durr:2010aw}, the
calculation of these constants is performed using 20 fully independent
$N_f=3$ simulations at the same four values of $\beta$ as the
$N_f=2+1$ production runs.

In order to compute the systematic uncertainties associated with
renormalization on our final results, we consider 6 different
procedures for the determination of $Z_S$ and 3  for $Z_A$, as
described in detail \cite{Durr:2010aw} and \cite{Durr:2011ap},
respectively. Here we simply outline the different procedures.

The renormalization of quark masses is performed in three
steps \cite{Durr:2010aw}. We first compute $Z_S$ in a MOM scheme at an
intermediate scale $\mu'$, which is low enough that discretization
errors on the renormalization constant are under control. We then run
the results nonperturbatively in that scheme up to a fully
perturbative scale $\mu=4$~GeV where they are converted
nonperturbatively to the usual massless, $N_f=3$, RI/MOM
scheme. Values in other schemes are then obtained using
renormalization-group-improved perturbation theory at
$O(\alpha_s^3)$ \cite{Chetyrkin:1999pq}, with negligible
uncertainty. These three steps lead to 6 procedures in the following
way. In step 1 we consider three different MOM schemes to determine
the uncertainties associated with the choice of an intermediate scale
$\mu'$ and with the chiral extrapolation required to define the RI/MOM
scheme. These correspond to the scale and quark-mass pairs,
${\mu'[\gev],m_\mathrm{ref}^\RGI[\mev]}$ $=$ $\{\{2.1,0\}$,
$\{2.1,70\}$, $\{1.3,70\}\}$. The additional factor of two comes from
the two ways in which we continuum extrapolate the nonperturbative
running and matching factors, either assuming that the $O(\alpha_s a)$
or $O(a^2)$ terms dominate.

$Z_A$ is a finite renormalization and therefore does not have a scale
or scheme dependence. Nevertheless, we must find a window, at large
values of the squared-momentum, $p^2\gg\lqcd$, of the quark
three-point function used to determine $Z_A$, in which this
correlation function is approximately constant. For such momenta the
correlation function is dominated by perturbation theory and allows
for a reliable extraction of $Z_A$. To estimate the uncertainties
associated with the choice of this window and with possible $(ap)^2$
discretization corrections, we fit our results for the relevant
three-point function to the functional form $Z_A+A(am_q)+B(ap)^2$ for
three different ranges in $p^2$. Here, $am_q$ is the common, $N_f=3$, bare PCAC
mass. For all four $\beta$ these ranges begin either at $p^2=3.35$,
4.37 or $5.52\,\gev^2$. These values of $p^2$ are large enough that we
are not sensitive to subleading OPE contributions proportional to
inverse powers of $p^2$. The upper bounds of the fit ranges are chosen
to be $1.5/a$ in all cases. This is below $\pi/(2a)$ which we found
in \cite{Durr:2010aw} is a region in which discretization errors on
the RI/MOM correlation functions are subdominant.

We provide in \tab{tab:aZSZA} illustrative results for $1/Z_S^\RGI$
and $Z_A$ for the four values of the lattice spacing used in our
study.  Their central values are the fit quality weighted averages of
the results from the different procedures and their statistical errors
are the variance of these central values over 2000 bootstrap
samples. Their systematic uncertainties are obtained from the variance
over the different procedures.

The results for $a$, $1/Z_S$ and $Z_A$ in \tab{tab:aZSZA} are only
illustrative, because they cannot be naively combined with the
observables given \tab{tab:data} to perform a fully self-consistent
analysis such as the one presented below. Indeed in our analysis, the
statistical and systematic uncertainties associated with these
quantities are propagated in a fully-consistent manner to our final
results by including them in our resampling and systematic error
loops. Such an analysis requires having the full statistical and
systematic error distributions of the quantities in
Tables~\reff{tab:data} and \reff{tab:aZSZA}, as well as their
correlations.

\subsection{Finite-volume corrections}
\labell{sec:fvcorr}

 Because our calculations are performed in large but finite
boxes, our results for $F_\pi$ and $M_\pi$ suffer from finite-volume
corrections. These effects have been determined at one loop in $SU(2)$
$\chi$PT in \cite{Gasser:1986vb}. In \cite{Colangelo:2005gd} they have
been computed to three loops for $M_\pi$ and two loops for $F_\pi$, up
to negligibly small exponential corrections. Since the expressions for
the latter involve
$O(p^4)$ LECs at two loops, some of
which we cannot self-consistently determine here, we prefer to rely on
the one-loop formulae, which can be written in terms of quantities
which we calculate directly. The difference is a correction on an
already small correction.

In the $\xi$-expansion, the one-loop finite-volume corrections are
given by \cite{Gasser:1986vb}:
\bea
 \frac{M_\pi^2(L)}{M_\pi^2}-1 & = & \frac12\xi\;\tilde{g}_1(M_\pi L)
+O\left(\xi^2\right)\labell{eq:Mpi2FV}\\
 \frac{F_\pi(L)}{F_\pi}-1 & = & -\xi\;\tilde{g}_1(M_\pi L)
+O\left(\xi^2\right)\labell{eq:FpiFV}
\eea
where $\xi$ is defined in \eq{eq:xi}. Analogous results are obtained
for the $x$-expansion.  The shape function
$\tilde{g}_1(x)$ has a well behaved large-argument expansion in terms
of Bessel functions of the second kind, which themselves can be
expanded asymptotically:
\bea
 \tilde{g}_1(x) &\stackrel{x\to\infty}{\sim}& \frac{24 K_1(x)}{x} + \frac{24 K_2(\sqrt{2}x)}{\sqrt{2}x} + \cdots  \\
 K_\nu(x) &\stackrel{x\to\infty}{\sim}& \left(\frac{\pi}{2x}\right)^{1/2}\exp(-x)\left[1 + \frac{4\nu^2-1}{8x}+\cdots\right]
\ . \eea

In a first instance, we include the corrections of \eq{eq:FpiFV} (and
the corresponding ones in the $x$-expansion), directly into the fit
functions given in \eq{eq:nloxipar} (and in \eq{eq:nloxpar} for the
$x$-expansion). We find that the subtraction of
finite-volume effects on $F_\pi$ significantly improves the fit
quality. The corrections on $M_\pi^2$, which are four times as small and
significantly smaller than statistical errors, 
do not improve the $\chi^2$ of the fits nor do they change the
results. 

For our simulation parameters, the one-loop finite-volume effects on
$F_\pi$ are typically 0.5\% and never exceed 1.1\%. Thus, higher-order
corrections are expected to be much smaller than our statistical
errors. To check this, we perform a second set of fits in which we
multiply the RHSs of each of the two equations in (\ref{eq:FpiFV})
(and the equivalent expressions in the $x$-expansion) by a coefficient
which is treated as an additional free parameter in these fits. Thus,
each of our $\xi$ and $x$-expansion fits have two additional
parameters. These parameters are 1 if the NLO estimate of
finite-volume corrections is exact.

In practice, for NLO fits in the important region $M_\pi\le
300\,\mev$, we find that the addition of these parameters does not
improve the quality of the fits. Moreover, the uncertainties on the
coefficients come out very large--between 80 and 90\%\ depending on
the quantity and the expansion--and the coefficients themselves are
consistent with 1 within at worst 1.2 standard deviations. Finally,
the results obtained for the LECs are consistent, within statistical
errors, with those obtained using the analytic finite-volume
expressions, and none of the conclusions that we draw below are
modified.

In light of these findings and of the expectation that higher-order,
finite-volume corrections are negligible compared to our statistical
errors, we have decided to fix the finite-volume corrections to their
NLO values in our analysis, so as not to artificially increase our
statistical errors by adding two irrelevant parameters.

\subsection{Illustrative bare results for the basic observables}

To conclude this section, we tabulate our simulation points, together
with the corresponding values of $Z_S\times am_{ud}$,
$a\Mss\equiv[2(aM_K)^2-M_\pi^2]^{1/2}$, $aM_\pi$ and
$aF_\pi/Z_A$. They are given in \tab{tab:data}. $Z_S\times am_{ud}$ is
the bare, subtracted value of the average up-down quark mass given by
the ratio-difference method described in \sec{sec:mass-and-decay},
before the final multiplicative renormalization. The quantities
in \tab{tab:data} are
the basic observables needed to study the chiral behavior of $M_\pi^2$
and $F_\pi$. Their central values are the fit quality weighted
averages of the results from the two different time-fit ranges of the
correlation functions and their statistical errors are the variance of
these central values over 2000 bootstrap samples. Their systematic
uncertainties are obtained from the variance over the two
procedures.

These values are only meant as illustrative. In particular, they do
not include a description of statistical and systematic error
correlations, including those with the lattice spacing. While this
seriously limits the reliability of any conclusion drawn from them, we
give them nonetheless so that the interested readers may get their own
sense of what sort of chiral behavior these results allow, after
combining them with the values of the lattice spacing $a$ and the
renormalization constants $1/Z_S$ and $Z_A$ given in \tab{tab:aZSZA}.

\begin{table}
\centering
{\footnotesize
\begin{tabular}{cccccccc} 
\hline
\hline
 $ \beta$ & $T\times L^3$ & $am_{ud}^\bare$ & $am_s^\bare$ & $Z_S (am_{ud})$ & $a\Mss$ & $a M_\pi$ & $aF_\pi/Z_A$ \\ \hline
& $48\times 24^3$ & -0.041 & -0.006 
& 0.01475(33) & 0.3415(5)(2) & 0.19188(50)(6) & 0.05491(34)(0) \\ 
& $48\times 24^3$ & -0.0437 & -0.006 
& 0.01188(27) & 0.3396(5)(2) & 0.17238(49)(3) & 0.05263(34)(0) \\ 
& $64 \times 24^3$ & -0.041 & -0.012 
& 0.01428(33) & 0.3175(95)(4) & 0.18790(90)(30) & 0.05384(84)(6) \\ 
& $64 \times 32^3$ & -0.0463 & -0.012 
& 0.00853(20) & 0.3134(10)(7) & 0.14440(70)(60) & 0.05004(62)(6) \\ 
3.5 & $64 \times 32^3$ & -0.048 & -0.0023
& 0.00726(17) & 0.3496(75)(5) & 0.13480(70)(20) & 0.04982(59)(1) \\ 
& $64 \times 32^3$ & -0.049 & -0.006
& 0.00579(15) & 0.3339(10)(5) & 0.12100(9)(3) & 0.04837(84)(3) \\ 
& $64 \times 32^3$ & -0.049 & -0.012
& 0.00560(14) & 0.3103(69)(9) & 0.11733(64)(3) & 0.04800(68)(2) \\ 
& $64 \times 48^3$ & -0.0515 & -0.012
& 0.00288(7) & 0.3079(9)(1) & 0.08410(60)(20) & 0.04628(58)(3) \\ 
& $64 \times 64^3$ & -0.05294 & -0.006
& 0.00149(5) & 0.3281(9)(5) & 0.06126(60)(9) & 0.04440(75)(6) \\ 
\hline
& $48 \times 32^3$ & -0.028 & 0.0045
& 0.01008(23) & 0.2955(6)(3) & 0.14852(49)(2) & 0.04408(34)(2) \\ 
& $48 \times 32^3$ & -0.03 & 0.0045
& 0.00808(18) & 0.2929(7)(3) & 0.13217(50)(9) & 0.04262(39)(1) \\ 
& $48 \times 32^3$ & -0.03 & -0.0042
& 0.00783(18) & 0.2602(7)(2) & 0.12943(59)(4) & 0.04207(39)(1) \\ 
3.61 & $48 \times 48^3$ & -0.03121 & 0.0045
& 0.00678(15) & 0.2926(6)(2) & 0.12096(30)(3) & 0.04234(25)(2) \\ 
& $48 \times 48^3$ & -0.033 & 0.0045
& 0.00490(12) & 0.2909(9)(3) & 0.10251(48)(7) & 0.04005(37)(1) \\ 
& $48 \times 48^3$ & -0.0344 & 0.0045
& 0.00344(8) & 0.2907(10)(4) & 0.08610(6)(2) & 0.03921(38)(6) \\ 
& $72 \times 64^3$ & -0.0365 & -0.003 
& 0.00096(3) & 0.2592(10)(5) & 0.04646(5)(3) & 0.03583(62)(0) \\ 
\hline
& $64 \times 32^3$ & -0.0208 & 0.0
& 0.00821(19) & 0.2276(13)(2) & 0.12455(11)(1) & 0.03660(57)(1) \\ 
& $64 \times 32^3$ & -0.0208 & 0.001
& 0.00831(19) & 0.2328(10)(2) & 0.12491(10)(1) & 0.03607(56)(1) \\ 
3.7 & $64 \times 32^3$ & -0.0208 & -0.005
& 0.00823(19) & 0.2082(7)(1) & 0.12489(64)(8) & 0.03588(42)(0) \\ 
& $64 \times 48^3$ & -0.0254 & 0.0
& 0.00354(8) & 0.2259(9)(3) & 0.08168(55)(2) & 0.03304(41)(2) \\ 
& $64 \times 48^3$ & -0.0254 & -0.005
& 0.00348(8) & 0.2043(6)(3) & 0.08046(4)(2) & 0.03270(51)(2) \\ 
& $64 \times 64^3$ & -0.027 & 0.0
& 0.00196(5) & 0.2235(52)(5) & 0.06029(30)(9) & 0.03303(44)(3) \\ 
\hline
& $64 \times 32^3$ & -0.0148 & 0.0
& 0.00915(21) & 0.1898(10)(3) & 0.12052(12)(3) & 0.03123(63)(1) \\ 
3.8 & $64 \times 48^3$ & -0.019 & 0.0
& 0.00422(9) & 0.1873(13)(2) & 0.08184(11)(0) & 0.02814(49)(1) \\ 
& $64 \times 48^3$ & -0.019 & 0.003 
& 0.00423(10) & 0.2010(11)(1)& 0.08261(14)(1) & 0.02784(47)(3) \\ 
& $144 \times 64^3$ & -0.021 & 0.0
& 0.00221(5) & 0.1879(80)(6) & 0.05981(25)(1) & 0.02688(59)(1) \\ 
\hline  
\hline  
  \end{tabular} 

}
\caption{Parameters of the simulations used in
  this work and illustrative results for the quantities $Z_S
  (am_{ud})$, $a\Mss$, $a M_\pi$ and $aF_\pi/Z_A$. In these results,
  the first error is statistical and the second is systematic, and
  they are obtained as described in the text.  For $Z_S
  (am_{ud})$,} the systematic error is 0 for the number of digits
  given and is not reported.  \labell{tab:data}
\end{table}

\clearpage

\section{Exploring the range of applicability of $SU(2)$ $\chi$PT for
  $M_\pi^2$ and $F_\pi$}
\labell{sec:su2validity}
In this section we explore the range of applicability of $SU(2)$
$\chi$PT, in $u$-$d$ and pion mass, for the various expansions discussed
in \sec{sec:intro}. We proceed in a systematic fashion. We begin by
assuming that $\chi$PT is valid around $M_\pi^\phys$, the experimental
value of $M_\pi$, where we have
our lightest points. We then study the $p$-values of the {\em combined,
fully-correlated} fit of the different chiral expansions to our
results for $F_\pi$ and $M_\pi^2$ with $m_{ud}\le m_{ud}^\max$
or $M_\pi\le M_\pi^\max$, as $m_{ud}^\max$ or
$M_\pi^\max$ is increased. Because our procedure correctly
accounts for all correlations in the lattice observables, the
$p$-value is a meaningful quantity whose value indicates the
probability that randomly chosen results consistent with the chiral
forms would give a worse fit. Thus we expect the $p$-value to drop as
$m_{ud}^\max$ or $M_\pi^\max$ is increased beyond the
range of applicability of a given $SU(2)$ $\chi$PT expansion for
$F_\pi$ and $M_\pi^2$. It is important to note, however, that the
sharpness of the drop and the conclusions which can be drawn depend on
the size of the error bars on the quantities studied.

In order to carry this program out on our simulation results, there
are two topics which we must address. The first is the dependence of
$M_\pi^2$ and $F_\pi$ on strange-quark mass. In our $N_f=2+1$
simulations, we vary $m_s$ in the vicinity of its real-world value to
allow us to tune it precisely to that value in our final results. To
parametrize this mass dependence we follow
\cite{Lellouch:2009fg} for instance, and expand the LECs of 
$SU(2)$ $\chi$PT in power series in the strange quark mass, or an
equivalent variable such as $\Mss^2\equiv 2M_K^2-M_\pi^2$, around the
physical strange quark point. Since these corrections are small, they
are usually only visible in the LO terms of the chiral expansion. For
instance, terms of order $x$ or $\xi$ times $(m_s-m_s^\phys)/M_\qcd$,
where $M_\qcd$ is a scale characteristic of QCD (e.g. the $\rho$-meson
mass $M_\rho$), are not detectable at our level of precision.  We will
retain only those terms whose coefficients differ from zero by more
than one standard deviation in our fits.

The second point that must be addressed is that of discretization
errors. At finite lattice spacing, results for $M_\pi$, $F_\pi$ and
the renormalized quark masses suffer from discretization errors which
are proportional to powers of $a$, up to logarithms. Because the
fermion action that we use is tree-level $O(a)$-improved, the leading
such errors are formally proportional to $\alpha_s(a)a$. However, at
our coarsest lattice spacing, terms proportional to $a^2$ may be
dominant. Thus, we consider both possibilities in our analysis.  We
find that our fits work better if we consider that discretization
errors are associated with a given lattice quantity and consistently
include the required corrections every time that quantity
appears. This is what is done in (\reff{eq:nloxpar}) for $m_{ud}$, for
instance. In fact, we performed an extensive study of these effects and found
that the only discretization corrections which our results are
sensitive to are $\alpha_s a$ or $a^2$ corrections in
$m_{ud}$. Attempts to add discretization corrections to $M_\pi$ or
$F_\pi$ always lead to coefficients which were consistent with zero
within less than one standard deviation. Thus, in the sequel, we keep
only discretization corrections on the light-quark mass.

\subsection{NLO and NNLO chiral fit strategy}
\labell{sec:fit-strategy}

Combining the strange-quark-mass and lattice-spacing dependencies,
discussed above, with the $SU(2)$ chiral expansions of \sec{sec:intro}
gives the desired NLO and NNLO parametrizations. At NLO in the
$x$-expansion, we obtain the following expressions for the lattice
quantities $(aM_\pi)^2/2(am_{ud})$, $(aF_\pi)$, $am_{ud}$,
$(a\Mss)^2$, $a$, $Z_A$ and $Z_S$:
\bea
  \frac{(aM_\pi)^2}{2(am_{ud})} &=& \frac{a^p}{Z_S^p}(1 - \gamma_1^a f(a^p) + \gamma_1^s(\Delta \Mss^2)^p)
  (B_\pi^{x-\NLO})(m_{ud}^p;B,F,\bar\ell_3)\,,\nonumber\\
\labell{eq:nloxpar}
  (aF_\pi) &=& \frac{a^p}{Z_A^p}(1
  + \gamma_2^s(\Delta \Mss)^2)F_\pi^{x-\NLO}(m_{ud}^p;B,F,\ell_4)\,,
\eea
$$
(am_{ud}) = a^pZ_s^p(1+\gamma_1^af(a^p)) m_{ud}^p\,,\;\;
(a\Mss)^2  = (a^p)^2 (\Mss^2)^p\,,$$
$$a=a^p\,,\;\; Z_A=Z_A^p\,,\;\; Z_S=Z_S^p, 
$$
where $(a\Mss)^2\equiv 2(aM_K)^2-(aM_\pi)^2$,
$(\Delta \Mss^2)^p\equiv(\Mss^2)^p-\Mss^\phys$ and
$f(a)=\alpha_s(a)\,a$ or $a^2$, depending on which discretization
errors are chosen as
leading. $B_\pi^{x-\NLO}(m_{ud}^p;B,F,\bar\ell_3)$ is $B$ times the
NLO part of the expression in brackets on he RHS of the first equation
in (\ref{eq:MF}) and
$F_\pi^{x-\NLO}(m_{ud}^p;B,F,\ell_4)$ are the NLO expressions
of \eq{eq:MF}. The relevant chiral parameters of the fit are the 2 LO
LECs, $B$ and $F$, and the 2 NLO LECs, $\bar\ell_3$ and
$\bar\ell_4$. There are also a discretization parameter, $\gamma_1^a$,
and the strange-mass dependence parameters $\gamma_1^s$ and
$\gamma_2^s$. An extensive study of these two effects showed that the
only relevant ones are those given by $\gamma_1^a$, $\gamma_1^s$ and
$\gamma_2^s$.

In \eq{eq:nloxpar}, variables with a superscript $p$ are also
parameters of the fit. These are associated with the corresponding
lattice quantities. As in our previous work, they are
added so that uncertainties and correlations in all lattice
quantities, including those which appear in nontrivial expressions involving the
parameters, can consistently be accounted for in the
$\chi^2$. Since there is one such variable per new observable added, the
total number of d.o.f.\ is unchanged.

For each $\beta$ we define the large lattice data vector
$y^T(\beta)=(a, Z_A, Z_S, am_{ud},(a\Mss)^2, 2(am_{ud})/(aM_\pi)^2,
(aF_\pi), \cdots)$ where the quantities
$am_{ud}$, $(a\Mss)^2$, $2(am_{ud})/(aM_\pi)^2$, $(aF_\pi)$ are
repeated for every simulation at that lattice spacing. We then use a
bootstrap to compute a correlation matrix $C_{ij}(\beta)$ for each
$\beta$ between different components $i$ and $j$ of the vector
$y$. Because simulations are independent, this matrix is essentially
block diagonal per simulation, in blocks corresponding to a set of
quantities $(am_{ud})^2,\cdots, (aF_\pi)$. There will be
large correlations within a given simulation block and smaller,
respectively much smaller, ones between these blocks and 
the lattice spacing, respectively the renormalization constants. We
then construct the fully correlated $\chi^2$ through
$\chi^2=\sum_{\beta} X^T(\beta) C^{-1}(\beta) X(\beta)$, where $X(\beta)$
is the vector constructed from the difference of $y(\beta)$ and the
expressions on the RHS sides of (\ref{eq:nloxpar}), appropriately
repeated for each simulation. This construction guarantees that the
$p$-value that we obtain for these fits accounts for all uncertainties
and correlations.

In \fig{fig:xnloeg} we show a typical NLO, $x$-expansion fit of
 $M_\pi^2$ and $F_\pi$. Points with $M_\pi\gsim 120\,\mev$
 (i.e. $m_{ud}\gsim 3.7\,\mev$) but less than
$M_\pi^\max=300\,\mev$ (i.e. $m_{ud}\sim 23.\,\mev$) are included in the combined, correlated
fit. Agreement of the NLO expressions with the lattice results is
excellent in this range. However the corresponding curves start deviating
significantly from the lattice results for larger values of $M_\pi$.

\begin{figure}[t]
 \centering 
\includegraphics[width=0.8\textwidth]{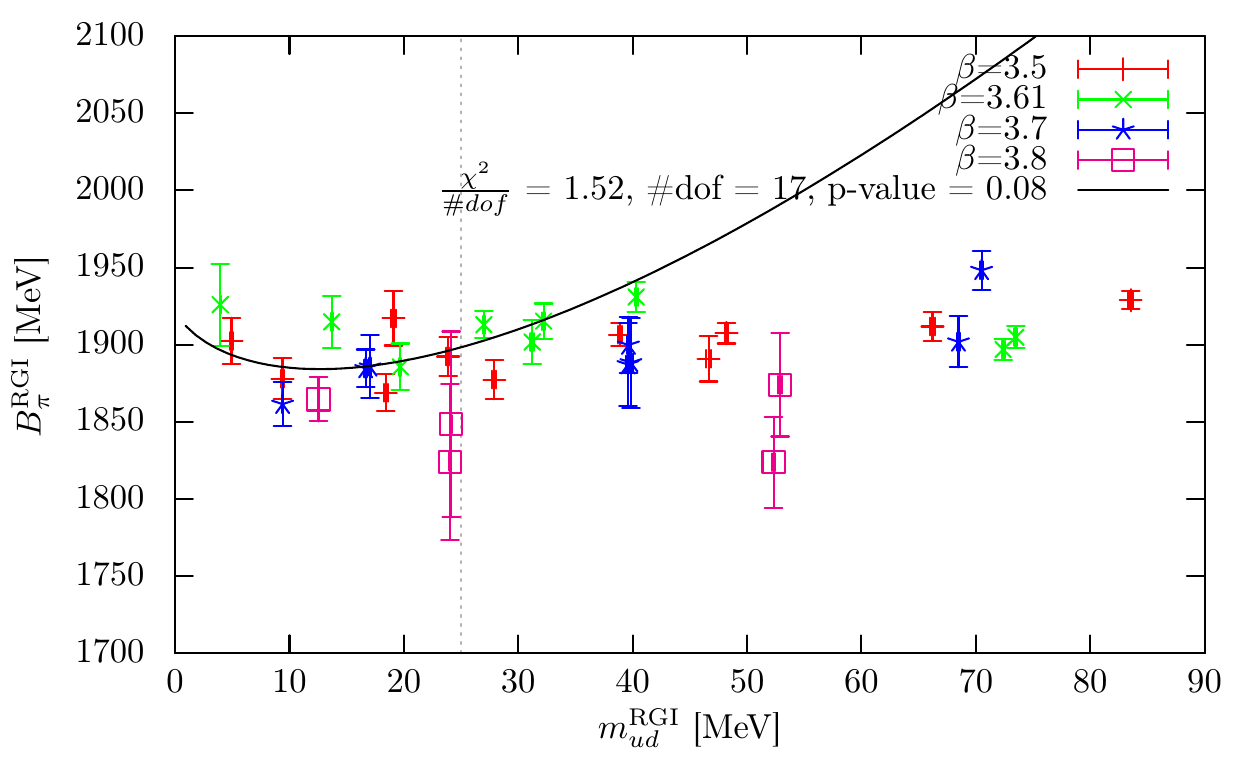}
\includegraphics[width=0.8\textwidth]{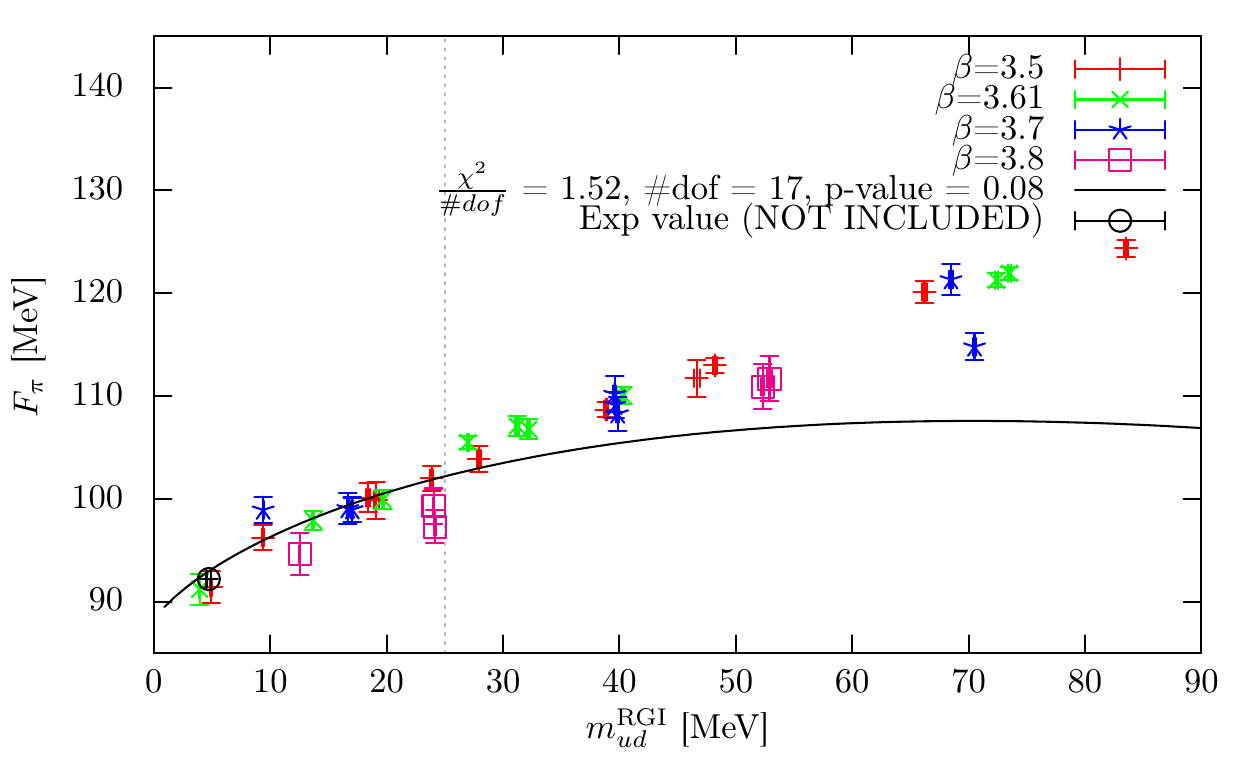} \caption{\sl Typical
  NLO $SU(2)$ $\chi$PT fit (curves) of our lattice results (points
  with error bars) for $B_\pi=M_\pi^2/(2m_{ud})$ and $F_\pi$ as functions of $m_{ud}$,
  in the $x$-expansion. These are fully correlated fits to the NLO
  expressions of (\ref{eq:nloxpar}), which also account for
  discretization and strange quark mass corrections. Only points with
  $M_\pi\le M_\pi^\max=300\,\mev$ (i.e. $m_{ud}\sim 23.\,\mev$) are included in the fits, i.e. those
  left of the dashed vertical line. The more massive points are shown for
  illustration. The lattice results in the figure are corrected for
  discretization and strange mass contributions, using the fit
  parameters obtained. Thus, they are continuum limit results at the
  physical value of $m_s$ and their only residual dependence is on
  $m_{ud}$. Nevertheless, results obtained at different lattice
  spacing are plotted with different symbols. The fact that they lie
  on a same curve indicates that residual discretization errors are
  negligible. Note that the corrections made to the more massive
  points may not be optimal as these points are not included in the fit and,
  as we will see, the applicability of NLO $\chi$PT is questionable
  for these points. Error bars on all points are statistical
  only. Also shown, but not included in the fits, is the experimental
  value of $F_\pi$ \cite{Beringer:1900zz}. Agreement with our results
  computed directly around the physical pion mass point is
  remarkable.} \labell{fig:xnloeg}
\end{figure}

For the NLO $\xi$-expansion, we perform a very similar
construction. Here, however, the lattice data are $(aM_\pi)^2$,
$(a\Mss)^2$, $2(am_{ud})/(aM_\pi)^2$, $(aF_\pi)$, $a$, $Z_A$ and
$Z_S$, and the corresponding NLO expressions are:
\bea
  \frac{2(am_{ud})}{(aM_\pi)^2} &=& \frac{Z_S^p}{a^p}(1 + \gamma_1^a f(a^p) + \gamma_1^s(\Delta \Mss^2)^p)/
  B_\pi^{\xi-\NLO}((M_\pi^2)^p;B,F,\bar\ell_3)\,, \nonumber \\
\labell{eq:nloxipar}
  (aF_\pi) &=& \frac{a^p}{Z_A^p}(1 + \gamma_2^s(\Delta \Mss)^2)F_\pi^{\xi-\NLO}((M_\pi^2)^p;B,F,\ell_4)\,,\\
\eea
$$
(aM_{\pi})^2 = (a^p)^2 (M_{\pi}^2)^p\,,\;\;     (a\Mss)^2  = (a^p)^2
(\Mss^2)^p\,,
$$
$$
a=a^p\,,\;\; Z_A=Z_A^p\,,\;\; Z_S=Z_S^p, 
$$
where $1/B_\pi^{\xi-\NLO}((M_\pi^2)^p;B,F,\bar\ell_3)$ is $1/B$ times
the NLO part of the expression in brackets on the RHS of the first
equation in
(\reff{eq:MpiFpi}). $F_\pi^{\xi-\NLO}((M_\pi^2)^p;B,F,\ell_4)$ is the
expression obtained by solving exactly the NLO part of the second
equation in (\reff{eq:MpiFpi}) for $F_\pi$, and keeping the physical
solution. This equation is quadratic in $F_\pi$ and the existence of a
physical solution is not guaranteed. The existence of such a solution
imposes a constraint on the LO and NLO $SU(2)$ $\chi$PT parameters,
which we take into account in our fits. We discuss these solutions and
constraints in more detail in \app{sec:xi-constraints}.

We show typical NLO $\xi$-expansion fits
in \fig{fig:xinloeg}. Again, only points with $M_\pi$ less than
$M_\pi^\max=300\,\mev$ are included. The behavior found here is quite
similar to the one found above for the NLO $x$-expansion, with the
fit curves agreeing well with the lattice results in the fit range,
but deviating more and more beyond that. However, the deviations
beyond $M_\pi^\max=300\,\mev$ are slightly less pronounced than in the
$x$-expansion. This is probably a demonstration of the statement made
in the Introduction, that the $\xi$-expansion resums some higher-order
physical contributions.

\begin{figure}[t]
 \centering 
\includegraphics[width=0.8\textwidth]{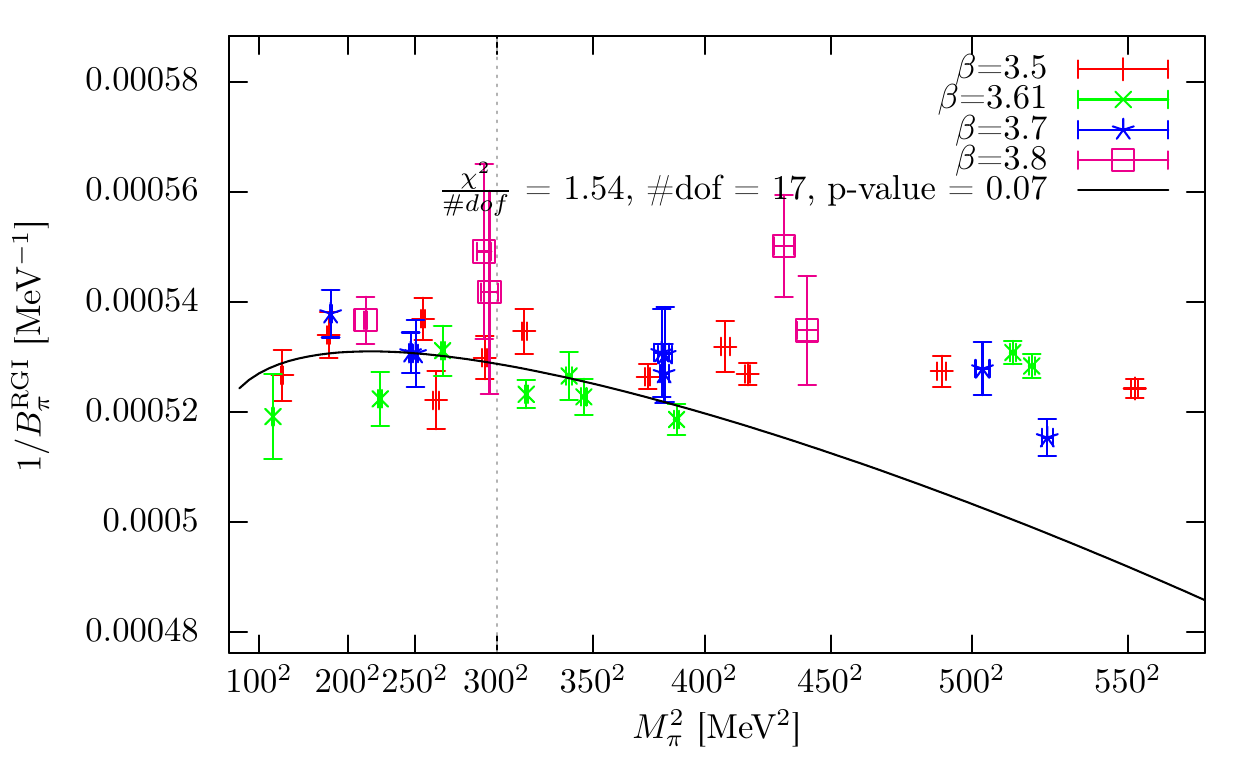} 
\includegraphics[width=0.8\textwidth]{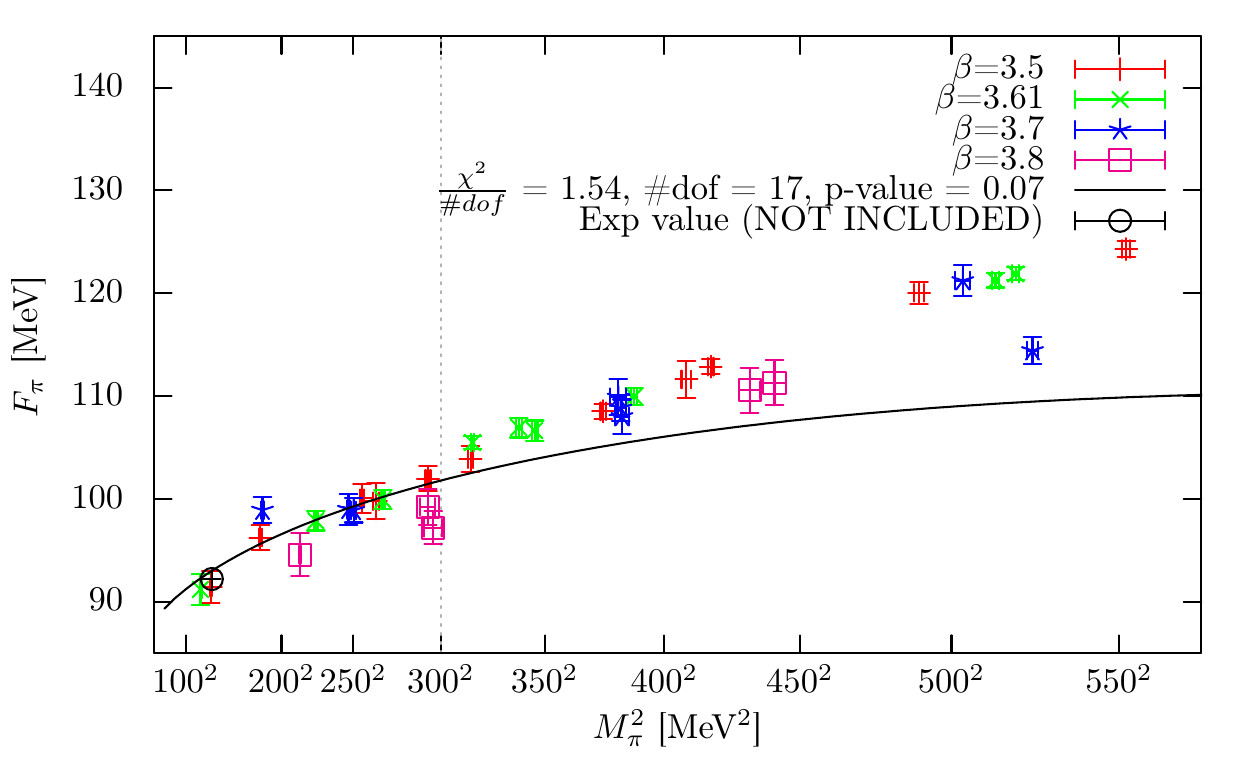} 
\caption{\sl Typical
  NLO $SU(2)$ $\chi$PT fit (curves) of our lattice results (points
  with error bars) for $1/B_\pi$ and $F_\pi$ as functions of $M_\pi^2$,
  in the $\xi$-expansion. Only points with
  $M_\pi\le M_\pi^\max=300\,\mev$ are included in the fits, i.e. those
  left of the dashed vertical line. The description is the same as
  in \fig{fig:xnloeg}, except that the functional forms used are those
  of (\ref{eq:nloxipar}).}  
\labell{fig:xinloeg}
\end{figure}

We now turn to NNLO fits. The procedure followed here is identical to
the one described above for NLO fits, except that the NLO expressions
in \eqs{eq:nloxpar}{eq:nloxipar} are replaced by the appropriate NNLO
expressions from \sec{sec:intro}. That is $B_\pi^{x-\NLO}(m_{ud}^p; B,
F, \bar\ell_3)$, $F_\pi^{x-\NLO}(m_{ud}^p;B, F, \bar\ell_4)$,
$B_\pi^{\xi-\NLO}((M_\pi^2)^p;B, F, \bar\ell_3)$ and
$F_\pi^{\xi-\NLO}((M_\pi^2)^p;B, F, \bar\ell_4)$ are replaced by
$B_\pi^{x-\NNLO}(m_{ud}^p;B, F, \bar\ell_3, \bar\ell_{12}, k_M)$,
$F_\pi^{x-\NNLO}(m_{ud}^p;B, F, \bar\ell_4, \bar\ell_{12}, k_F)$,
$B_\pi^{\xi-\NNLO}((M_\pi^2)^p;B, F, \bar\ell_3, \bar\ell_{12}, c_M)$
and $F_\pi^{\xi-\NNLO}((M_\pi^2)^p;B, F, \bar\ell_4, \bar\ell_{12},
c_F)$. Thus, in addition to the 4 $\chi$PT parameters required in the
NLO fits, the NNLO expressions contain 5 additional chiral parameters:
$\bar\ell_{12}$, $k_M$ and $k_F$ for the $x$-expansion and
$\bar\ell_{12}$, $c_M$ and $c_F$ for the $\xi$-expansion.

$F_\pi^{\xi-\NNLO}((M_\pi^2)^p;B, F, \bar\ell_4, \bar\ell_{12}, c_F)$
is the expression obtained by solving exactly the quartic, second
equation in (\ref{eq:MpiFpi}) for $F_\pi$, and by keeping the physical
solution. Again, the existence of a physical solution imposes
constraints on the LO, NLO and now NNLO $SU(2)$ $\chi$PT parameters.
We take these constraint into account in our
fits. In \app{sec:xi-constraints} we give the physical solution and
discuss the conditions for its existence in more detail.

\begin{figure}[t]
 \centering
 \includegraphics[width=0.8\textwidth]{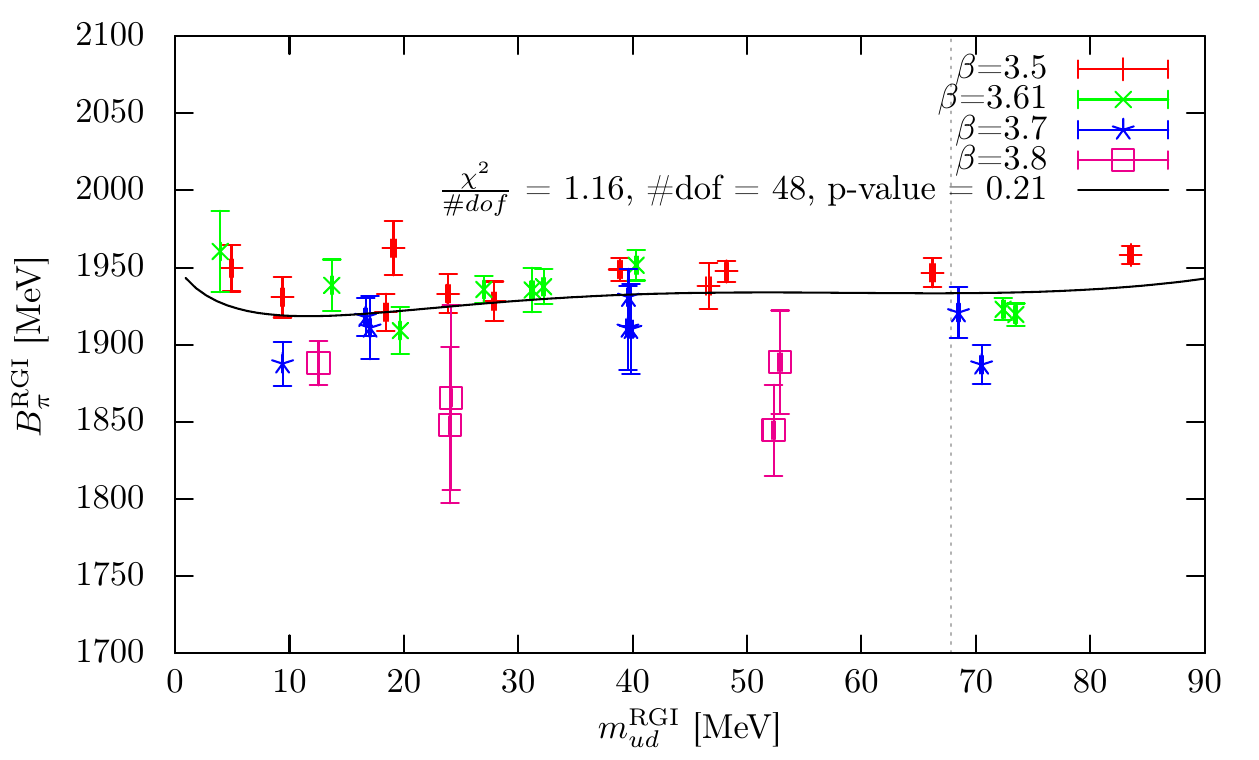}
 \includegraphics[width=0.8\textwidth]{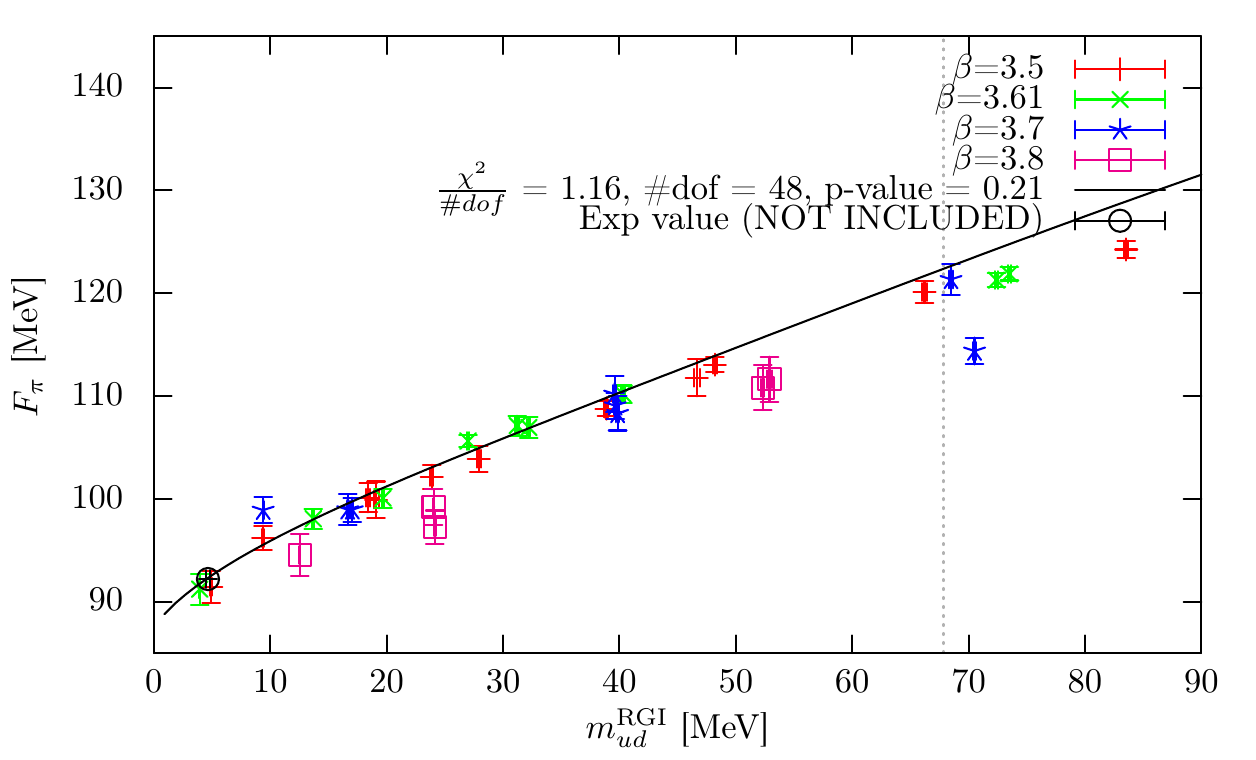}
\caption{\sl Typical
 NNLO $SU(2)$ $\chi$PT fit (curves) of our lattice results (points
  with error bars) for $B_\pi$ and $F_\pi$ as functions of $m_{ud}$,
  in the $x$-expansion. Only points with
  $M_\pi\le M_\pi^\max=500\,\mev$ (i.e. $m_{ud}^\max\lsim 65.\,\mev$) are included in the fits, i.e. those
  left of the dashed vertical line. The description is the same as
  in \fig{fig:xnloeg}, except that the functional forms used are those
  of (\ref{eq:nloxpar}) with $B_\pi^{x-NLO}$ and $F_\pi^{x-NLO}$
  replaced by $B_\pi^{x-NNLO}$ and $F_\pi^{x-NNLO}$, respectively.}  
 \labell{fig:xnnloeg}
\end{figure}

Defining the $\chi^2$ as we do for the NLO fits, we perform fully
correlated, NNLO $x$ and $\xi$-expansion fits to $M_\pi^2$ and
$F_\pi$, with $M_\pi^\max$ between 400 and 550~MeV. A typical example
of such a fit is shown in \fig{fig:xnnloeg} for the $x$-expansion, and
in \fig{fig:xinnloeg} for the $\xi$-expansion, both for
$M_\pi^\max=500\,\mev$. The $p$-values of these fits are
excellent. The agreement with the lattice results is also visibly very
good and extends better beyond $M_\pi^\max$ than in the NLO case. In
both the $x$ and $\xi$-expansions, the NNLO serves to cancel the
curvature of the NLO forms to give a more linear behavior of the mass
dependence of $M_\pi^2$ and $F_\pi$.

\begin{figure}[t]
 \centering
  \includegraphics[width=0.8\textwidth]{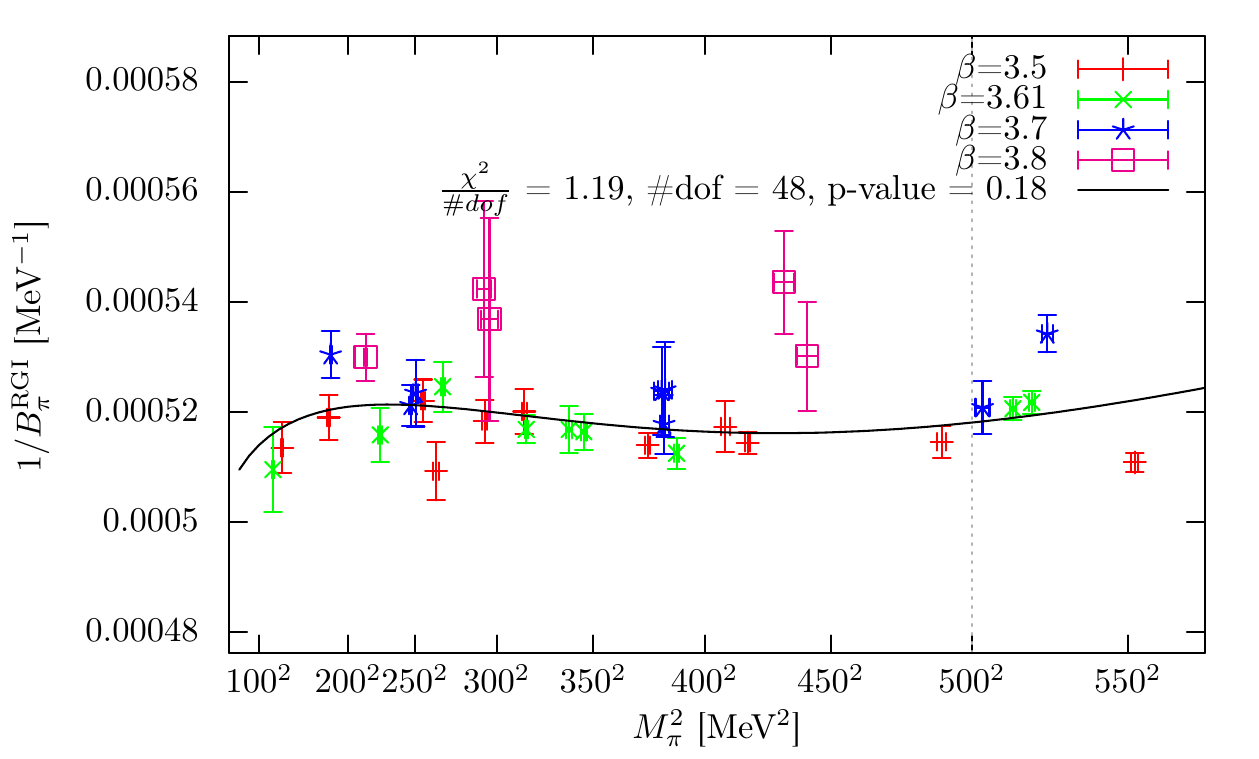}  
 \includegraphics[width=0.8\textwidth]{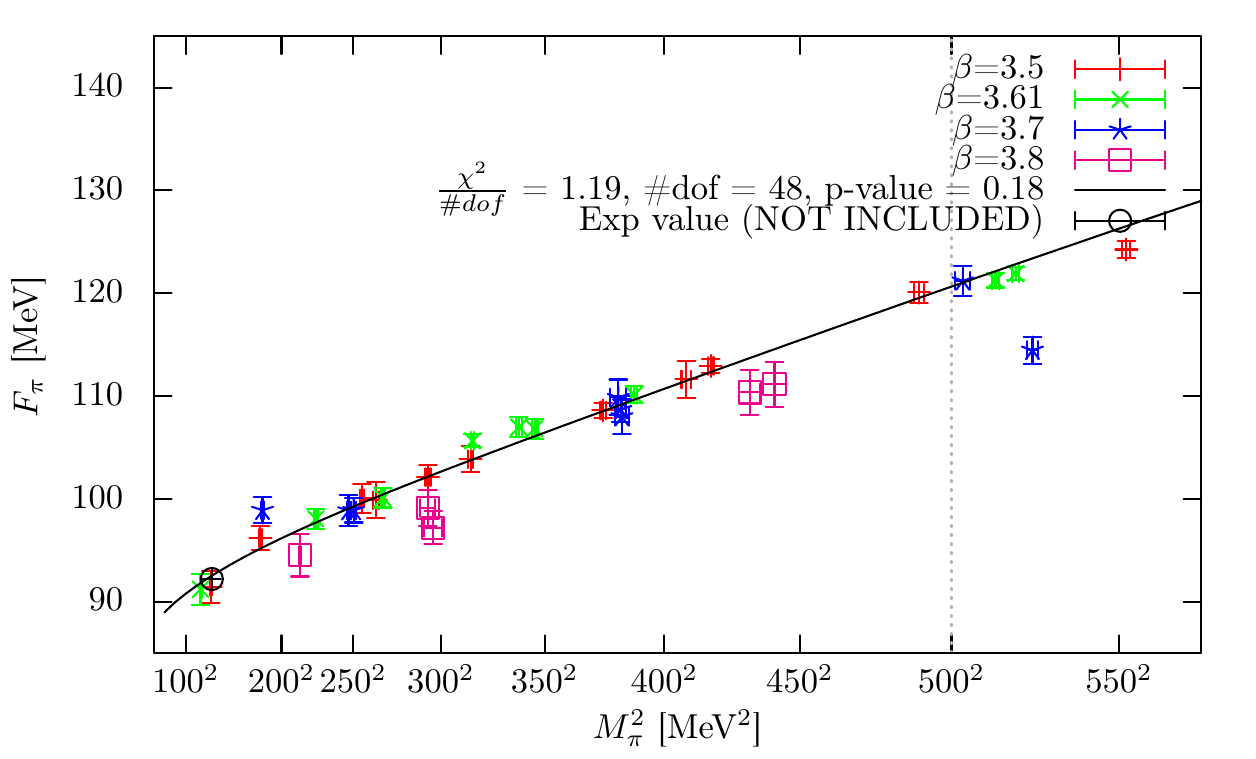}
\caption{\sl Typical
 NNLO $SU(2)$ $\chi$PT fit (curves) of our lattice results (points
  with error bars) for $1/B_\pi$ and $F_\pi$ as functions of $M_\pi^2$,
  in the $\xi$-expansion. Only points with
  $M_\pi\le M_\pi^\max=500\,\mev$ are included in the fits, i.e. those
  left of the dashed vertical line. The description is the same as
  in \fig{fig:xnloeg}, except that the function forms used are those
  of (\ref{eq:nloxipar}) with $B_\pi^{\xi-\NLO}$ and $F_\pi^{\xi-\NLO}$
  replaced by $B_\pi^{\xi-\NNLO}$ and $F_\pi^{\xi-\NNLO}$, respectively.}  
 \labell{fig:xinnloeg}
\end{figure}

\clearpage

\subsection{Fit quality and LECs in terms of maximum pion mass for 
NLO  $\chi$PT}
\labell{sec:nlo-fitquality}

We now turn to our systematic study of the range of applicability of
$SU(2)$ chiral perturbation theory to the quark-mass dependence of
$M_\pi^2$ and $F_\pi$. We implement the fully correlated, combined
fits described above, including lattice results extending from our
smallest pion mass of around 120~MeV up to a maximal value,
$M_{\pi}^\max$. We then study the $p$-value of these fits as a
function of $M_{\pi}^\max$. We consider NLO $x$ and $\xi$-expansion
fits in this section and NNLO ones in the following. For the
$x$-expansion fits, the cut is made at a value of $m_{ud}$
corresponding to $M_\pi^\max$ such that the same lattice results are
include as would be with a cut at $M_\pi^\max$ in the $\xi$-expansion
fits.

For each value of $M_{\pi}^\max$ and for each functional form tried,
we compute the fit quality, including a systematic error. Indeed, we
want to make sure that the $p$-value which we quote is not peculiar to
a particular choice of analysis procedure. This is particularly
important in fits, such as those performed here, where the observables
considered have significant correlations and small changes can make
large changes in the fit quality. The
$p$-values are obtained from the $p$-value-weighted distributions of
results from $2\times 2 \times 3 \times 6 =72$ different analysis
procedures for a given $M_{\pi}^\max$. These procedures correspond to
$2$ time-fit intervals for the two-point functions, 2 mass cuts in the
scale setting, 3 ways of doing RI/MOM renormalization for $Z_A$ and 6
for $Z_S$, as described in \sec{sec:latticedetails}. The central value
of the fit quality for a given $M_{\pi}^\max$ is chosen as the mean of
the corresponding distribution and its systematic error obtained from
this distribution's variance.

\begin{figure}[t]
 \centering \includegraphics[width=0.9\textwidth]{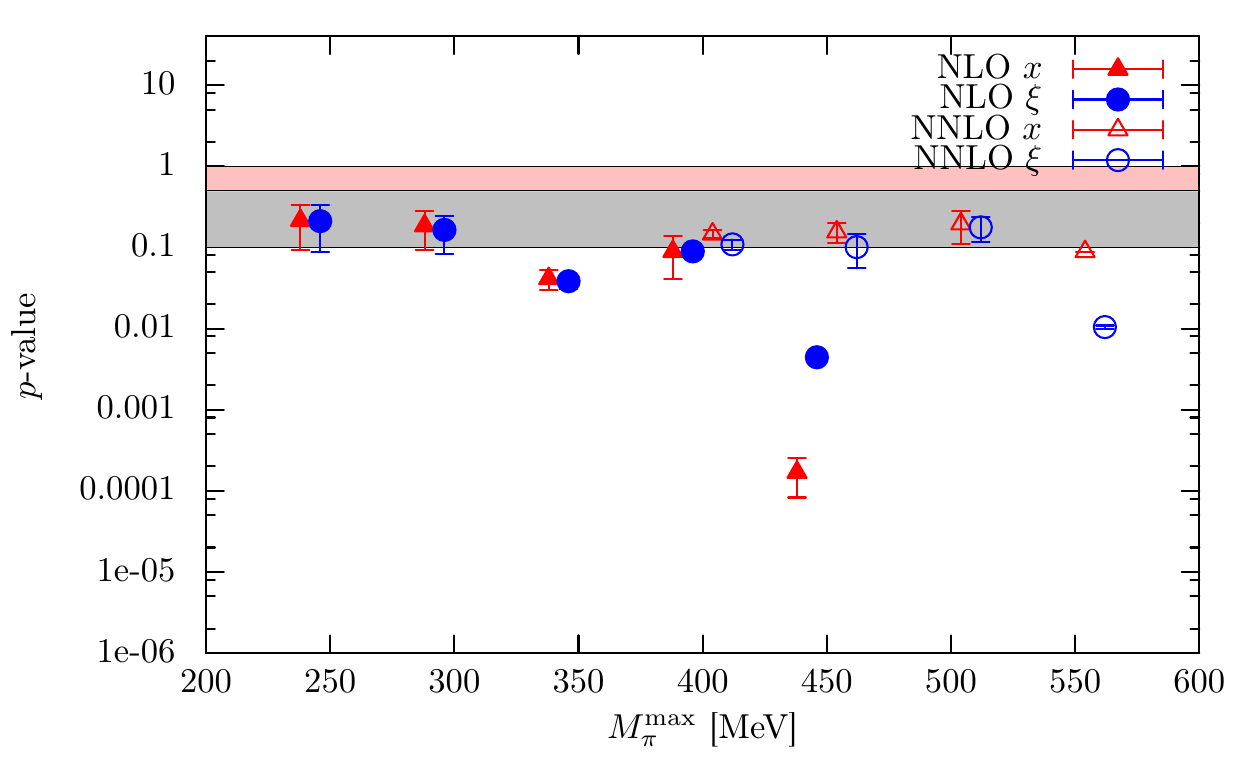} 
\caption{\sl Fit
 quality of the fully correlated $SU(2)$ $\chi$PT fits to our lattice
 results for $B_\pi$ and $F_\pi$ described
 in \sec{sec:fit-strategy}. The fits to our lattice results for these
 quantities include points whose pion mass is in the range
 $[120\,\mev,\,M_\pi^\max]$. The $p$-values shown are those of NLO and
 NNLO fits in the $x$ and $\xi$-expansions. They are are plotted as a
 function of $M_\pi^\max$. For $M_\pi^\max\le 350\,\mev$, only the
 $p$-values of NLO fits are plotted as these ranges do not contain
 enough data to constrain NNLO chiral expressions. For
 $M_\pi^\max\in[400,\,450]\,\mev$, the $p$-values of both NLO and NNLO
 fits are shown. For larger $M_\pi^\max$ only NNLO results are shown,
 as the $p$-values of NLO fits are negligibly small. The gray band
 corresponds to the $p$-value interval of 10 to 50\% and the red one
 to that of 50 to 100\%. Error bars on each point are the systematic
 uncertainties described in the text. For the sake of clarity, results
 are shifted about the values of $M_\pi^\max=250,\cdots\,\mev$, at
 which they are obtained. }
\labell{fig:pvalue}
\end{figure}

The results for the the $p$-values of our NLO and NNLO, $x$ and
$\xi$-expansion fits are shown together in \fig{fig:pvalue}. For the
NLO fits we consider values of $M_\pi^\max$ between 250 and
450~MeV. Below 250~MeV the number of lattice points which we have
starts becoming too small to reliably constrain the NLO form. Above
450~MeV, these fits have tiny $p$-values.

As \fig{fig:pvalue} shows, the NLO $x$ and $\xi$-expansion fits work
very well for $M_\pi^\max\le 300\,\mev$. There is a first drop in
$p$-value for $M_\pi^\max$ in the region of 350 to 400~MeV in which
fit qualities are in the 1 to 10\% range. Between 400 and 450~MeV the
fit quality drops enormously and keeps on doing so beyond that point
(not shown). We have checked that these changes are not the artefact
of a single stray point in these intervals. This discussion
suggests that, for $M_\pi^2$ and $F_\pi$, the range of validity of
$SU(2)$ extends safely up to 300~MeV and may be stretched up to around
400~MeV. Beyond that point it clearly breaks down. Of course, these
conclusions only hold within the statistical accuracy of our
calculation, which is described in more detail in \sec{sec:rel-xpt-contribs}.

It is worth noting that the break-down is less pronounced for the NLO
$\xi$-expansion. This may be ascribed in part to a difference in size
in the relative uncertainties on $M_\pi^2$ and $m_{ud}$. It also seems to
corroborate the observation, made in \sec{sec:fit-strategy}, that the
$\xi$-expansion range of applicability may extend to slightly
larger quark-mass values because it resums some higher-order physical
contributions.

In order to further verify the conclusions drawn up to now, we also
monitor the values of the fitted LECs, as a function of
$M_\pi^\max$. We begin with the LO LECs $B$ and $F$. Their values as a
function of $M_\pi^\max$ are shown in \fig{fig:lo-lec-vs-mpicut} for
the NLO $x$ and $\xi$-expansions. These values include full
statistical and systematic errors, obtained with the same collection
of analyses as those used in determining the $p$-values. For each
quantity, we weigh the result given in each procedure by its
$p$-value. This yields a distribution of results for each
quantity. The central value for each quantity is chosen to be the mean
of the distributions. Its systematic uncertainty is obtained by
computing the variance with respect to the mean, over this
distribution. Finally, the statistical error is obtained by repeating
the construction of the distributions for the 2000 bootstrap samples,
and considering the variance of these means around the central value.

As the
plots show, the LO LECs obtained from NLO fits jump for $M_\pi^\max$
between 300 and 350~MeV, but appear to remain consistent within
errors. However, because the values of the LECs for two different pion
mass cuts are obtained from data sets which have significant overlap,
they are correlated, which may give a false impression of
agreement. In order to eliminate the effect of these correlations in
the comparison, we study the quantities $\Delta B^\RGI$ and $\Delta F$,
which are the differences of the LECs at the given value of
$M_\pi^\max$ minus the ones obtained for $M_\pi^\max=300\,\mev$. The
latter is chosen because it is clearly within the range of
applicability of $SU(2)$ $\chi$PT, at the level of accuracy considered
here. We compute the statistical and systematic errors directly on
these differences, both within our bootstrap resampling and systematic
error analysis loops. The errors on these differences determine
directly the significance of the deviations of the values of the LECs
obtained for a given $M_\pi^\max$ with that obtained for
$M_\pi^\max=300\,\mev$. These differences are plotted
in \fig{fig:lo-lec-vs-mpicut}, in a panel below the corresponding
LEC. By definition, $\Delta B^\RGI$ and $\Delta F$ are exactly zero at
$M_\pi^\max=300\,\mev$.

\begin{figure}[t]
 \centering 
\includegraphics[width=0.8\textwidth]{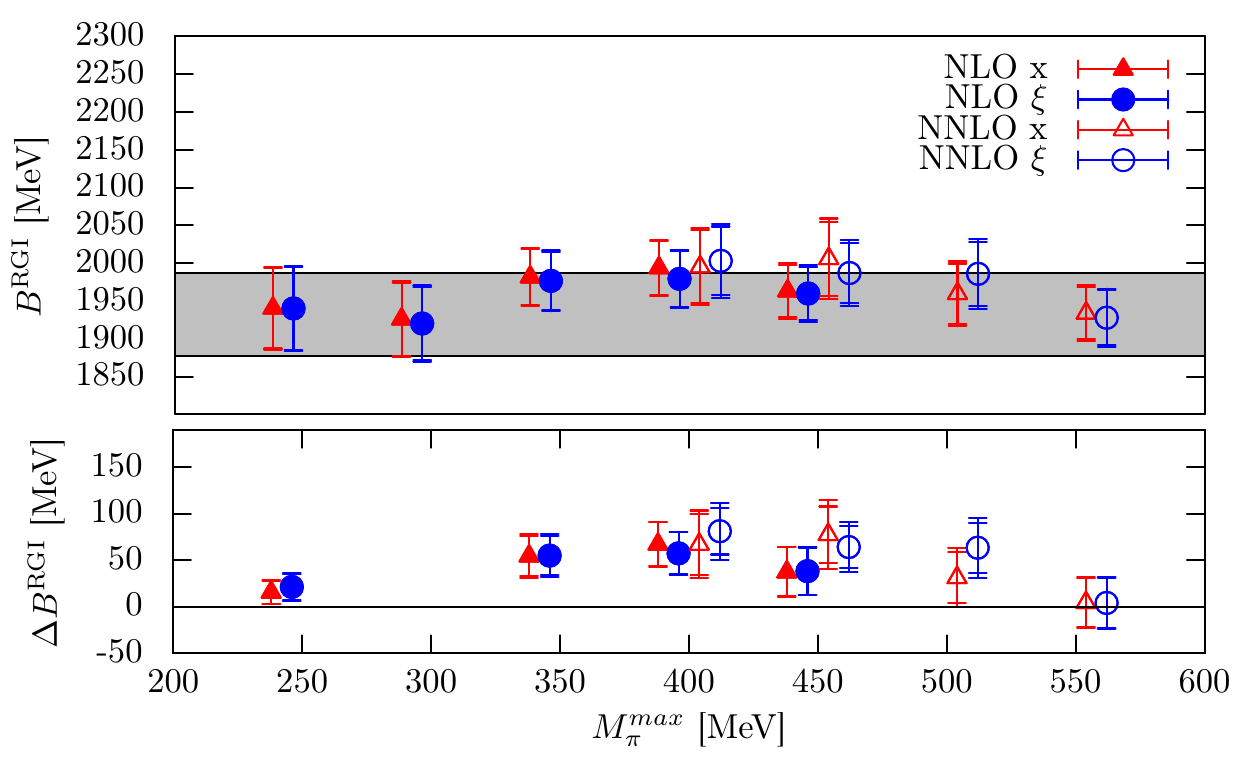} 
\includegraphics[width=0.8\textwidth]{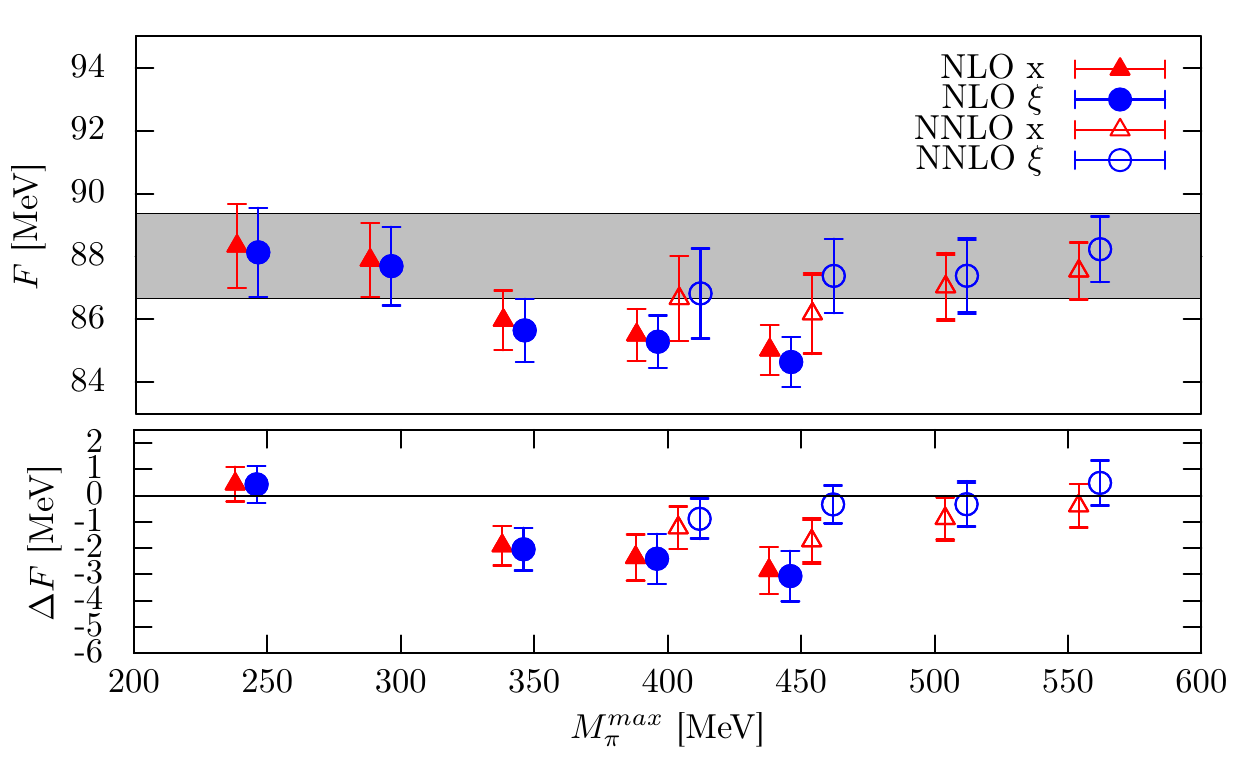} 
\caption{\sl LO
 LECs as a function of $M_\pi^\max$ obtained from the $SU(2)$ $\chi$PT
 fits to our lattice results for $B_\pi$ and $F_\pi$ in the pion-mass
 range $[120\,\mev,\,M_\pi^\max]$, as described
 in \sec{sec:fit-strategy}. Results are shown for NLO and NNLO fits in
 the $x$ and $\xi$-expansions (see the caption
 of \protect\fig{fig:pvalue} for additional details). In the top panel
 of each of the two figures, it is the LEC in physical units which is
 shown. The horizontal gray band denotes our final result for the
 corresponding LEC, given in \tab{tab:final-lec-results}, and obtained
 as described in \sec{sec:lecresults}. In the lower panel of each
 figure it is the difference of the LEC obtained from a fit with
 $M_\pi\in[120\,\mev,\,M_\pi^\max]$ to that obtained from the NLO fit
 in the range $[120,\,300]\,\mev$, in the corresponding expansion. As
 argued in the text, this reference domain is in the range of
 applicability of NLO $\chi$PT at our level of accuracy. Error bars on
 each point are the statistical and the quadratically combined
 statistical-plus-systematic uncertainties. For the sake of clarity,
 results are shifted about the values of
 $M_\pi^\max=250,\cdots\,\mev$, at which they are obtained. }
\labell{fig:lo-lec-vs-mpicut}
\end{figure}

The plots of these differences show that the seeming agreement deduced
from a direct comparison of the values of the LECs at two different
pion mass cuts is misleading. While one finds that the values of the
LO LECs for $M_\pi^\max=250$ and 300~MeV agree within one standard
deviation, this is no longer true for values of $M_\pi^\max\ge
350\,\mev$. Indeed, the values of $\Delta B^\RGI$ and $\Delta F$ are
almost 2 standard deviations away from 0 and more for $M_\pi^\max=
400\,\mev$. Beyond that point, the values of the
LECs obtained from NLO fits are not meaningful, because the quality of
the fits becomes so poor. The results on these
differences sharpen the earlier conclusion that NLO, $SU(2)$ $\chi$PT
starts breaking down above 300~MeV, for the precisions reached here.

\begin{figure}[t]
\centering
 \includegraphics[width=0.8\textwidth]{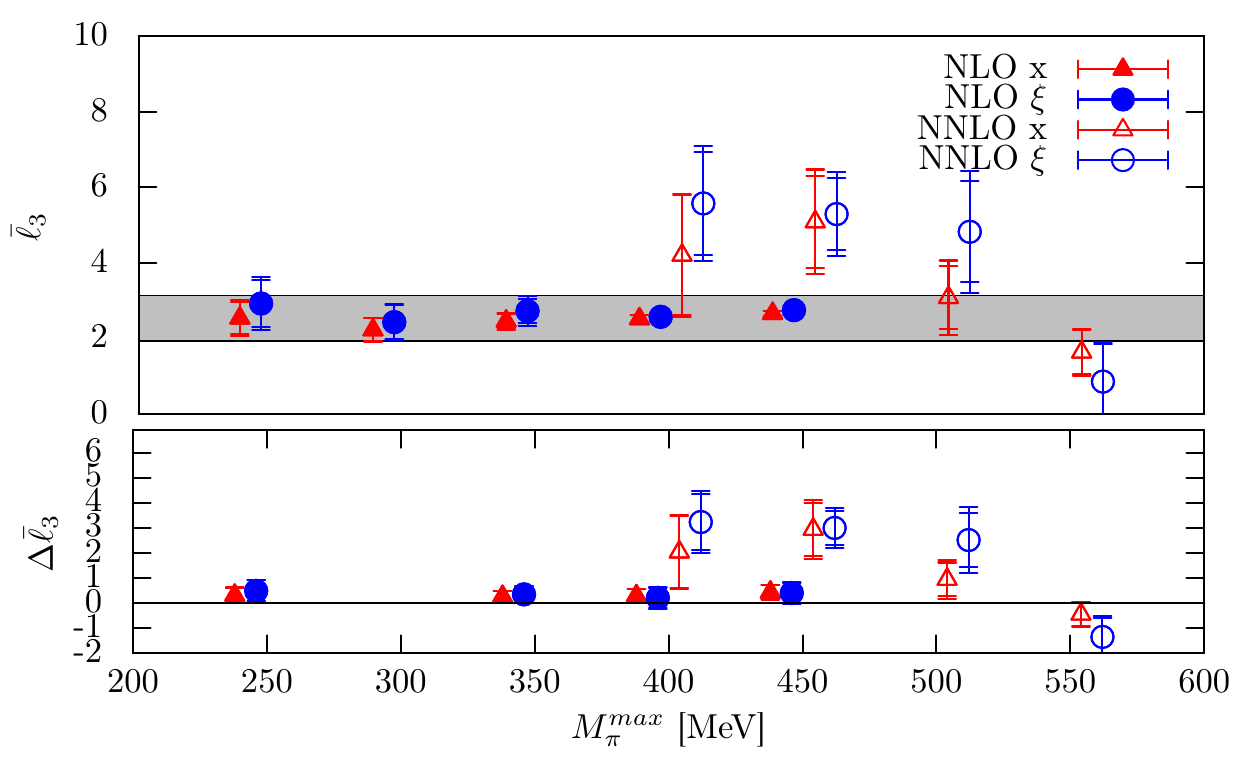} 
 \includegraphics[width=0.8\textwidth]{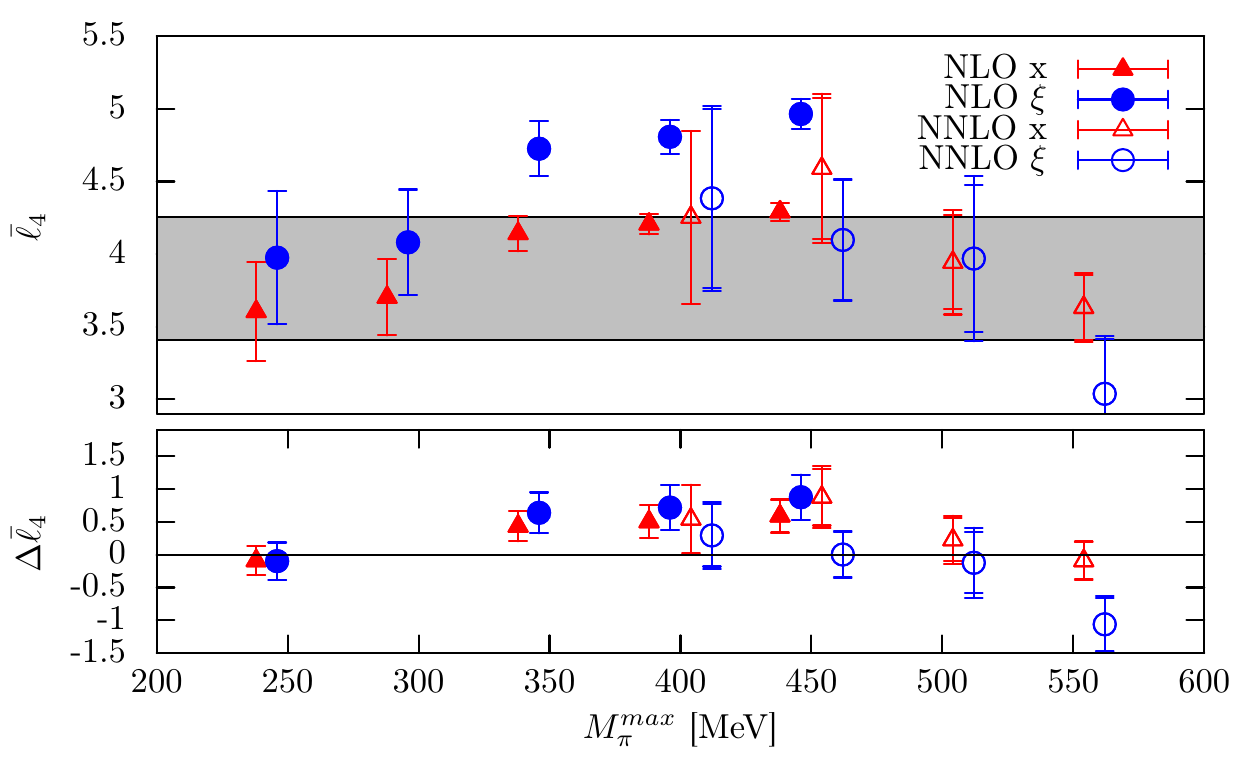}  
 \caption{\sl NLO
 LECs as a function of $M_\pi^\max$ obtained from the $SU(2)$ $\chi$PT
 fits to our lattice results for $B_\pi$ and $F_\pi$ in the pion mass range
 $[120\,\mev,\,M_\pi^\max]$, as described
 in \sec{sec:fit-strategy}. This figure is the same
 as \protect\fig{fig:lo-lec-vs-mpicut}, but for NLO instead of LO LECs.}
 \labell{fig:nlo-lec-vs-mpicut}
\end{figure}

We perform a very similar analysis for the NLO LECs, $\bar\ell_3$ and
$\bar\ell_4$, extracted from our combined, correlated NLO fits. In
particular, we define the differences $\Delta\bar\ell_3$ and
$\Delta\bar\ell_4$ in full analogy with $\Delta B^\RGI$ and $\Delta
F$. These LECs and their differences with respect to their values for
$M_\pi^\max=300\,\mev$ are plotted as a function of $M_\pi^\max$
in \fig{fig:nlo-lec-vs-mpicut}. The jump between $M_\pi^\max=300$ and
350~MeV observed in the $p$-values and in $\Delta B^\RGI$ and $\Delta
F$ is still present in $\Delta\bar\ell_4$, but less so in
$\Delta\bar\ell_3$. It is also interesting to note that for
$M_\pi^\max\ge 350\,\mev$, the values of $\bar\ell_4$ obtained from
the $x$ and $\xi$-expansion fits are no longer compatible, a clear
sign that higher order contributions are becoming relevant. Thus these
NLO LEC results are compatible with the conclusions drawn so far as to
the range of applicability of NLO, $SU(2)$ $\chi$PT.

\clearpage

\subsection{Fit quality and LECs in terms of maximum pion mass for 
NNLO  $\chi$PT}
\labell{sec:nnlo-fitquality}

We now turn to the study of NNLO $SU(2)$ $\chi$PT. The analysis we
perform here parallels the one discussed above for NLO $\chi$PT. In
particular, we study the dependence of the $p$-value and the LECs as a
function of $M_\pi^\max$. Here there are 5 additional LECs that have
to be considered. These are
$\bar\ell_{12}$, $k_M$ and $k_F$, in the case of the $x$-expansion,
and $\bar\ell_{12}$, $c_M$ and $c_F$ for the $\xi$-expansion. The
lowest value of $M_\pi^\max$ that we consider is 400~MeV, because NLO
fits work reasonably well up to around that point and because we need
more lever-arm and data to fix the 3 additional parameters required at
NNLO in each expansion.

The results for the fit quality as a function of $M_\pi^\max$ are
shown in \fig{fig:pvalue}, together with the results from NLO fits. As
these show, the introduction of NNLO terms brings the $p$-values
back up to acceptable values up to $M_\pi^\max\simeq
500\,\mev$. Beyond that point the $p$-values of NNLO fits also
drop. These observations suggest that the NNLO, $SU(2)$ chiral
expansion of $M_\pi^2$ and $F_\pi$ may extend up to 500~MeV, at least
for the statistical precision reached in this work and described
in \sec{sec:rel-xpt-contribs}.

To check this statement, we turn to the study of the LECs as a
function of $M_\pi^\max$. The results for the LO LECs, $B^\RGI$ and
$F$, are show in \fig{fig:lo-lec-vs-mpicut} and those for the NLO
LECs, $\bar\ell_3$ and $\bar\ell_4$, are given
in \fig{fig:nlo-lec-vs-mpicut}, together with the results obtained from
the NLO fits discussed in the previous
section. 

The results for $F$ and $\bar\ell_4$ appear to confirm the conclusions
drawn from the behavior of the $p$-values, at least for the $\xi$-expansion. In that case, the addition of NNLO terms for $M_\pi^\max\ge 400\,\mev$
brings the values of $F$ and $\bar\ell_4$, associated with $F_\pi$,
back in line with those obtained at NLO, with
$M_\pi^\max=300\,\mev$. This suggests that the NNLO terms are just
what is needed to accommodate the tensions which appear in the NLO
fits for $M_\pi^\max\gsim 350\,\mev$. However this picture is
not fully borne out by the LECs associated with the quark-mass
dependence of $M_\pi^2$. Indeed the jump in $B^\RGI$, 
observed in NLO fits in the region of pion-mass cuts between 300 and
350~MeV, remains present for $M_\pi^\max\sim 400$ to 450~MeV, despite
the addition of NNLO terms. Similar features are observed in the
$x$-expansion, though the addition of NNLO terms reduces the jump in
$F$ and $\bar\ell_4$ less than it does in the $\xi$-expansion. 

\begin{figure}[t]
 \centering
 \includegraphics[width=0.6\textwidth]{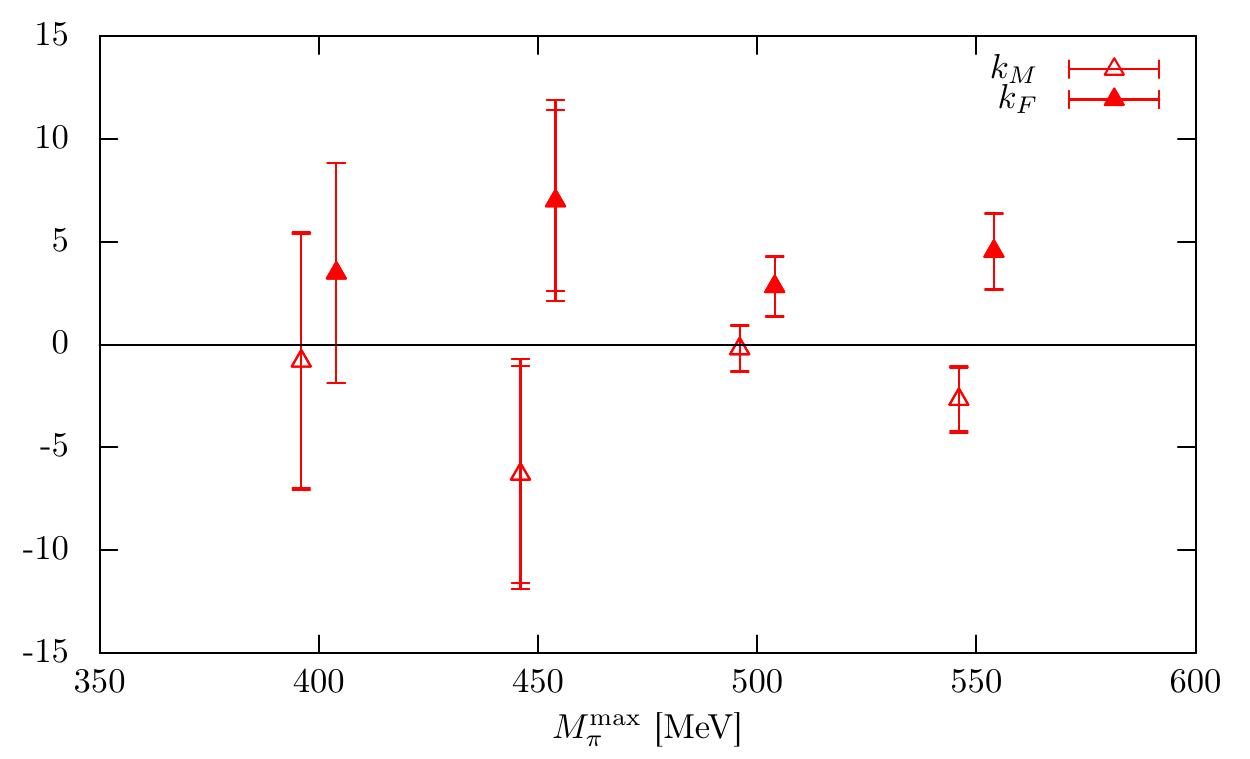} 
\includegraphics[width=0.6\textwidth]{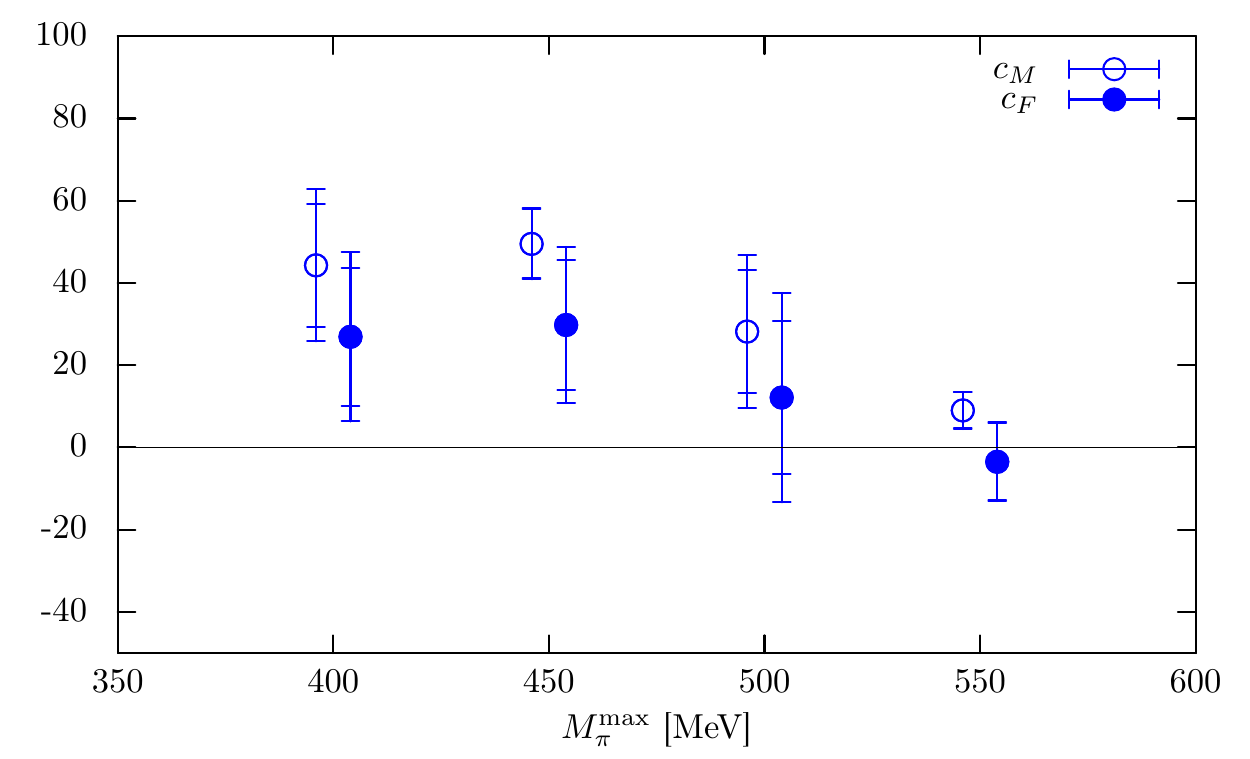} \\ 
\includegraphics[width=0.6\textwidth]{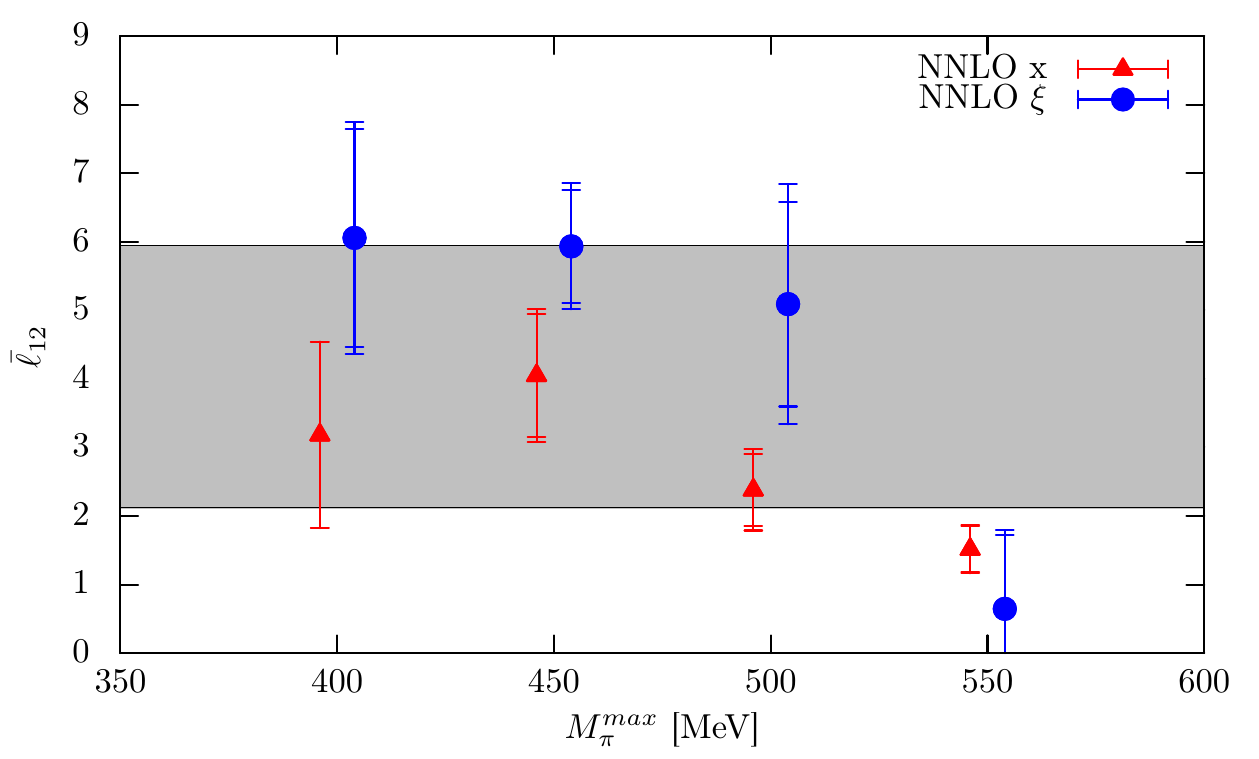}  
\caption{\sl  LECs which appear at NNLO in the $SU(2)$ $\chi$PT
 expansions of $B_\pi$ and $F_\pi$ given
 in \eqs{eq:MF}{eq:MpiFpi}. The top figure shows the results obtained
 for $k_M$ and $k_F$ from NNLO fits in the $x$-expansion to our
 lattice results with $M_\pi\in[120\,\mev,\,M_\pi^\max]$. The results
 are plotted as functions of $M_\pi^\max$. In the middle figure are
 plotted the NNLO LECs $c_M$ and $c_F$, which appear in the
 $\xi$-expansion. The NLO LEC combination
 $\bar\ell_{12}=(7\bar\ell_1+8\bar\ell_2)/15$ appears in the the $x$
 and $\xi$-expansions of $B_\pi$ and $F_\pi$ at NNLO. The results that
 we obtain for this LEC in each of the expansions are plotted in the
 bottom panel. The horizontal gray band denotes our final result for
 $\bar\ell_{12}$, obtained as described in \sec{sec:lecresults}. Error
 bars on each point are the statistical and the quadratically combined
 statistical-plus-systematic uncertainties. For the sake of clarity,
 results are shifted about the values of
 $M_\pi^\max=250,\cdots\,\mev$, at which they are obtained.}
\labell{fig:nnlo-lec-vs-mpicut}
\end{figure}

In view of this discussion, we conclude that the addition of NNLO
terms appears to allow a description of the mass dependence of $F_\pi$
up to a pion mass of around 500~MeV, which is consistent with NLO fits
in a smaller range of pion masses. This is more true for the expansion
in $\xi$ than it is for the one in $x$. However, this apparent
extension of the applicability range does not carry over to the study
of the chiral behavior of $B_\pi$, suggesting that the NNLO chiral
expansion of this quantity begins to fail for $M_\pi^\max$ in the
region of 300 to 350~MeV, for the accuracies reached here. Moreover,
it is important to remember that $B_\pi$ and $F_\pi$ share common LECs
and lattice data, and are fitted together. Thus, there is limited
sense in suggesting that the range of applicability of $\chi$PT for
these two quantities differs.

For completeness, in \fig{fig:nnlo-lec-vs-mpicut} we show results for
the NNLO $x$-expansion LECs, $k_M$ and $k_F$, as well as results
for the NNLO $\xi$-expansion LECs, $c_M$ and $c_F$, as functions of
$M_\pi^\max$. At NNLO these fits also allow the determination of the
linear combination of the NLO LECs $\bar\ell_1$ and $\bar\ell_2$ given
by $\bar\ell_{12}$ that is defined
after \eq{eq:LMLF}. This combination is also shown
in \fig{fig:nnlo-lec-vs-mpicut} as a function of $M_\pi^\max$. The
uncertainties on all of these coefficients are large, since the
precision of our results is barely sufficient to determine these
higher order contributions, at least for $M_\pi^\max\le
450\,\mev$. The coefficients $k_M$ and $k_F$ of the $x$-expansion show little
dependence on $M_\pi^\max$ all he way up to 550~MeV. This is only the
case up to 500~MeV for $k_M$ and $k_F$ of the $\xi$-expansion. In both
expansions, $\bar\ell_{12}$ drop beyond
$M_\pi^\max=500\,\mev$. However, it is worth
noting that the $x$-expansion gives a value of $\bar\ell_{12}$ which
is consistent with the determination of \cite{Colangelo:2001df}
discussed below in \sec{sec:lecresults}, for
$M_\pi^\max\le 500\,\mev$. The $\xi$-expansion yields values which are
larger.

\clearpage

\subsection{Relative contributions of different orders in 
$\chi$PT and conclusions on its range of applicability}
\labell{sec:rel-xpt-contribs}

As a final indication on the range of applicability of $SU(2)$
$\chi$PT to $B_\pi$ and $F_\pi$, we consider the size of NLO and NNLO
contributions relative to the LO ones, as functions of $m_{ud}$ and
$M_\pi^2$. We do so for two purposes. The first is to verify
that the corrections obtained in the NLO fits, which we perform for
$M_\pi^\max\le 400\,\mev$ (i.e.\ $\mudrgimax\le 41.\,\mev$), remain
reasonable over the mass range
$M_\pi\in[120\,\mev, M_\pi^\max]$ (i.e.\ $m_{ud}^\RGI\in[3.7\,\mev,
\mudrgimax]$). The second reason for investigating the size of these
corrections is to further assess the validity of our NNLO fits which include 
points up to $M_\pi^\max\simeq 500\,\mev$
(i.e. $\mudrgimax\simeq 65\,\mev$).

\begin{figure}[t]
\centering \includegraphics[width=0.47\textwidth]{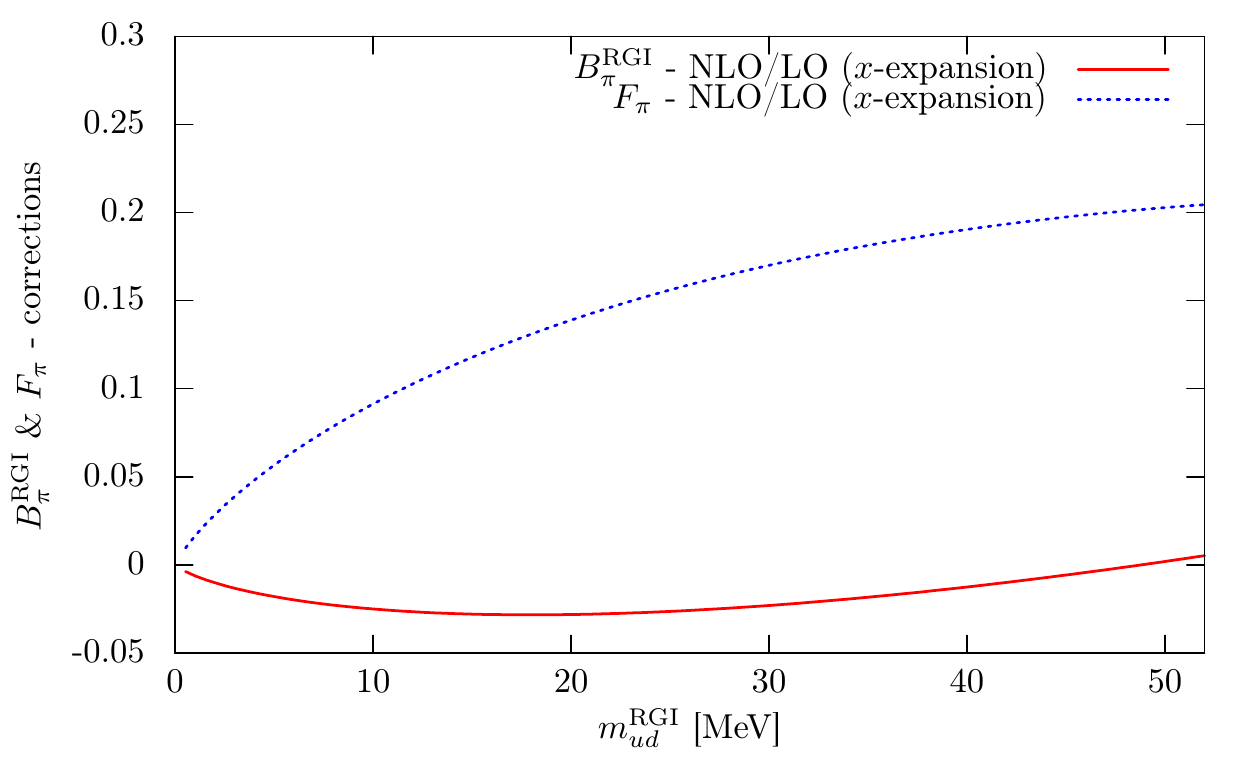}
\centering \includegraphics[width=0.49\textwidth]{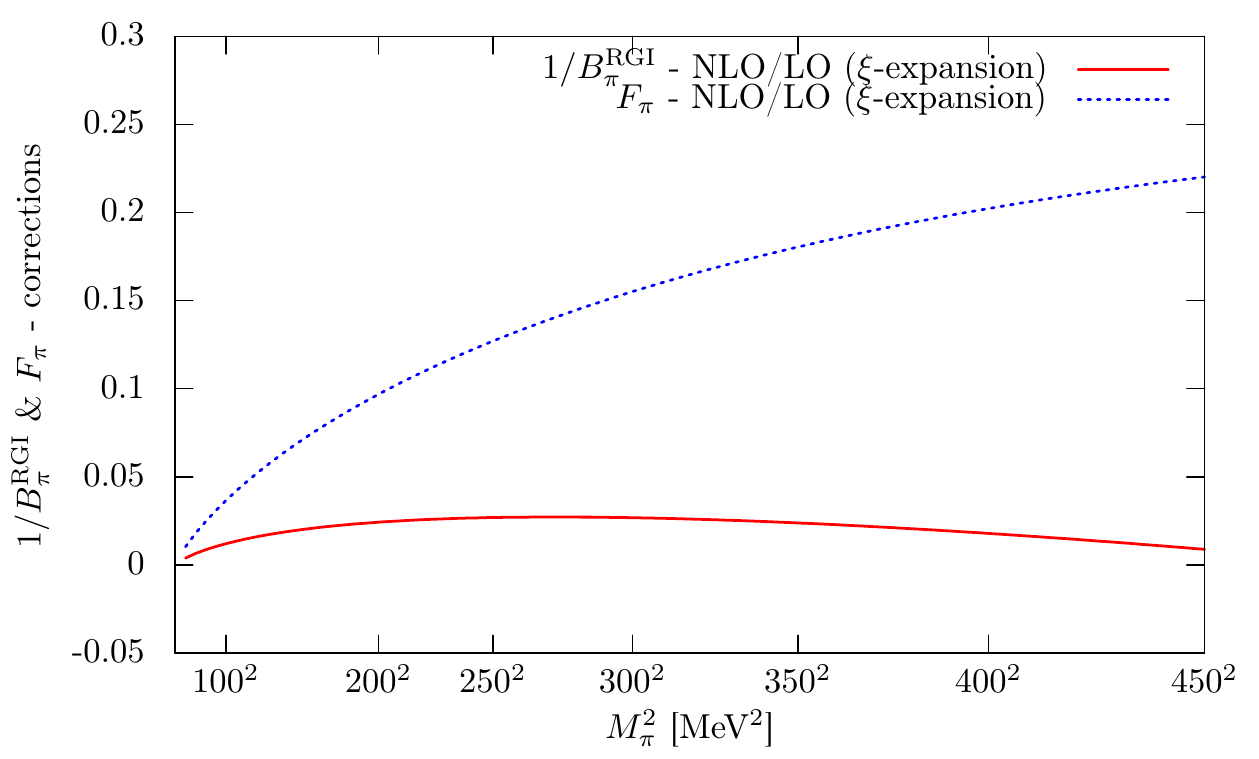}
\caption{\sl Ratios of the NLO contributions to $B_\pi$ and $F_\pi$ with respect to
the LO ones, as a function of $m_{ud}$ in the $SU(2)$ chiral
$x$-expansion (left panel) and of
$M_\pi^2$, in the $\xi$-expansion (right panel). The values of the
LECs used are those given in \tab{tab:indiv-exp-lec-results} for the
respective expansions.}
\labell{fig:nlo-corr-all}
\end{figure}  

In \fig{fig:nlo-corr-all} we plot together the NLO corrections to
$B_\pi$ and $F_\pi$ in the $x$-expansion with those of $1/B_\pi$ and
$F_\pi$ in the $\xi$-expansion, for $m_{ud}^\RGI\le 52.\,\mev$,
respectively $M_\pi\le 450\,\mev$. As the plots show, the NLO
corrections on $F_\pi$ remain less than about 10\% for $M_\pi\le
200\,\mev$ (i.e. $m_{ud}^\RGI\le 10.\,\mev$), less than about 15\% for
$M_\pi\le 300\,\mev$ (i.e. $m_{ud}^\RGI\le 23.\,\mev$) and less than
about 20\% for $M_\pi\le 400\,\mev$ (i.e. $m_{ud}^\RGI\le
41.\,\mev$). The NLO corrections on $B_\pi$ are significantly
smaller. They remain significantly less than 5\% all the way up to
$M_\pi= 450\,\mev$ (i.e. $m_{ud}^\RGI\le 52.\,\mev$). However they
exhibit non-monotonic behavior, with a turnover around $M_\pi\sim
280\,\mev$ (i.e. $m_{ud}^\RGI\sim 20.\,\mev$). All of this is entirely
consistent with the picture, drawn earlier, that our results with
errors on the order of a percent start becoming sensitive to NNLO
effects for $M_\pi\sim 300\,\mev$ and require their presence beyond
$M_\pi\gsim 400\,\mev$.

\begin{figure}[t]
\centering \includegraphics[width=0.47\textwidth]{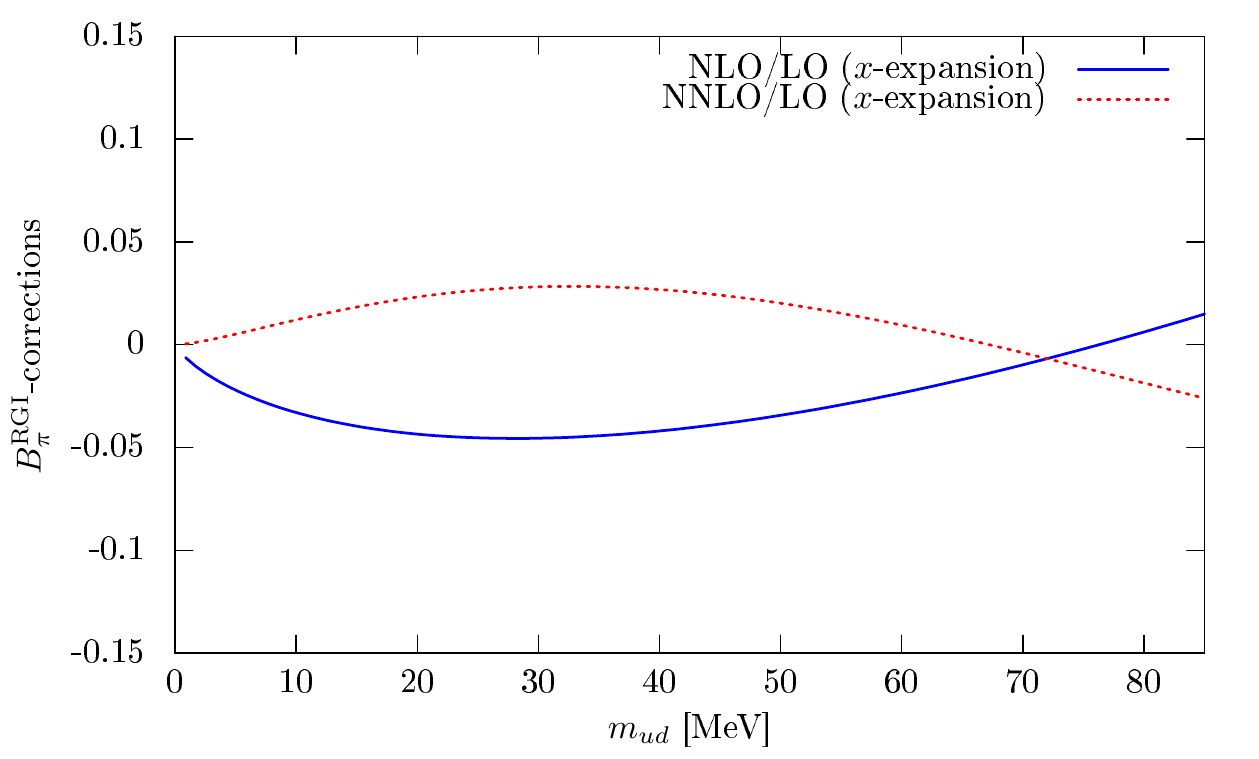}
\centering \includegraphics[width=0.49\textwidth]{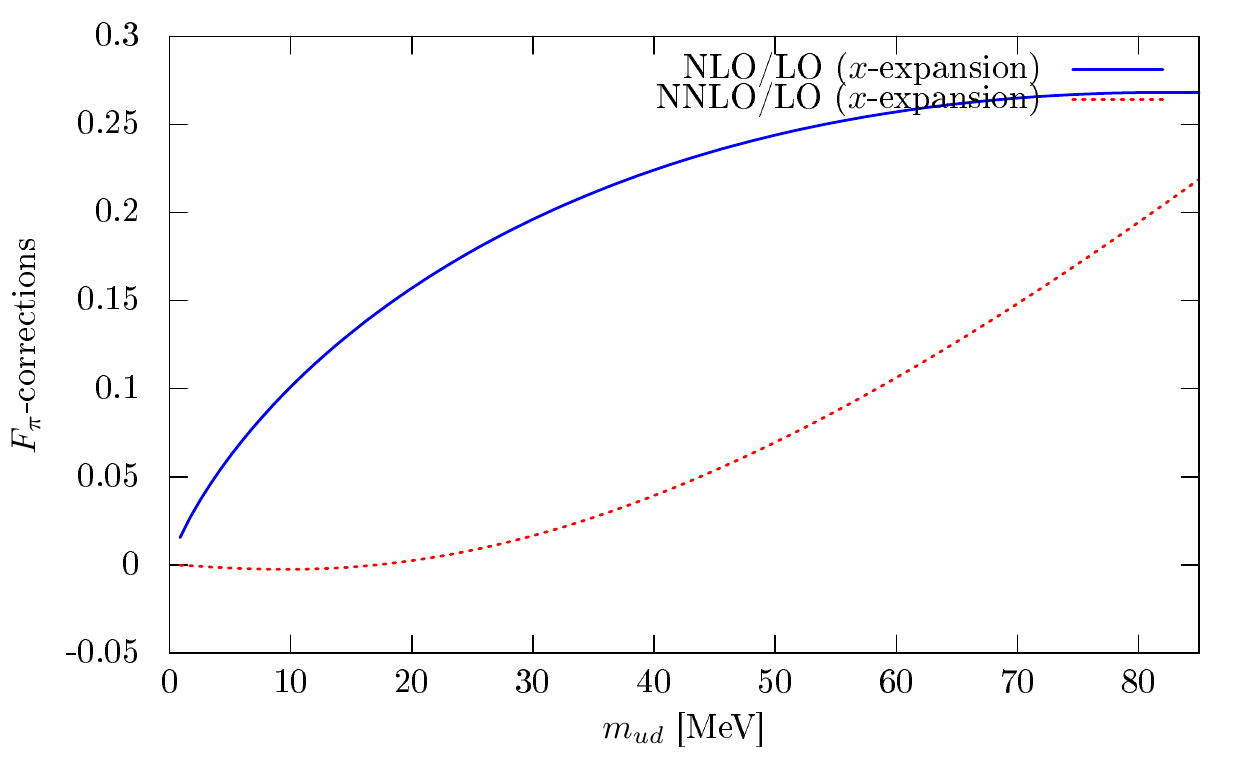}
\caption{\sl Typical ratios of the NLO and NNLO contributions to
$B_\pi$ (left panel) and $F_\pi$ (right panel) with respect to
the LO ones, as a function of $m_{ud}$ in the $SU(2)$ chiral
$x$-expansion. The values of the
LECs used are those obtained from the fit shown in \protect\fig{fig:xnnloeg}.}
\labell{fig:xcorreg}
\end{figure}  

Now let us investigate the size of the NLO and NNLO corrections in our
NNLO fits. For this we consider the same typical NNLO fits that were
shown in \fig{fig:xnnloeg} for the $x$-expansion
and \fig{fig:xinnloeg} for the $\xi$-expansion. We plot the relative size of
the NLO and NNLO corrections to $B_\pi$ and $F_\pi$ as a function of
$m_{ud}^\RGI$ in \fig{fig:xcorreg} for the $x$-expansion and
in \fig{fig:xicorreg} as a function of $M_\pi^2$ for the
$\xi$-expansion. Although the $p$-values of our NNLO fits remain good
up to $M_\pi\sim 500\,\mev$ (i.e.\ $m_{ud}^\RGI\sim 65.\,\mev$), at
that value of $M_\pi$ the NNLO corrections to $F_\pi$ are a significant
fraction of the NLO corrections, raising doubts as to the legitimacy
of neglecting NNNLO terms in these fits. This is more than confirmed
by the corrections to $B_\pi$ for which the NNLO corrections are
already a significant fraction of the NLO corrections for $M_\pi\sim
300\,\mev$ or $m_{ud}^\RGI\sim 23.\,\mev$. Moreover, these NLO and
NNLO corrections have here opposite signs, implying cancellations
which may be affected by the inclusion of higher-order terms at larger
pion-mass values. 

\begin{figure}[t]
\centering \includegraphics[width=0.47\textwidth]{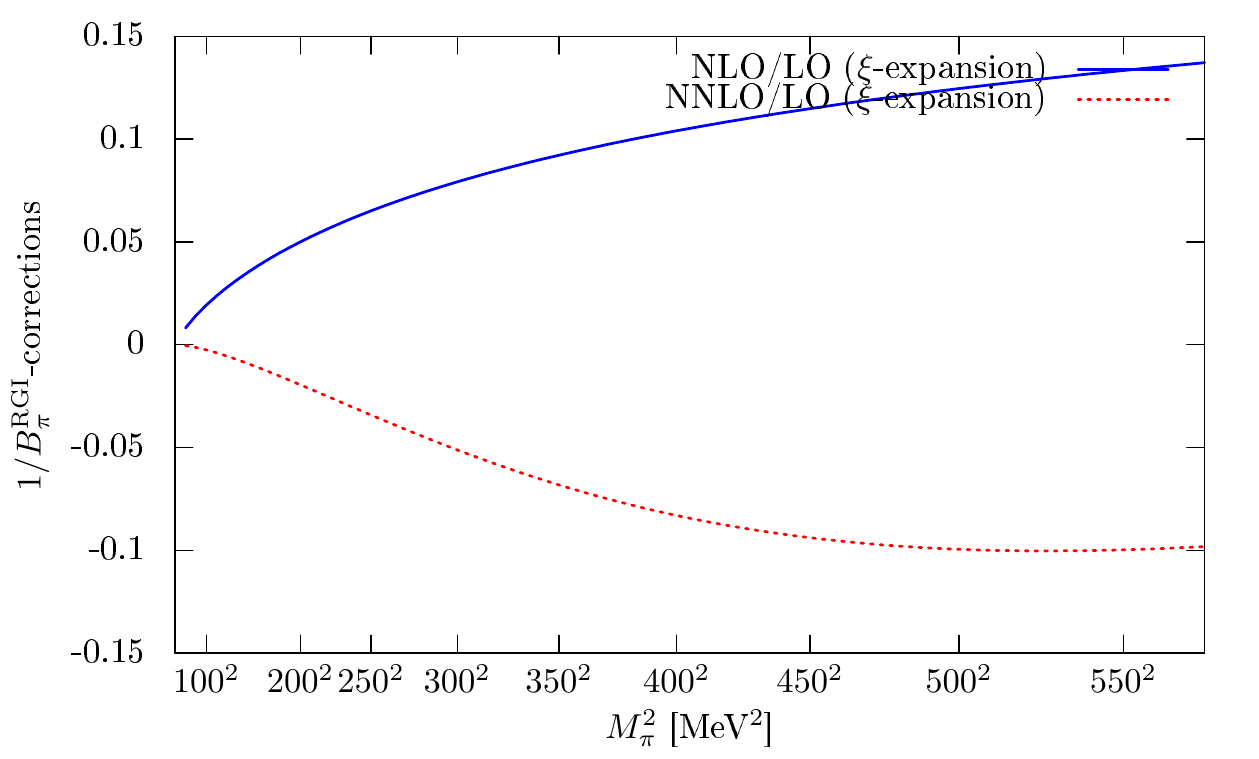}
\centering \includegraphics[width=0.49\textwidth]{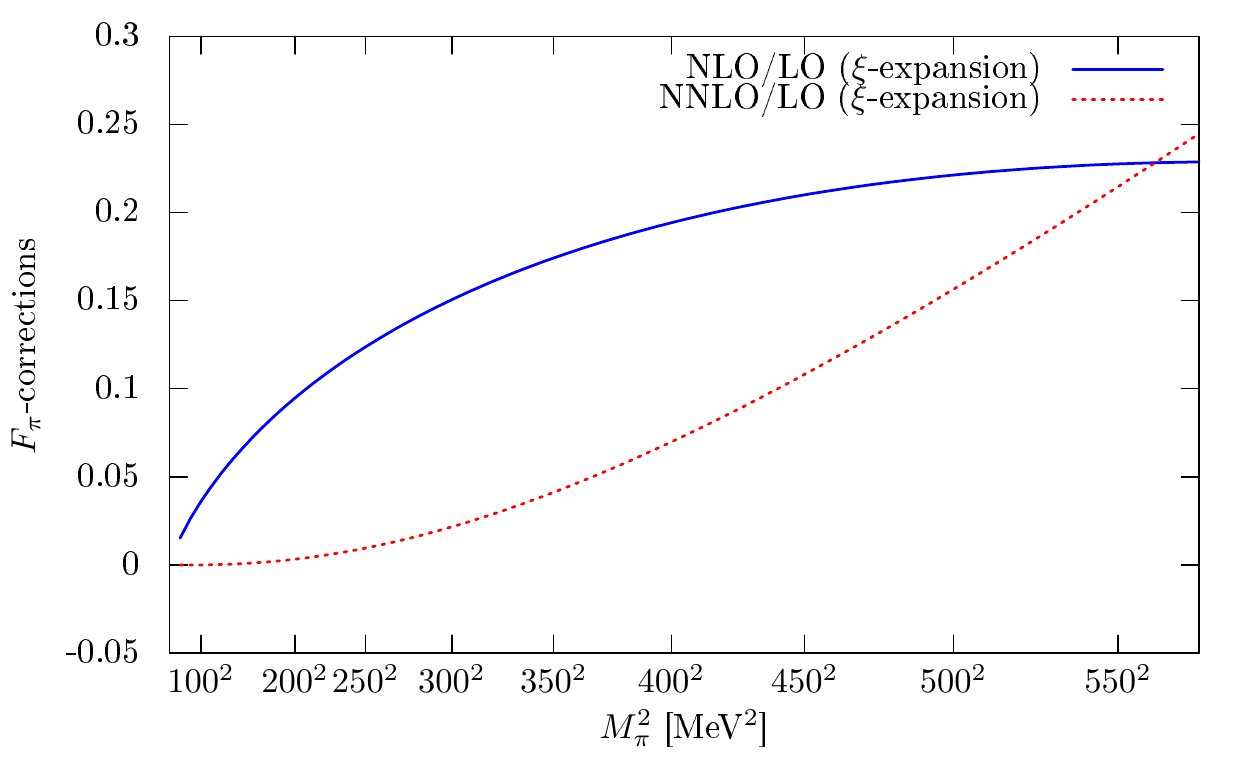}
\caption{\sl Typical ratios of the NLO and NNLO contributions to
$1/B_\pi$ (left panel) and $F_\pi$ (right panel) with respect to
the LO ones, as a function of $M_\pi^2$ in the $SU(2)$ chiral
$\xi$-expansion. The values of the
LECs used are those obtained from the fit shown in \protect\fig{fig:xinnloeg}.}
\labell{fig:xicorreg}
\end{figure}  

It is worth noting that the expansion appears better behaved for
$F_\pi$ than for $B_\pi$, since the hierarchy of corrections for the
former remains acceptable up to $M_\pi\sim 450\,\mev$ or
$m_{ud}^\RGI\sim 52.\,\mev$. The situation is quite different with the
chiral expansion of $B_\pi$. Unlike $F_\pi$, $B_\pi$ has very little
mass dependence. Thus, the role of the NLO and NNLO analytic terms in
the expansion of $B_\pi$ is to cancel as much as possible the mass
dependence brought by the non-analytic terms. When this is done
correctly in an NLO fit, adding an NNLO term destabilizes the balance
between analytic and non-analytic terms, therefore requiring a
retuning of the LECs.

Putting together all of the information discussed up until now, we
draw the following conclusions as to the range of applicability of
$SU(2)$ $\chi$PT for $N_f=2+1$ QCD. Note that conclusions may
differ when considering applications to $N_f=2$ QCD, since the latter
is missing the relatively light degrees of freedom associated with
the strange quark. As indicated in \tab{tab:data}, our results for
$F_\pi$ have statistical uncertainties typically in the range of 0.5\%
to 2.2\%, with a median error over our simulations of 1.2\% and a
standard deviation of 0.5\%. Those for $B_\pi$ are in the range of
0.3\% to 3.3\%, with a median and a standard error 0.8\% and
0.7\%. Similarly, the statistical uncertainties on $m_{ud}$ and
$M_\pi^2$ are in the ranges of 0.2\% and 1.9\% and of 0.4\% and 3.2\%,
with medians and standard errors of $(0.4\%,0.4\%)$ and
$(0.9\%,0.7\%)$, respectively. For such results, we find that NLO
$\chi$PT begins showing signs of failure for $M_\pi$ beyond 300~MeV
and breaks down completely around 450~MeV for both expansions. Adding
NNLO terms allows one to describe consistently the mass dependence of
$F_\pi$ in the $\xi$-expansion, up to around 500~MeV, at the expense
of NNLO corrections which are approaching those of the NLO ones. This
is only marginally true in the $x$-expansion, as $F$ and $\bar\ell_4$
begin deviating from the values given by the NLO fits with
$M_\pi^\max\leq 300\,\mev$ in that expansion. However in both
expansions, the addition of NNLO terms in $B_\pi$ does not allow a
description of that quantity beyond 300-350~MeV that is consistent
with the NLO description at the level of around one standard
deviation.

\clearpage

\section{Results for LECs and other physical quantities}
\labell{sec:lecresults}
Having explored the range in which one can describe the
mass-dependence of the quantities $B_\pi$ and $F_\pi$ in $SU(2)$
$\chi$PT, we are now in a position to determine the corresponding
LECs. We observe a small but significant change of behavior if
we include points with pion masses above 300~MeV, which suggests that
the NLO $\chi$PT expansion is beginning to break down beyond that
point. Moreover, the inclusion NNLO terms does not seem to allow one
to extend the range of applicability of $\chi$PT beyond that point, in
particular for $B_\pi$. Thus, we will consider only NLO fits to
determine the LO and NLO LECs, as well as quantities such as $F_\pi$
or the condensate. Moreover, we will not include results with
$M_\pi^\max>300\,\mev$.

\begin{table}[t]
\centering
 
\begin{tabular}{ccc}
\hline
\hline
 &  $x$-expansion & $\xi$-expansion \\ 
 \hline
  & \multicolumn{2}{c}{LO} \\  \hline\\[-0.4cm]
$B^{\RGI}$ [GeV] & $1.93 \pm 0.06 \pm 0.02$   & $1.93 \pm 0.05 \pm 0.02$ \\  
F [MeV] & $88.1 \pm 1.3 \pm 0.3$   & $87.9 \pm 1.4 \pm 0.3$ \\  
$[\Sigma^{\RGI}]^{1/3} $ [MeV] & $246.7 \pm 3.5 \pm 0.9 $  & $246.2 \pm 3.6 \pm 1.0$ \\  
\hline & \multicolumn{2}{c}{NLO} \\ \hline\\[-0.4cm]
$\bar\ell_3$  & $2.38 \pm 0.4 \pm 0.3$   & $2.7 \pm 0.6 \pm 0.4$ \\  
$\bar\ell_4$  & $3.65 \pm 0.32 \pm 0.06$   & $4.03 \pm 0.43 \pm 0.07$ \\ \hline 
& \multicolumn{2}{c}{Other quantities} \\ \hline
$F_\pi$ [MeV]  & $92.9 \pm 0.9 \pm 0.2 $  & $ 92.9 \pm 0.9 \pm 0.2 $ \\  
$F_\pi/F$  & $1.054 \pm 0.006 \pm 0.002 $  & $ 1.058 \pm 0.007 \pm 0.002 $ \\ \hline 
     \end{tabular} 

\caption{\sl Results for LO and NLO LECs
obtained from NLO, $SU(2)$ $\chi$PT fits in the $x$ and
$\xi$-expansion. We also give results for $F_\pi$ and its ratio to
$F$. The relevant $\chi$PT expressions are fitted to our lattice
results for $B_\pi$ and $F_\pi$ with pion masses in the range
$[120,\,300]\,\mev$. In these results, the first error is statistical and
the second is the systematic error in each expansion, computed as
described in the text.}
\labell{tab:indiv-exp-lec-results}
     \end{table} 

We begin by considering separately the results for the LECs and other
physical quantities of interest in the $x$ and $\xi$-expansion. They
are given in \tab{tab:indiv-exp-lec-results}. As described
in \sec{sec:latticedetails}, we consider all sources of systematic
error. In particular, we consider 2 initial fit times in the two-point
functions to account for possible excited state contributions
((8,9,11,13) / (9,11,13,15)), 2 mass cuts for the scale setting (380 /
480 MeV), 3 ways of performing the RI/MOM renormalization for $Z_A$
and 6 for $Z_S$ and different mass cuts in chiral fits (250 / 300
MeV). This implies a total of $2\times 2 \times 3 \times 6 \times 2=
144$ procedures for determining each quantity. We then weigh the
result of each procedure by its $p$-value. This yields a distribution
of results for each quantity. The distributions for the LO and NLO
LECs are shown in \fig{fig:LO-dist} and \fig{fig:NLO-dist},
respectively. The central value for each quantity is chosen to be the
mean of the distributions. Its systematic uncertainty is
obtained by computing the variance. Finally, the statistical error is
determined by repeating the construction of distributions for 2000
bootstrap samples, and considering the variance of their means around
the central value.

\begin{figure}[t]
 \centering \includegraphics[width=0.49\textwidth]{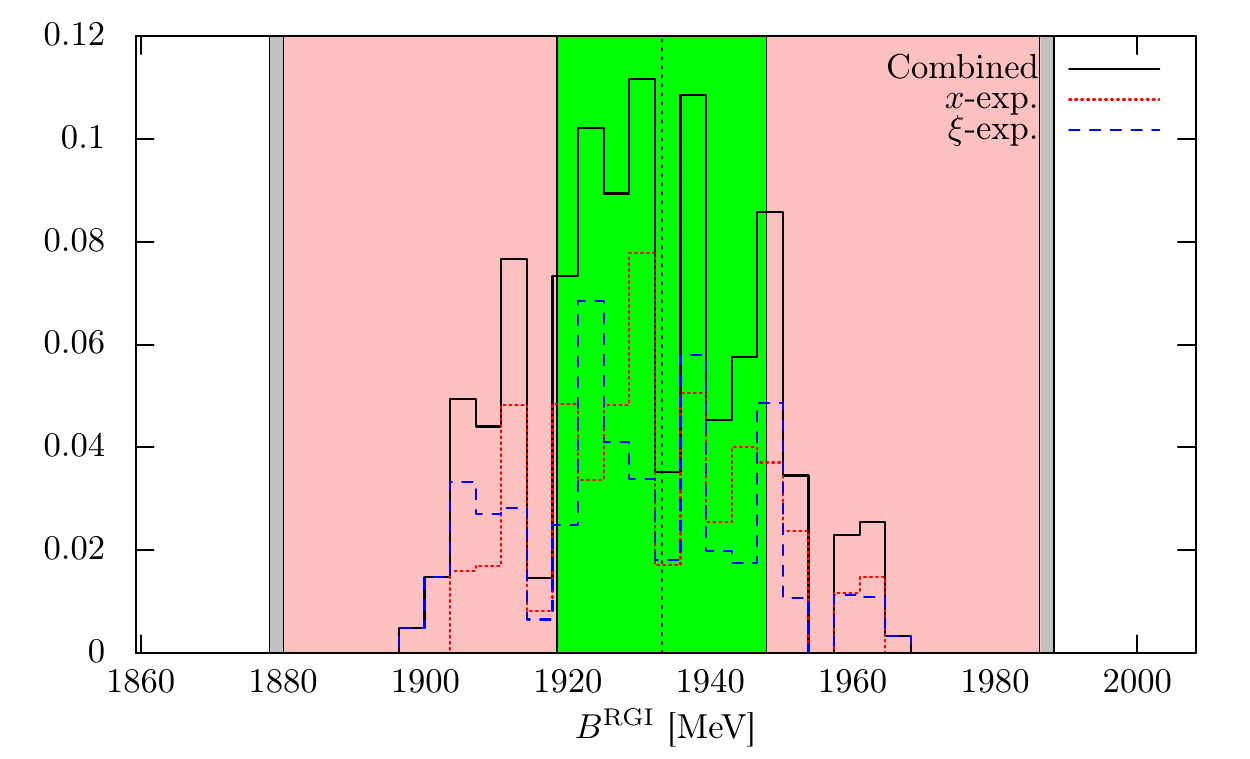} \includegraphics[width=0.49\textwidth]{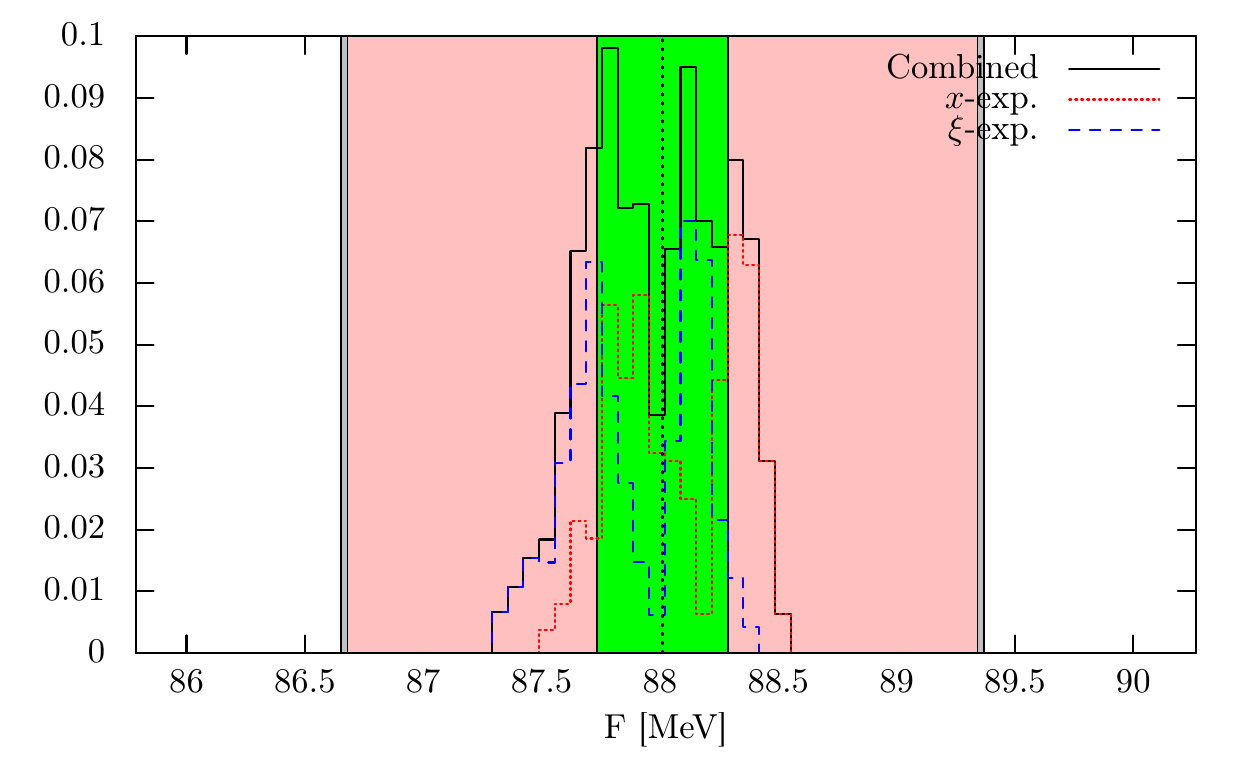} \caption{\sl Systematic
  error distributions for the LO LECs. These are obtained by varying
  the analysis procedure, as described in the text. The total
  distribution is delineated by the solid black line. It is the
  sum of the distributions corresponding to the analyses performed in
  the $x$ and $\xi$-expansion. These are shown as a red dotted line
  and a blue dashed line, respectively. Where only the $x$ or
  $\xi$-expansion distributions contribute, they partially hide the line
  corresponding to the total distribution. In the plots, the central,
  vertical, dotted line is the mean of the total distribution, i.e. our final
  central value. The central, vertical green band denotes the
  systematic error, the larger pink one, the statistical error and the
  largest gray one, the sum in quadrature of these two
  errors.}  \labell{fig:LO-dist}
\end{figure}

In our approach, it is possible to decompose the systematic
uncertainty into its various components. This is done by constructing
systematic error distributions as above, but instead of considering a
single distribution per observable, one constructs a separate
distribution for each analysis variation associated with a given
source of systematic uncertainty. For instance, for each quantity we
have two distributions to estimate the uncertainty associated with the
choice of pion mass range, one for $M_\pi^\max=250\,\mev$ and another
for 300~MeV. We then compute the mean of each of these
distributions. The error associated with this source of systematic
uncertainty is obtained from the variance of these means.

\begin{figure}[t]
 \centering \includegraphics[width=0.49\textwidth]{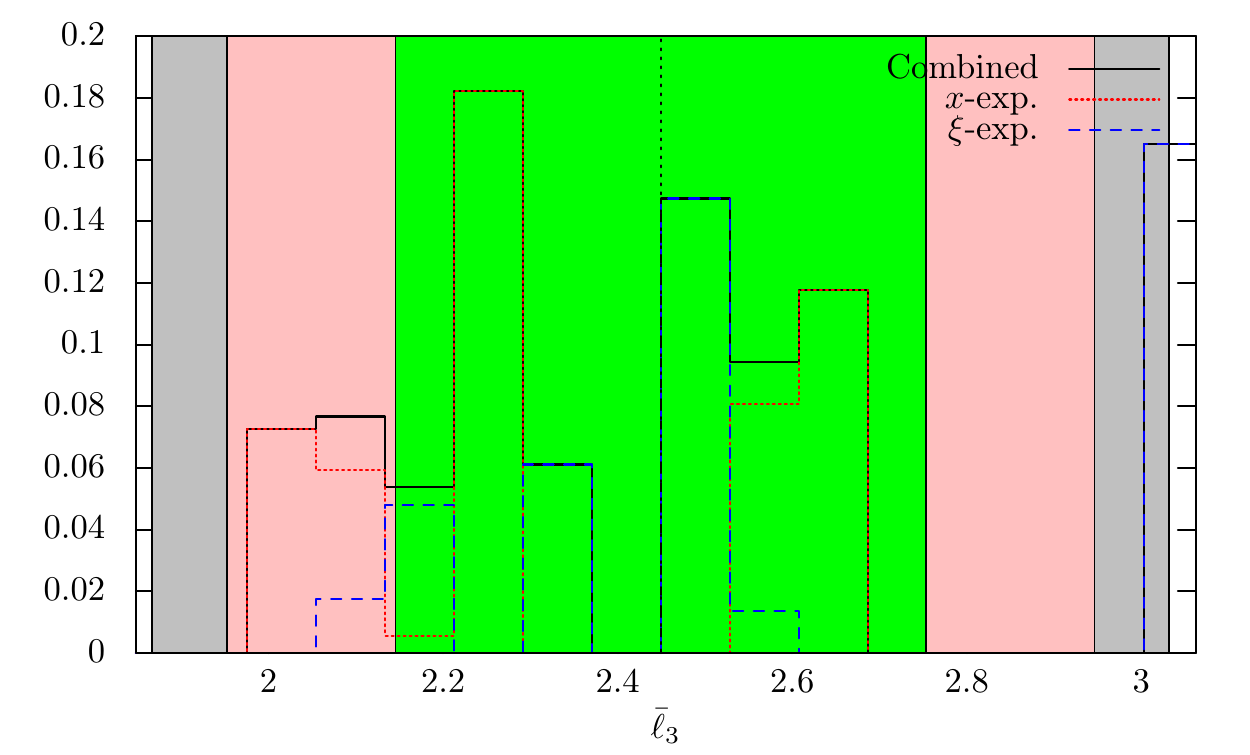} 
\includegraphics[width=0.49\textwidth]{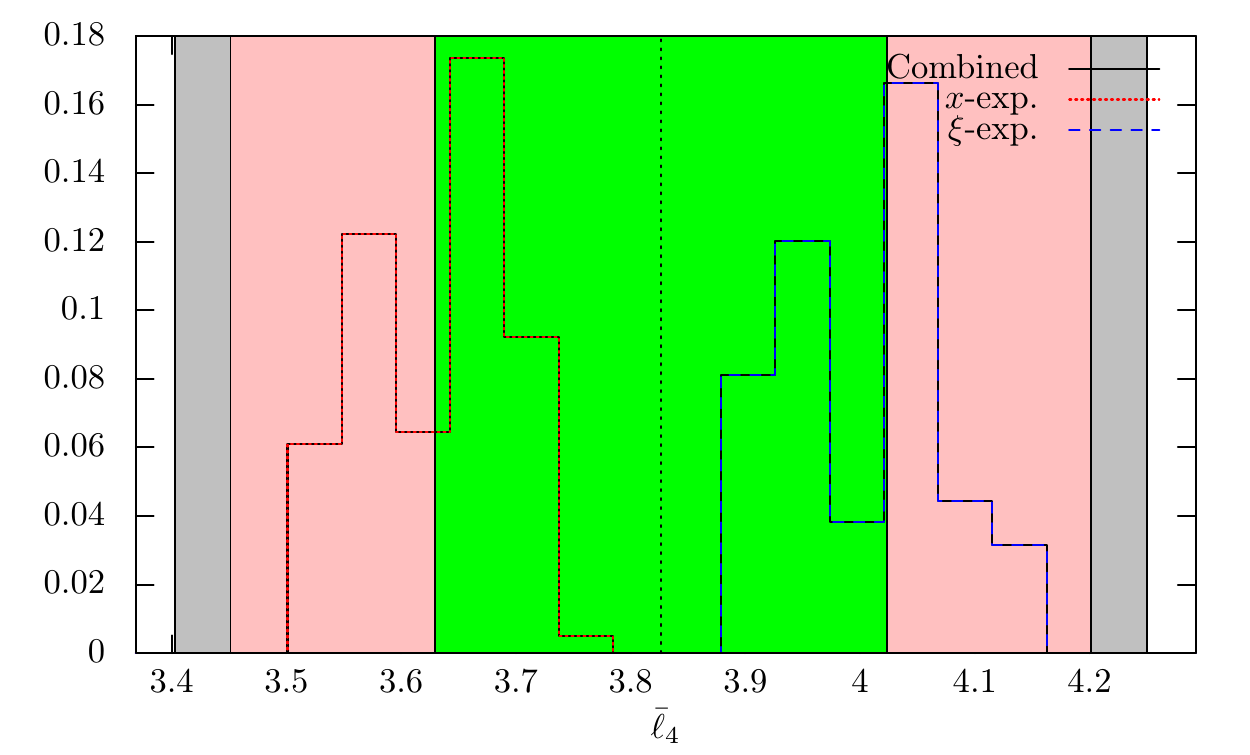} 
\caption{\sl Systematic
  error distributions for the NLO LECs. The different components of
  the graphs have the same meaning as
  in \fig{fig:LO-dist}.}  \labell{fig:NLO-dist}
\end{figure}

As \tab{tab:indiv-exp-lec-results} shows, the uncertainties on our
results are dominated by statistical errors. This means that the
numerical values of the contributions of each source of systematic
uncertainty are not particularly relevant here. Nevertheless, for
completeness, we provide a rough hierarchy of these contributions
here. The dominant source for $F$, $\Sigma$ and $F_\pi$ is the
pion-mass cut, followed by $Z_S$. The pion-mass cut also dominates the
systematic error in $\bar\ell_3$, but is followed by the one
associated with the choice of expansion ($x$ versus $\xi$). The latter
dominates in $\bar\ell_4$.

Let us now turn to a discussion of the results themselves. In both
expansions, we determine the LO LECs with total uncertainties in the
range of 1.5 to 2.9\%. The pion decay constant is obtained even more
precisely, with a total uncertainty of less than 1\% and the
uncertainty on $F_\pi/F$ is as small as 0.7\%. Of course the NLO LECs
are obtained with significantly less precision: $\bar\ell_4$ has
a total uncertainty of approximately 10\% while for $\bar\ell_3$ it is
around 25\%.

The agreement of the results obtained from the $x$ and
$\xi$-expansions is striking. This is an additional confirmation that
NLO $SU(2)$ $\chi$PT correctly describes $M_\pi^2$ and $F_\pi$ up to
$M_\pi\simeq 300\,\mev$. Indeed, the two expansions differ by higher
order terms. This difference also explains why the agreement is better
for LO LECs and $F_\pi$ than it is for NLO LECs: the smaller, less
constrained NLO contributions are more affected by changes made at
higher orders.

Because of the consistency of the results in the two expansions, we
combine them in the first column of \tab{tab:final-lec-results} to
obtain our final results. This combination is performed in a way which
is entirely consistent with our determination of systematic
errors. The two expansions ($x$ / $\xi$) are treated as an additional
alternative in our determination of LECs and other quantities. Thus,
our final results are obtained from a total of $144\times 2 = 288$
different analyses for each quantity. The corresponding systematic
error distributions for the LECs are shown
in \figs{fig:LO-dist}{fig:NLO-dist}, together with our final results
for these quantities.  It should be noted, however, that in performing
fine comparisons between lattice studies, one may wish to compare them
separately in each expansion.

The LO LECs and $F_\pi$ do not change visibly compared to those
obtained from the individual expansions. The systematic uncertainties
on the NLO LECs increase slightly as a result of the variation induced
by the use of the two expansions. For comparison, we give in the
second column of \tab{tab:final-lec-results} the averages for these
quantities obtained by FLAG \cite{Colangelo:2010et} and/or the
PDG \cite{Beringer:1900zz}.

\begin{table}[t]
\centering

\begin{tabular}{ccc}
\hline
\hline
 &  Combined & FLAG \&\ PDG  \\ 
 \hline
  & \multicolumn{2}{c}{LO} \\
  \hline
   $B^\RGI$ [GeV]            &  $ 1.93 \pm 0.06 \pm 0.02 $   & \\
   $B^\msbar(2\,\gev)$ [GeV] &  $ 2.58 \pm 0.07 \pm 0.02 $   & \\
    $F$ [MeV]                   &  $ 88.0 \pm 1.3 \pm 0.3 $  & $86.4\pm 0.7$~\cite{Colangelo:2010et,Beringer:1900zz}  \\ 
    $[\Sigma^\RGI]^{1/3}$ [MeV] &   $ 246.5 \pm 3.5 \pm 0.9 $   & $244\pm 16$  \\ 
    $[\Sigma^\msbar(2\,\gev)]^{1/3}$ [MeV] &  $ 271 \pm 4 \pm 1 $    & $269\pm 18$   ~\cite{Colangelo:2010et}    \\ 
    \hline
    & \multicolumn{2}{c}{NLO} \\ \hline
    $\bar\ell_3$              & $ 2.5 \pm 0.5 \pm 0.4 $  & $3.2\pm 0.8$~\cite{Colangelo:2010et}         \\
    $\bar\ell_4$                    & $ 3.8 \pm 0.4 \pm 0.2 $  & $4.4\pm  0.2$~\cite{Colangelo:2001df}    \\  \hline
    & \multicolumn{2}{c}{Other quantities} \\ \hline
     $F_\pi$ [MeV]             & $ 92.9 \pm 0.9 \pm 0.2 $  & $92.21\pm      0.02\pm 0.14$~\cite{Beringer:1900zz}    \\ 
      $F_\pi/F$               & $ 1.055 \pm 0.007 \pm 0.002 $ & $1.073\pm0.015$~\cite{Colangelo:2001df} \\  \hline
      \end{tabular}

\caption{\sl Our final
 results for LO and NLO LECs, as well as for $F_\pi$ and its ratio to
 $F$. They are obtained by combining the results leading to those
 given for the individual $x$ and $\xi$-expansion, as described in the
 text. In these results, the first error is statistical and the second
 is systematic. The computation of these errors is described in the
 text. The conversion of RGI numbers to those in the $\msbar$ scheme
 at 2~GeV is performed using the results of \cite{Durr:2010aw}. For
 comparison, we give in the second column the estimates of the FLAG
 review \cite{Colangelo:2010et} for the LECs and $F_\pi/F$, and of the
 PDG~\cite{Beringer:1900zz} for $F_\pi$.  Note that an update of the FLAG review is planned, of which a preliminary version can be found at \cite{flag-update}.}
\labell{tab:final-lec-results}
      \end{table}

We now turn to a comparison of our results with those of other
collaborations who have performed $N_f\ge 2+1$
studies \cite{Allton:2008pn,Aoki:2008sm,Bazavov:2009fk,Aoki:2010dy,Bazavov:2010hj,
Bazavov:2010yq,Beane:2011zm,Borsanyi:2012zv,Arthur:2012opa,
Baron:2010bv,Baron:2011sf}. Note that amongst those, the only study
which includes simulations all the way down to the physical value of
the pion mass is the staggered fermion one
in \cite{Borsanyi:2012zv}. That study computes the LO quantities
$2Bm_{ud}^\phys$ and $F_\pi/F$, and the NLO LECs $\bar\ell_3$ and
$\bar\ell_4$. Thus, in addition to the physical value of $M_\pi$, it
requires $F_\pi$ to determine the LO LEC $F$ and the renormalized
quark mass, $m_{ud}^\phys$, to determine the other LO LEC, $B$, or
alternatively the quark condensate. It takes the former
from \cite{Beringer:1900zz} and the latter
from \cite{Durr:2010vn,Durr:2010aw}, which make use of the same Wilson
quark simulations as employed in the present paper, and is thus not
fully de-correlated from the results presented here. Moreover, the use
of outside input for $F_\pi$ and $m_{ud}^\phys$ forbids predicting
these two quantities and thus making valuable crosschecks of the
calculation.  It may also be noted that the smallest lattice spacing
in that work is 0.1~fm.

We find agreement with \cite{Borsanyi:2012zv} on the LO LECs $F$ and
$B$.  MILC \cite{Bazavov:2010yq} obtains a condensate which is more
than one standard deviations larger than ours while
RBC/UKQCD \cite{Arthur:2012opa} find a value which is more than two
standard deviations smaller than ours. As for $F$, it is not studied
by RBC/UKQCD, but agreement with MILC \cite{Bazavov:2010yq} is
excellent, while ETM \cite{Baron:2010bv}, in an $N_f=2+1+1$
computation, find a value which is more than 1.5 combined standard
deviations smaller than ours. Regarding $F_\pi/F$, which measures the
chiral corrections to $F_\pi$ at $M_\pi^\phys$, our result is in good
agreement with that of Borsanyi {\em et al.} \cite{Borsanyi:2012zv},
NPLQCD \cite{Beane:2011zm} and MILC \cite{Bazavov:2010yq}. However,
ETM's $N_f=2+1+1$ result \cite{Baron:2010bv} is almost 2.5 standard
deviations way from ours.

It is interesting to note that the deviations from
ETM's \cite{Baron:2010bv} results gradually decrease as we increase
$M_\pi^\max$ above 300~MeV. This is clearly visible in the right panel
of \fig{fig:lo-lec-vs-mpicut} which shows that $F$ decreases by more
than one standard deviation when lattice results with $M_\pi\gsim
350\,\mev$ are included. Though we have not shown the $M_\pi^\max$
dependence of $F_\pi/F$, it undergoes a very similar increase, instead
of decrease. This suggests that the discrepancy that we observe with
ETM \cite{Baron:2010bv} on $F$ and $F_\pi/F$ may be due to the fact
that ETM's lightest pion is 270~MeV and that they include points up to
510~MeV in their NLO fits. This observation is further corroborated by
the discussion in \sec{sec:xpt-misuse}, where we investigate the
effect of removing lattice data at the low-$M_\pi$ end.

For completeness we note that $B$ undergoes a more than one standard
deviation increase when lattice results with $M_\pi\gsim 350\,\mev$
are included. The net effect is that $\Sigma$ remains essentially
stable as $M_\pi^\max$ is increased.

We now discuss NLO LECs. Our results for $\bar\ell_3$ and $\bar\ell_4$
are systematically smaller than those obtained in other recent $N_f\ge
2+1$
computations \cite{Allton:2008pn,Aoki:2008sm,Bazavov:2009fk,Aoki:2010dy,Bazavov:2010hj,
Bazavov:2010yq,Beane:2011zm,Borsanyi:2012zv,Arthur:2012opa,
Baron:2010bv,Baron:2011sf}, the effect being more pronounced in the
$x$-expansion which is the one used in other studies. Though the
discrepancy is generally marginal, it is marked with the $N_f=2+1+1$
ETM results \cite{Baron:2010bv}. Their results for $\bar\ell_3$ and
$\bar\ell_4$ are almost two combined standard deviations above
ours. As \fig{fig:nlo-lec-vs-mpicut} shows, these larger values are
compatible with those which we obtain including points with
$M_\pi\gsim 350\,\mev$. Thus, the possible explanation for the
discrepancy with ETM's LO LECs also applies for NLO LECs. The only
other results obtained with simulations down to the physical pion
mass \cite{Borsanyi:2012zv} are also larger than ours, though the
difference here is within a standard deviation.

We conclude this section with a discussion of NNLO LECs. The results
presented here should be taken with a grain of salt. The first reason
is that we are only sensitive to them if we include points with
$M_\pi\ge 400\,\mev$. While NNLO $\chi$PT for $F_\pi$ may be
applicable for such masses, this is not the case for
$B_\pi$. Moreover, the statistical uncertainties on these results are
very large. Nevertheless, because very little is known about these
LECs, we believe that the information brought by our analysis is
useful. We obtain these estimates very much in the same way as we
determine the LO and NLO LECs. The only difference is that instead of
considering $M_\pi^\max=250$ and 300~MeV, we estimate systematic
errors associated with the neglect of higher-order terms using
$M_\pi^\max=400$, 450 and 500~MeV. Note that for these ranges, the
$p$-values of the NNLO fits are good, as shown in \fig{fig:pvalue}.

 The results that we obtain are, for the $x$-expansion, 
$k_M=-2.4 \pm 5.3 \pm 2.8$ and
$k_F=4.4 \pm 4.3 \pm 2.1 $, and $c_M=37. \pm 12. \pm 13.$ and $c_F=20. \pm 15. \pm 17.$ for the $\xi$-expansion. The
only other lattice study in which $k_M$ and $k_F$ are considered
is \cite{Borsanyi:2012zv}. As already noted this study uses the
physical value of $F_\pi$ as input. Moreover, the NNLO fits are
constrained with a prior on $\bar\ell_{12}$, and in some cases on
$k_M$ and $k_F$. Considering only the fits in which $k_M$ and $k_F$
are not constrained, they find $k_M\sim 2$ and $k_F\sim 1$.

As already mentioned, our NNLO fits are sensitive to the combination
of NLO LECs, $\bar\ell_{12}=(7\bar\ell_1+8\bar\ell_2)/15$. 
We determine it in the same way as the NNLO LECs, finding
$\bar\ell_{12}=3.0 \pm 1.1 \pm 0.9$ and $5.5 \pm 1.3 \pm 0.9$ in the
$x$ and $\xi$-expansion, respectively.  
The $\xi$-expansion leads to
a larger value of that LEC, the discrepancy probably indicating a
sensitivity to the treatment of higher-order terms. Since we have no
reason to favor the result of one expansion over that from the other,
we include the results from both in our final estimate of
$\bar\ell_{12}$. 
In this way, we find $\bar\ell_{12}=3.9 \pm 1.1 \pm 1.5$. 
For comparison we can use the LECs $\bar\ell_1$ and $\bar\ell_2$
obtained from the fitting of NLO expansions of $\pi\pi$ scattering
amplitudes to experimental data \cite{Colangelo:2001df}. Combining the
results for $\bar\ell_1$ and $\bar\ell_2$
from \cite{Colangelo:2001df}, one obtains $\bar\ell_{12}=2.1\pm
0.3$. It should be noted that the results in \cite{Colangelo:2001df}
only include uncertainties coming from the phenomenological input and
not possibly-significant uncertainties coming from neglected
higher-order terms in the relevant chiral expansion. Though our
determination from NNLO fits have much larger errors, it is compatible
with the value from $\pi\pi$ scattering.

We have also performed NNLO fits imposing a Gaussian constraint on
$\bar\ell_{12}$. Instead of taking $\bar\ell_{12}=2.1\pm 0.3$ as done
in \cite{Borsanyi:2012zv}, we more than triple the error and consider
$\bar\ell_{12}=2.1\pm 1.0$. The fits still have good
$p$-values. However, even such a loose prior has a significant impact
on the LECs present at NNLO. Instead of the values given above, with
this prior we find $k_M=-0.1 \pm 1.3 \pm 0.9$, $k_F=3.0 \pm 1.8 \pm
0.4$ and $\bar\ell_{12}=2.15\pm0.05\pm 0.11$ for the $x$-expansion,
and $c_M=3. \pm 4. \pm 3.$, $c_F=14. \pm 9. \pm 4.$
$\bar\ell_{12}=2.15\pm0.03\pm 0.03$ for the $\xi$-expansion. Perhaps
more surprisingly, this prior also affects the NLO LECs extracted from
NNLO fits. Determining these LECs from the three pion-mass intervals
with $M_\pi^\max=400$, 450 and 500~MeV leads to
$(\bar\ell_3,\bar\ell_4)=$ $(2.8\pm0.4\pm0.3,3.83\pm 0.28\pm 0.05)$
for the $x$-expansion and $(\bar\ell_3,\bar\ell_4)=$
$(2.5\pm0.6\pm0.4,3.3\pm 0.5\pm 0.2)$ in the $\xi$-expansion with the
prior. This is to be compared with $(\bar\ell_3,\bar\ell_4)=$ $(3.9\pm
1.3\pm1.2,4.2\pm0.5\pm0.4)$ for the $x$-expansion and
$(\bar\ell_3,\bar\ell_4)=$ $(5.1\pm 1.2\pm0.8,4.1\pm0.5\pm0.3)$ in the
$\xi$-expansion obtained without prior. Not surprisingly, the
difference observed in the $\xi$-expansion also carries over to
$F_\pi/F$ which is significantly lower with the constraint. More
generally, while the $x$-expansion results with and without prior are
consistent within errors, those in the $\xi$-expansion are not. This
is due to the fact that, without a Gaussian constraint, our NNLO,
$\xi$-expansion fits favor a larger value of $\bar\ell_{12}$. Needless
to say that a more stringent constraint on $\bar\ell_{12}$ or forcing
the NNLO LECs to vanish within a few units will have an even larger
impact. Thus, while we cannot exclude the use
of priors based solely on the absolute quality of the fits which
include them, we take the differences that we observe when they are
added as a warning. The use of even loose priors may induce one to believe
that data has more resolution power than it actually has and may bias
the results obtained.

\clearpage

\section{On the presence of chiral logarithms and the 
possible misuse of $\chi$PT}
\labell{sec:logsandmisuse}
\subsection{On the presence of NLO chiral logarithms}
\labell{sec:presence-chilogs}

Having studied the range of applicability of the NLO expansions, we now
explore the extent to which chiral logarithms are required to
describe our results. We do so by fitting, to our results for
$B_\pi$ and $F_\pi$, the NLO expressions in \eq{eq:nloxpar} and
\eq{eq:nloxipar}, with the logarithms omitted. As in our study of the
range of applicability of $SU(2)$ $\chi$PT, we include in these fully
correlated fits all points with $m_{ud}\le m_{ud}^\max$ or $M_\pi\le
M_\pi^\max$, and study the behavior of the $p$-value as the cut is
increased. We also monitor the value of $F_\pi$ at $M_\pi^\phys$.

\begin{figure}[t]
 \centering
\includegraphics[width=0.9\textwidth]{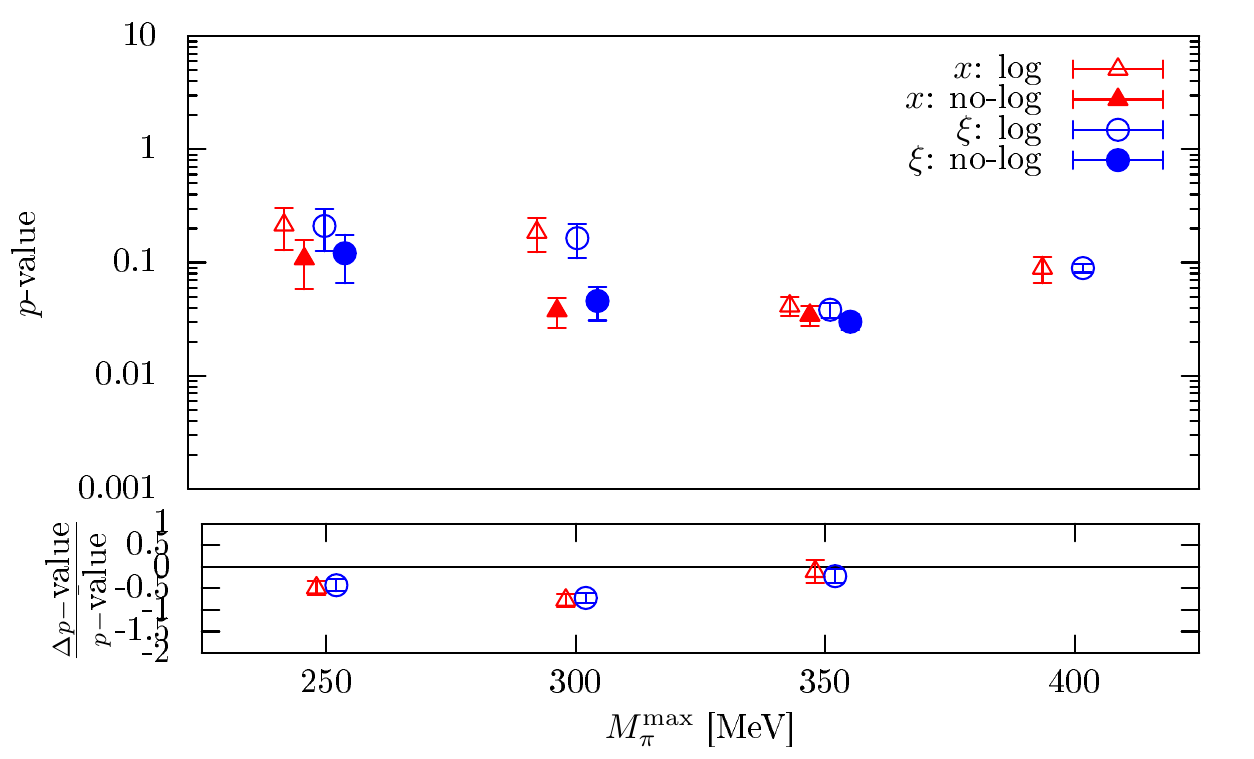}
 \caption{\sl Comparison of the $p$-values obtained in fits of NLO
   $SU(2)$ expansion, including and omitting the chiral
   logarithms. These fully correlated fits to our lattice results for
   $B_\pi$ and $F_\pi$ include points whose pion mass is in the range
   $[120\,\mev,\,M_\pi^\max]$. Results are shown for the $x$ and
   $\xi$-expansion. In the top panel the individual $p$-values are
   shown. Those of fits including the logarithms are the same as the
   ones given in \fig{fig:pvalue}. In the lower panel it is the
   difference of the $p$-value obtained omitting logarithms minus the
   $\chi$PT one, normalized by the latter. Error bars on each point
   are the systematic uncertainties discussed in
   \sec{sec:nlo-fitquality}. For the sake of clarity, results are
   shifted about the values of $M_\pi^\max=250,\cdots\,\mev$, at which
   they are obtained.}  \labell{fig:pvalue-nolog}
\end{figure}

In \fig{fig:pvalue-nolog} we compare these $p$-values of NLO fits
without logarithms to those of the NLO $\chi$PT fits performed in
\sec{sec:su2validity}, both in the $x$ and $\xi$-expansion. The
$p$-values obtained when logarithms are omitted are consistently lower than for
the $\chi$PT fits, though they remain acceptable for $M_\pi^\max\le
350\,\mev$. Beyond that point they become very bad. To determine the
significance of the preference for the presence of logarithms, we
compute the difference of the $p$-values obtained omitting the chiral
logarithms to those including them, normalized by the latter. These
are shown in the lower panel of \fig{fig:pvalue-nolog}. As the
figure shows, in the range of applicability of NLO $\chi$PT,
i.e. $M_\pi^\max\le 300\,\mev$, the presence of logarithms is favored
by about five to eight standard deviations in the $p$-value in both expansions.

\fig{fig:fpi-nolog} shows the $M_\pi^\max$ dependence of the value of
$F_\pi$ at physical $M_\pi$, obtained in fits with and without
logarithms. Both fits give very similar results in the range of
applicability of NLO $\chi$PT, where $M_\pi^\max\le 300\,\mev$. Thus,
at our level of accuracy, a simple linear interpolation would allow us
to obtain $F_\pi$. However, the bottom panel of \fig{fig:fpi-nolog}
shows that this will no longer be true when the total uncertainty on
$F_\pi$ reaches a few tenths of an MeV.

To conclude this discussion, our lattice results clearly favor the
presence of logarithms in the range of applicability of NLO $SU(2)$
$\chi$PT, though the values of $F_\pi$ obtained without them are
compatible with those obtained in $\chi$PT at the present level of accuracy.

\begin{figure}[t]
 \centering
\includegraphics[width=0.9\textwidth]{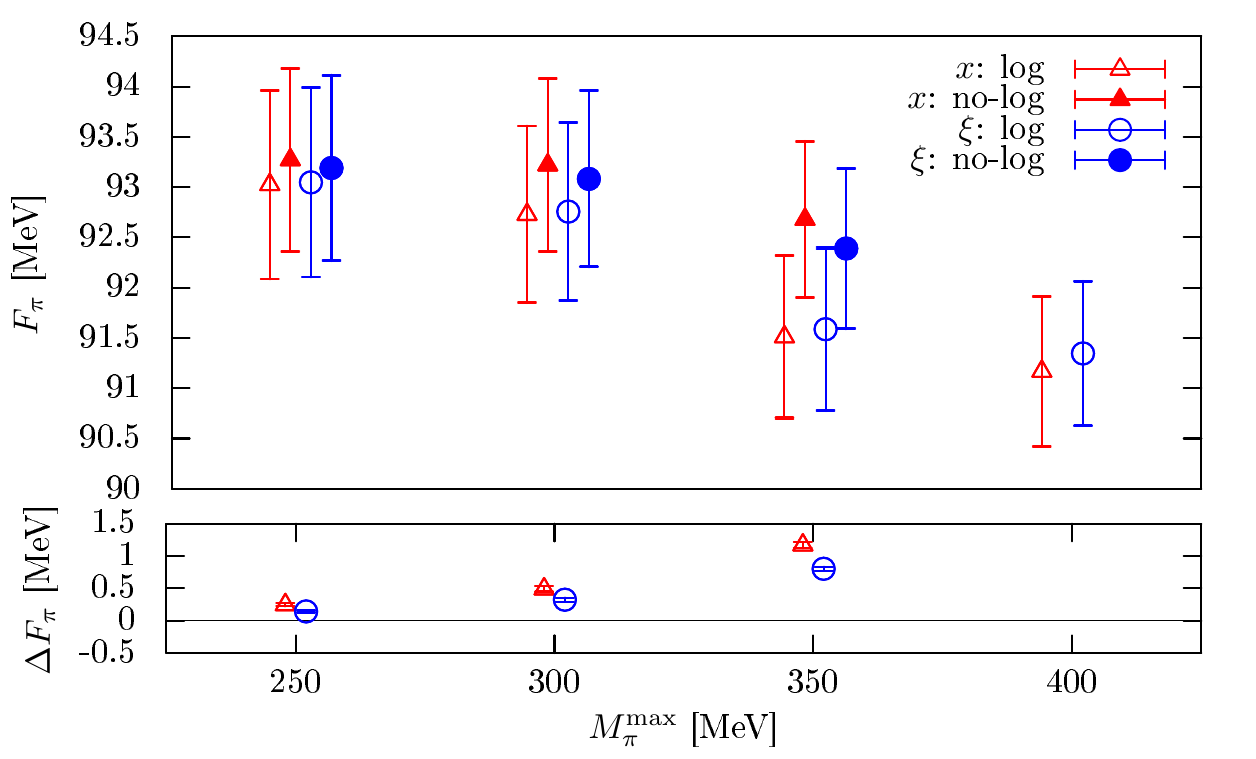}
 \caption{\sl $F_\pi$ as a function of $M_\pi^\max$, obtained from NLO
   fits with and without and logarithms, in the $x$ and
   $\xi$-expansion (upper panel). In the lower panel it is the
   difference (no log. minus log.) of these highly correlated results
   that are shown. Error bars on each point are the statistical and
   the quadratically combined statistical-plus-systematic
   uncertainties. For the sake of clarity, results are shifted about
   the values of $M_\pi^\max=250,\cdots\,\mev$, at which they are
   obtained.}  \labell{fig:fpi-nolog}
\end{figure}

\subsection{On the possible misuse of $\chi$PT}
\labell{sec:xpt-misuse}

In this section we examine the role of lattice results near the
physical value of $M_\pi$, for the determination of LECs. For this
purpose we fix the maximum value of $M_\pi$ to $M_\pi^\max=450\,\mev$
and study the dependence of the $p$-value and of the LECs as a
function of the lower bound, $M_\pi^\min$, that we place on the lattice
results included in the fit. We consider fully correlated NLO, $SU(2)$
$\chi$PT fits, both in the $x$ and $\xi$-expansion. We compare the
results obtained to those given by NLO fits in our canonical range,
$M_\pi\in[120, 300]\,\mev$. We perform the comparison by subtracting
these canonical results for the LECs from the new ones, under our
systematic and bootstrap error loops. Thus we obtain fully controlled
statistical and systematic errors on these differences.

\begin{figure}[t]
 \centering
\includegraphics[width=0.9\textwidth]{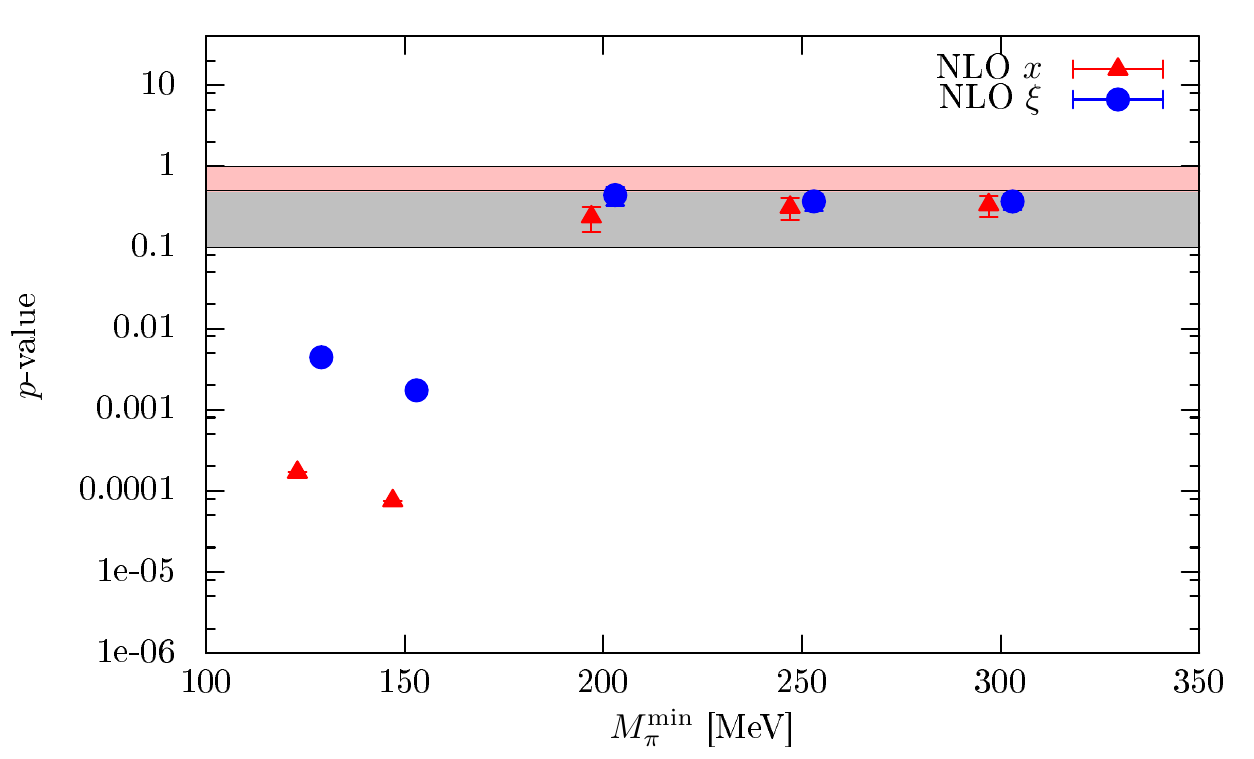}
 \caption{\sl $p$-value as a function of $M_\pi^\min$. The $p$-values
   are obtained by performing fully correlated NLO, $SU(2)$ $\chi$PT
   fits to lattice results for $B_\pi$ and $F_\pi$ with pion masses in
   the range $[M_\pi^\min,\, 450\,\mev]$. Both the $x$ and
   $\xi$-expansion are considered. The points with
   $M_\pi^\min=120,\cdots\,\mev$ are the same as those with
   $M_\pi^\max=450,\cdots\,\mev$ in \fig{fig:pvalue}. The horizontal
   bands have the same meaning as in \fig{fig:pvalue}. Error bars on
   each point are the systematic uncertainties discussed in
   \sec{sec:nlo-fitquality}. For the sake of clarity, results are
   shifted about the values of $M_\pi^\min=120,\cdots\,\mev$, at which
   they are obtained. } \labell{fig:pvalue-mpimin}
\end{figure}

In \fig{fig:pvalue-mpimin} we plot the $p$-value of these NLO fits as
a function of $M_\pi^\min$ with full systematic
errors. We find acceptable values for $M_\pi^\min\ge 200\,\mev$, which
may give the erroneous impression that NLO, $SU(2)$ $\chi$PT is
applicable in the range $M_\pi\in[200, 450]\,\mev$. However, as we
showed in \sec{sec:nlo-fitquality}, NLO $\chi$PT is not applicable up to
450~MeV.

\begin{figure}[t]
 \centering
\includegraphics[width=0.8\textwidth]{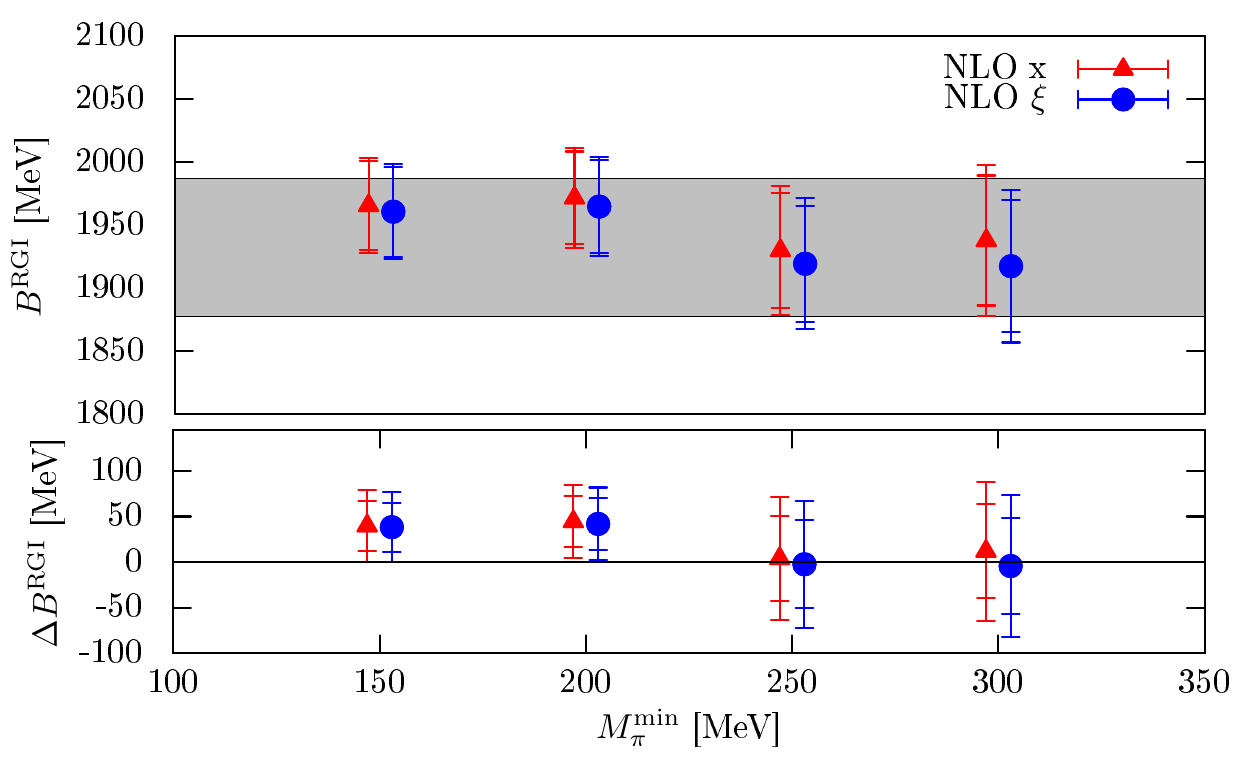}
\includegraphics[width=0.8\textwidth]{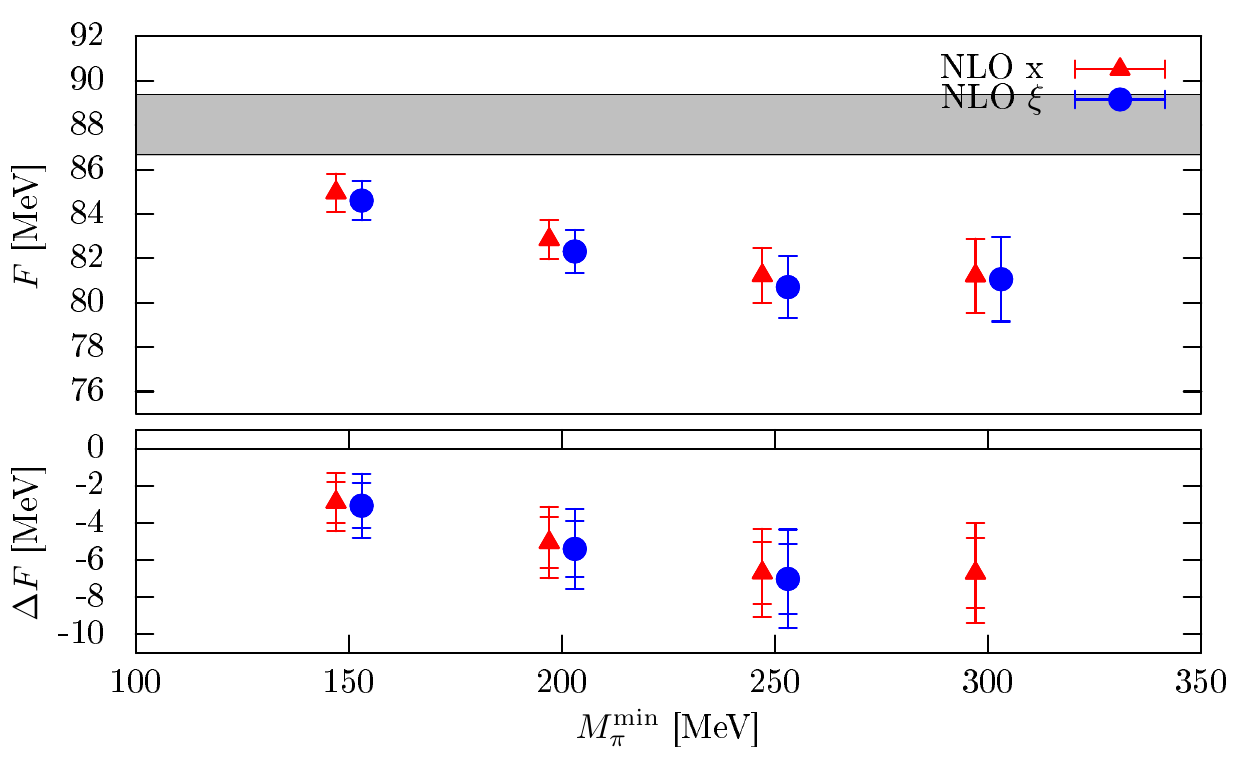}
 \caption{\sl LO LECs as a function of $M_\pi^\min$ (upper panel of
   each plot). The LECs are obtained from the fits described in
   \fig{fig:pvalue-mpimin}. The horizontal gray band denotes our final
   result for the corresponding LEC, given in
   \tab{tab:final-lec-results}, and obtained as described in
   \sec{sec:lecresults}. In the lower panel corresponding to each LEC,
   it is the difference of this LEC with the one obtained from fits in
   our canonical range, $M_\pi\in[120, 300]\,\mev$. Error bars on each
   point are the statistical and the quadratically combined
   statistical-plus-systematic uncertainties. For the sake of clarity,
   results are shifted about the values of
   $M_\pi^\min=150,\cdots\,\mev$, at which they are obtained. }
 \labell{fig:lo-mpimin}
\end{figure}

To give an idea of how one might be misled in the determination of
LECs and physical quantities, in \fig{fig:lo-mpimin} we plot the LO
LECs and $F_\pi$ as a function of $M_\pi^\min$ for $M_\pi^\min\in
[150,\,300]\,\mev$, for both the $x$ and $\xi$-expansion. As in
\figs{fig:lo-lec-vs-mpicut}{fig:nlo-lec-vs-mpicut}, we also plot, in
the lower panel, the difference of these quantities with the
corresponding results obtained in our canonical range $M_\pi\in
[120,\,300]\,\mev$. While $B$ remains
close to its physical value, $F$ and $F_\pi$ drop significantly below
their correct values, by as much as 7\%. The net result on the
condensate, $\Sigma$, is even larger since $\Sigma=F^2 B$.

\begin{figure}[t]
 \centering
\includegraphics[width=0.8\textwidth]{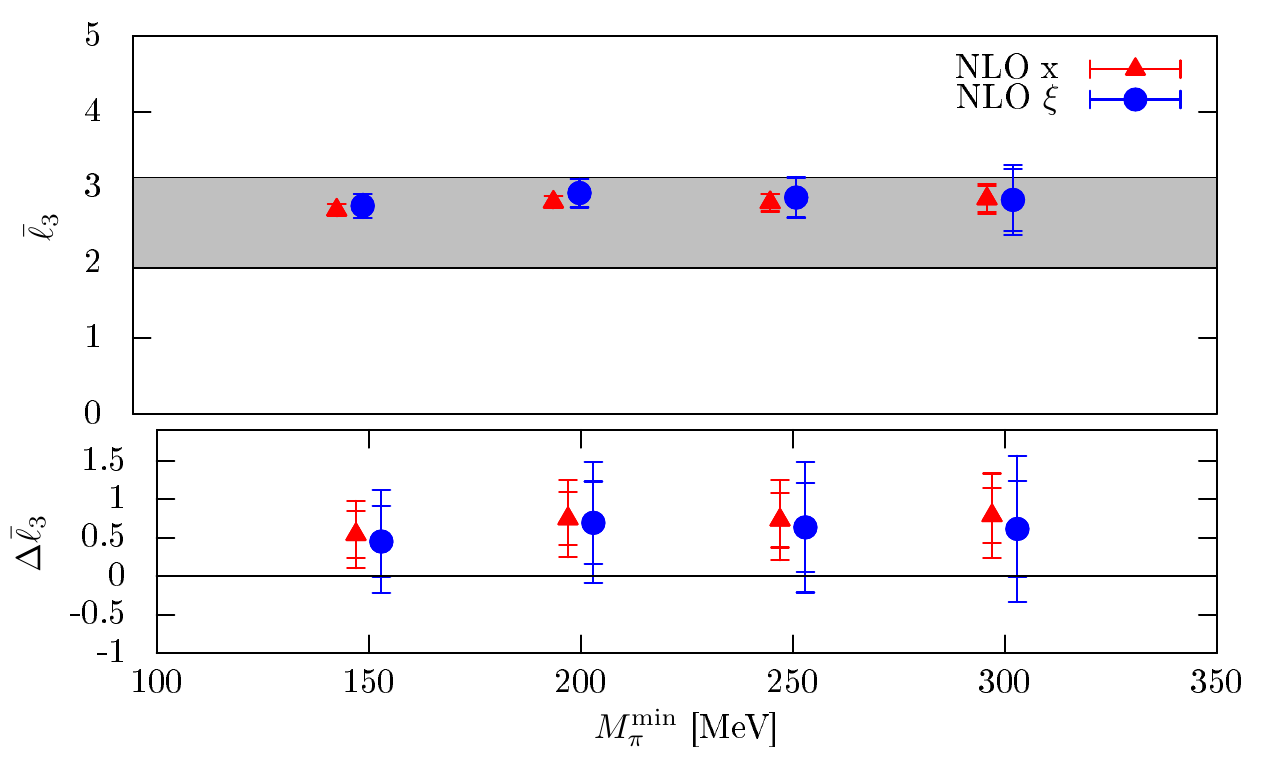}
\includegraphics[width=0.8\textwidth]{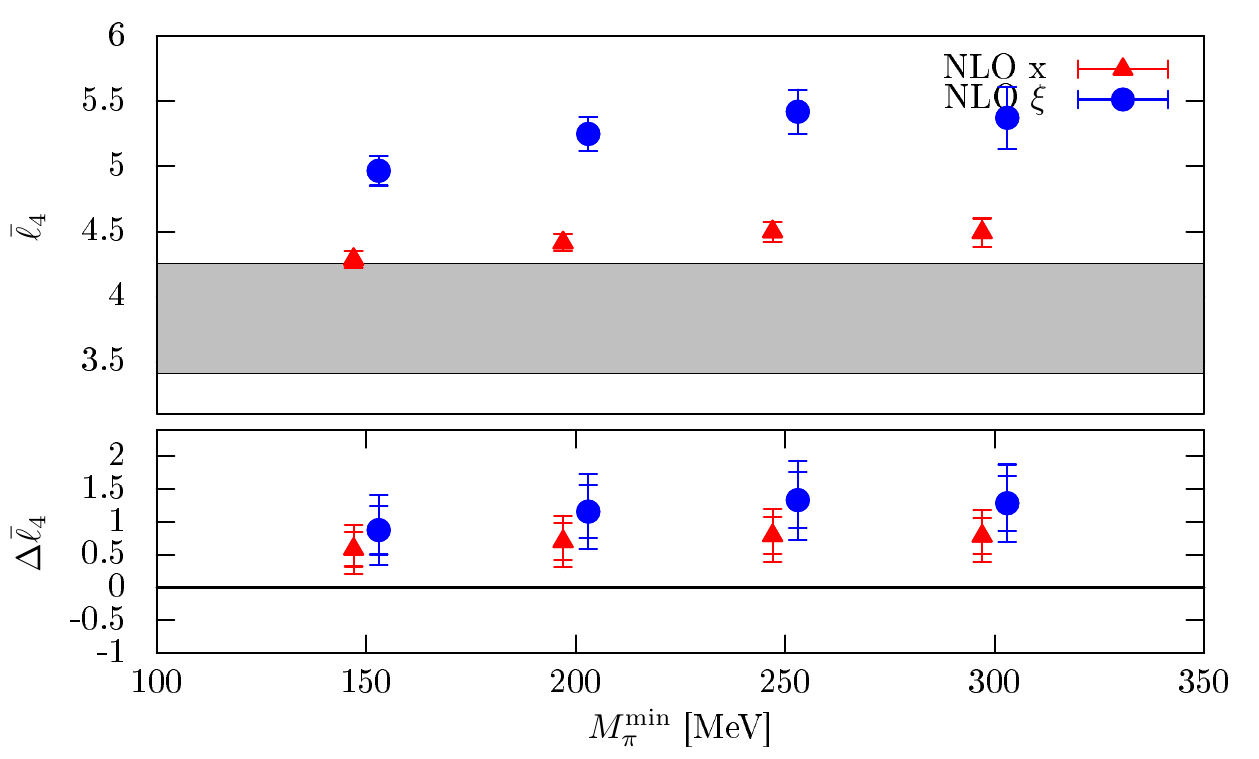}
 \caption{\sl Same as \fig{fig:pvalue-mpimin}, but for NLO LECs.}
 \labell{fig:nlo-mpimin}
\end{figure}

\fig{fig:nlo-mpimin} displays the same study, but for NLO LECs. While
the value of $\bar\ell_3$ remains compatible with its physical value,
$\bar\ell_4$ increases steadily as $M_\pi^\min$ is increased,
especially in the $\xi$-expansion. These are
the NLO expression of the observations made at LO. In particular, the
larger values of $\bar\ell_4$, or equivalently of the scale
$\Lambda_4$, indicate that as lattice results at lower $M_\pi$ are
removed, the downward trend of the chiral logarithm in $F_\pi$, as the
chiral limit is approached, is allowed to begin at larger values of
$M_\pi$. The end result is lower values of $F$ and $F_\pi$ for
larger $M_\pi^\min$. These results fully corroborate the observations
that we made, in \sec{sec:lecresults}, about the values of LECs
obtained by groups whose simulations do not reach down to small
$M_\pi<200\,\mev$. 

To summarize, if NLO $SU(2)$ $\chi$PT is applied to results for
$B_\pi$ and $F_\pi$ up to $M_\pi^\max=450\,\mev$, one obtains a good
description if one does not have results very close to the physical
point, i.e. with $M_\pi<200\,\mev$. Thus, one may be led to believe
that one is in the range of applicability of NLO $\chi$PT. However, as
we show, the description of $F_\pi$, in particular, is significantly
different from that obtained around the physical point, with values of
$F$ and $F_\pi$ which are too small and of $\bar\ell_4$ which are too
large. Said differently, results for $F_\pi$ close to the physical
point show less downward curvature than results at larger values of
$M_\pi$ suggest.

\clearpage

\section{Conclusion}
\labell{sec:conclusion}
We have performed a detailed, fully-correlated study of the chiral
behavior of the pion mass and decay constant, based on 2+1 flavor
lattice QCD simulations. These calculations are implemented using
tree-level, $O(a)$-improved Wilson fermions all the way down to
$M_\pi\simeq 120\,\mev$. This coverage of the low-mass region allows
to probe deeply into the chiral regime. Quark masses and decay
constants undergo fully-controlled nonperturbative
renormalization. Moreover, our fine lattice spacings down to 0.054~fm
and large volumes up to 6~fm enable us to accurately perform the
relevant continuum and infinite-volume extrapolations. We set the
scale of our calculations with the $\Omega$ baryon mass, which is
independent of the quantities of interest here. This allows us to make
valuable tests of our calculation. The first is an {\it ab initio}
computation of $F_\pi$, whose result agrees well with
experiment \cite{Beringer:1900zz} within our 1\% error bar. The second
is a determination of $m_{ud}$ that is fully compatible with the FLAG
value \cite{Colangelo:2010et}. In fact, it is nearly identical to the
result of \cite{Durr:2010vn,Durr:2010aw}, which is not surprising as
our treatment of quark masses is carried over from that work. 

We begin the study presented in this paper with a systematic
investigation of the range of applicability of $SU(2)$ $\chi$PT. We
consider two expansions. The first, which is that used in previous
$N_f\ge 2+1$
studies \cite{Allton:2008pn,Aoki:2008sm,Bazavov:2009fk,Aoki:2010dy,Bazavov:2010hj,
Bazavov:2010yq,Beane:2011zm,Borsanyi:2012zv,Arthur:2012opa,
Baron:2010bv,Baron:2011sf}, is in quark mass ($x$-expansion). The
second is in squared pion mass ($\xi$-expansion) and has not, as far
as we know, been investigated before. The study of the later has led
us to find constraints on the NLO LEC $\bar\ell_4$ in terms of the LO
LEC $F$ and bounds on the NNLO LEC $c_F$ in terms of the $F$ and the
NLO LECs $\bar\ell_4$ and $\bar\ell_{12}$ defined in and
around \eq{eq:barelldef}. These bounds are derived and discussed
in \app{sec:xi-constraints}.

To explore the range of applicability of $SU(2)$ $\chi$PT we consider
a number of criteria. These include a study of the $p$-value of our
combined, fully-correlated $\chi$PT fits, to $M_\pi^2$ and $F_\pi$, as
a function of $M_\pi^\max$, where $[120\,\mev,M_\pi^\max]$ is the range
of the masses of the lattice pions which we include in our fits. We
also study the values of the LO, NLO and NNLO LECs obtained in these
fits, as a function of $M_\pi^\max$. We further investigate the
relative size of contributions of different orders in the $\chi$PT
expansion for different pion masses. While our study of NLO expansions
is well controlled, we find that we do not really have enough
precision to make definite statements about NNLO.

Our systematic investigation leads to the following conclusions. We
find that NLO $\chi$PT for $M_\pi^2$ and $F_\pi$ begins showing signs
of failure for $M_\pi$ beyond 300~MeV and breaks down completely
around 450~MeV for both expansions. Adding NNLO terms allows one to
describe consistently the mass dependence of $F_\pi$ in the
$\xi$-expansion, up to around 500~MeV, at the expense of NNLO
corrections which are approaching those of the NLO ones. This is only
marginally true in the $x$-expansion, as $F$ and $\bar\ell_4$ begin
deviating from the values given by the NLO fits with $M_\pi^\max\leq
300\,\mev$ in that expansion. However in both expansions, the addition
of NNLO terms in $B_\pi$ does not allow a description of that quantity
beyond 300-350~MeV that is consistent with the NLO description at the
level of around one standard deviation. This behavior is
consistent with the fact that these are asymptotic expansions. Since
conclusions about applicability of $SU(2)$ $\chi$PT depends not only
on the range of pion masses, but also on the precision of the results
to which it is applied, it is important that the latter be
specified. This is discussed in detail
in \sec{sec:rel-xpt-contribs}. Here we only remind the reader that the
typical precision of our lattice results is around 1\%. Note
also that conclusions may differ when considering applications of
$SU(2)$ $\chi$PT to $N_f=2$ QCD, since the latter is missing the
relatively light degrees of freedom associated with the strange
quark.

Having established the range of applicability of $SU(2)$ $\chi$PT,
which is very similar for both expansions, we use lattice results in
that range to determine the theories' LECs. In particular, we use our
combined, fully-correlated NLO $\chi$PT fits to lattice results for
$M_\pi^2$ and $F_\pi$ with $M_\pi^\max\le 300\,\mev$, to compute $F$,
$B$, $\bar\ell_3$ and $\bar\ell_4$, as well as the quark condensate
and $F_\pi$, with fully controlled uncertainties. Our final results
are summarized in \tab{tab:final-lec-results} and those for the
individual $x$ and $\xi$-expansions
in \tab{tab:indiv-exp-lec-results}. A detailed comparison with the
$N_f\ge 2+1$ studies
of \cite{Allton:2008pn,Aoki:2008sm,Bazavov:2009fk,Aoki:2010dy,Bazavov:2010hj,
Bazavov:2010yq,Beane:2011zm,Borsanyi:2012zv,Arthur:2012opa,
Baron:2010bv,Baron:2011sf} is given in \sec{sec:lecresults}. Here we
note that while our results for $\bar\ell_3$ and $\bar\ell_4$ are
consistent with those obtained from lattice $N_f\ge 2+1$ simulations
with pion masses below
200~MeV \cite{Bazavov:2010yq,Borsanyi:2012zv,Arthur:2012opa}, they are
systematically smaller, particularly those obtained in the
$x$-expansion, which is used by all other collaborations. It is also
interesting to note that our result for the quark condensate has an
uncertainty which is almost 5 times smaller than the latest FLAG
compilation of \cite{Colangelo:2010et}.

We investigate the application of NNLO $SU(2)$ $\chi$PT to our
lattice results. There we find that we have to include results with
$M_\pi$ at least up to 400~MeV to have enough information to stabilize
these fits without imposing arbitrary priors. Unfortunately, our
studies suggest that, at such masses, we are already reaching beyond
the range of applicability of NNLO $SU(2)$ $\chi$PT. Nevertheless,
since little is known about NNLO LECs, we still attempt to determine
them, with results given at the end of \sec{sec:lecresults}. As noted
there, these results should be taken with a grain of salt and are only
meant as indicative.

In \sec{sec:logsandmisuse} we explore the presence of NLO chiral
logarithms in our lattice results. We show that this presence is
significantly favored in the region of applicability of NLO $SU(2)$
$\chi$PT. While the inclusion of logarithms does not make a
significant difference on the value of $F_\pi$ obtained at the present
level of accuracy, we find that it will when the total uncertainty on
$F_\pi$ reaches a few tenths of an MeV.

In this same section, we examine the role of lattice results near the
physical value of $M_\pi$, in particular for the determination of
LECs.  We find that one obtains perfectly good NLO fits of lattice
results for $M_\pi^2$ and $F_\pi$ in the range
$[M_\pi^\min,450\,\mev]$ with $M_\pi^\min\ge 200\,\mev$. This might
lead one to believe that NLO $SU(2)$ $\chi$PT is applicable in this
range. However, our systematic study of the range of applicability of
this theory already showed that the theory failed for $M_\pi\gsim
450\,\mev$. Moreover, while the value of $B$ and $\bar\ell_3$ are not
strongly affected by considering higher pion mass ranges, this is not
true of $F$, $\bar\ell_4$, the pion decay constant and the quark condensate.

\section*{Acknowledgments}

We thank Alberto Ramos for his help in the early stages of this
project. Jérôme Charles and Marc Knecht are thanked for helpful
conversations. Computations were performed using HPC resources
provided by GENCI-[IDRIS] (grant 52275) and FZ J\"ulich, as well as
using clusters at Wuppertal and CPT. This work was supported in part
by the OCEVU Excellence Laboratory, by CNRS grants GDR $n^0$2921 and
PICS $n^0$4707, by EU grants FP7/2007-2013/ERC $n^0$208740,
MRTN-CT-2006-035482 (FLAVIAnet) and by DFG grants FO 502/2, SFB-TR 55.

\clearpage
\begin{appendices}
\section{Solution for $F_\pi$ in the $\xi$-expansion and ensuing constraints}
\labell{sec:xi-constraints}
As mentioned in \sec{sec:fit-strategy}, the expressions for $F_\pi$ in
the $\xi$-expansion are obtained by solving the second equation in
(\reff{eq:MpiFpi}) for $F_\pi$. At NLO this equation is quadratic and,
at NNLO, it is quartic. Therefore, it has either up to 2 or 4
solutions and there is no guarantee that any of them are physical. In
this section we investigate the conditions under which a physical
solution exists. At fixed order in $\chi$PT, we find that these
conditions impose non-trivial constraints on the LECs. Of course, if
higher orders are allowed, these constraints eventually disappear.

The second equation in (\reff{eq:MpiFpi}) can be rewritten as
\be
\labell{eq:xi-fpi-eq-nnlo}
f(r)\equiv r^4 - r^3 - C r^2-D=0
\ ,
\ee
with
\be
\labell{eq:CDdef}
C=X\ln\left(\frac{\Lambda_4}{M_\pi}\right)^2, \qquad
D=\frac{X^2}{4}\left\{\left[\ln\left(\frac{\Omega_F}{M_\pi}\right)^2\right]^2-4 c_F\right\}
\ ,\ee
and
\be
\label{eq:rFxiFdef}
r=\frac{F_\pi}{F},\qquad X=\left(\frac{M_\pi}{4\pi F}\right)^2
\ .\ee

At NLO, $D=0$ and, since $F_\pi=0$ is not physical, \eq{eq:xi-fpi-eq-nnlo}
reduces to the quadratic equation
\be
\labell{eq:xi-fpi-eq-nlo}
r^2 - r - C=0
\ .
\ee
This equation has real solutions iff $C\ge -1/4$ or
\be
\labell{eq:l4-cstrt-nlo}
\bar\ell_4\ge \ln\left(\frac{M_\pi}{\hat M_{\pi^+}}\right)^2
- \left(\frac{2\pi F}{M_\pi}\right)^2 
\ .\ee
Since we want $F_\pi\ge F/2$, the physical solution is the larger of
the two, i.e.
\be
\labell{eq:Fpi-soln-nlo}
F_\pi = \frac{F}{2}\left[1+\sqrt{1+4 C}\right]
\ .\ee

Note that $F_\pi$ is greater than $F$ iff $C>0$ or, equivalently for
$M_\pi\ge\hat M_{\pi^+}$, $\bar\ell_4$ is positive and its
contribution in (\reff{eq:MpiFpi}) dominates over that of the chiral
logarithm.  Thus, the constraint in \eq{eq:l4-cstrt-nlo} is weaker
than requiring that $F_\pi>F$. On the other hand, the validity of NLO
$\chi$PT would generically require that $|C|\ll 1$. From that
perspective, the constraint of \eq{eq:l4-cstrt-nlo}, $C\ge -1/4$ is a
little more specific, since it tells us that a positive NLO correction
in (\reff{eq:MpiFpi}), whose magnitude is more than 25\%, is not
allowed if one assumes that the NLO $\xi$-expansion of $F_\pi$ is
exact. Assuming that this is the case, as we do when we
fit our lattice results to this expression, \eq{eq:xi-fpi-eq-nnlo}
imposes a constraint on the NLO LEC, $\bar\ell_4$, in terms of the LO
LEC, $F$, and of the pion mass at which the NLO $\xi$-expression is
applied. Note that the RHS of \eq{eq:l4-cstrt-nlo} is a monotonically
increasing function of $M_\pi$, indicating that the constraint on
$\bar\ell_4$ becomes more and more stringent as one tries to apply NLO
$\xi$-expressions to more and more massive pions. In particular, if we
assume that the expansion must hold up to a value of
$M_\pi=M_\pi^\max$, the lower bound on $\bar\ell_4$ that must be
enforced is the value of the RHS at $M_\pi^\max$. We impose this lower
bound dynamically in the NLO, $\xi$-expansion fits which are described
in \sec{sec:fit-strategy}. 

For illustration, in \fig{fig:l4-cstrt} we plot this bound and its
uncertainty as a function of $M_\pi^\max$. The curves correspond to
our final result for $F$, given in \tab{tab:final-lec-results}. This
bound is rather weak. It requires that $\bar\ell_4$ must be positive
if one wants a physical solution above $M_\pi\sim 400\,\mev$ at NLO in
the $\xi$-expansion and larger than 4 only for $M_\pi^\max\gsim
1.1\,\gev$. The latter indicates that the NLO fit of our data that we
perform for $M_\pi^\max=300\,\mev$ cannot be extended up to
1.1~GeV. While our study shows that there are many other important
reasons for why this is the case, it is still interesting that
fixed-order $\xi$-expansions have a built-in maximum pion-mass range.

 \begin{figure}[t]
  \centering 
  \includegraphics[width=0.8\textwidth]{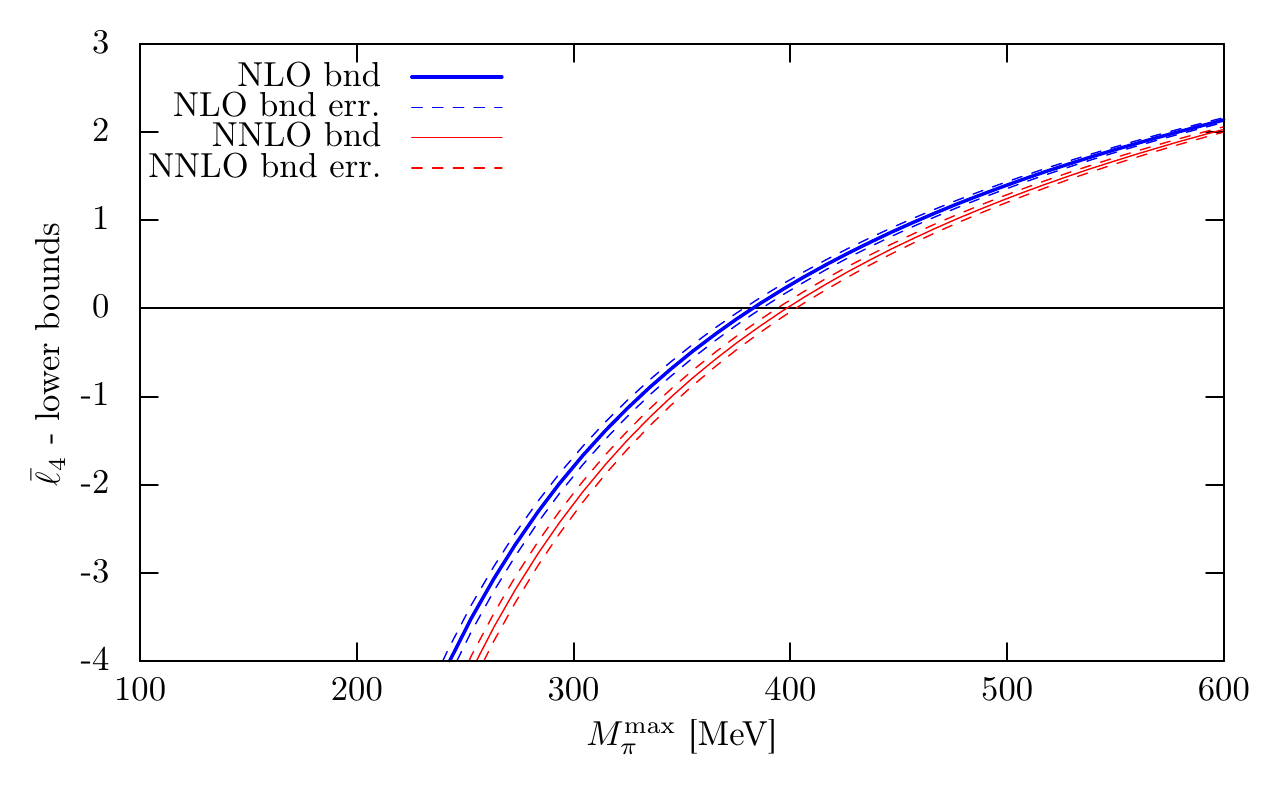} 
\caption{\sl Illustration of the 
NLO and NNLO lower bounds on $\bar\ell_4$ coming from the requirement
  that there is a physical solution for $F_\pi$ assuming that the NLO
  or the NNLO $\xi$-expansion expressions of (\reff{eq:MpiFpi}) hold
  exactly. $\bar\ell_4$ must lie above the given curve for each order
  in the expansion. To plot these curves, we use our final result for
  $F$ given in \protect\tab{tab:final-lec-results}. The dashed curves
  delimit the 1$\sigma$ error band on each bound arising from
  the total uncertainty on $F$.}
\labell{fig:l4-cstrt}
 \end{figure}

At NNLO, 
\eq{eq:xi-fpi-eq-nnlo} for $F_\pi$ is quartic and therefore has up to 4
solutions. Moreover, it is easy to show that $f(r)$ has 3 extrema, one
of which is at $r=0$. There are two other real extrema iff
\be
\labell{eq:l4-cstrt-nnlo}
\bar\ell_4\ge \ln\left(\frac{M_\pi}{\hat M_{\pi^+}}\right)^2
- \frac98\left(\frac{2\pi F}{M_\pi}\right)^2
\ ,\ee
which is slightly less constraining than \eq{eq:l4-cstrt-nlo}. Thus,
for any pion-mass range, the NNLO $\xi$-expansion admits slightly
smaller values of $\bar\ell_4$ than does the NLO expansion. This is
not surprising as we know that bounds on the LECs must disappear in the
limit of infinite order. However, finding such a value would imply
that the NLO expansion is only applicable in a smaller mass range than the
NNLO one. In turn, this would be a sign that $\chi$PT is having trouble.

Now let us consider the possibility that $r=0$ is the only real
extremum, i.e. that $C<-9/32$. Because of the signs of the terms in
$f(r)$, it must be a minimum. Since we want a solution
to \eq{eq:xi-fpi-eq-nnlo} such that $F_\pi>F$, we must have
$|D|>|C|$. But for this to be true, the NNLO term in the
$\xi$-expansion must be larger than the NLO term. In that case the
$\xi$-expansion has clearly broken down, which is not an option of
interest here. Thus we assume that (\reff{eq:l4-cstrt-nnlo}) is
satisfied, so that $f(r)$ has 3 real extrema. It is then
straightforward to convince oneself that the absolute minimum of
$f(r)$ is at
$r_+=(3+\sqrt{9+32C})/8$. Therefore, \eq{eq:xi-fpi-eq-nnlo} will have
at least one real solution for $F_\pi$ iff $f(r_+)\le 0$. This
translates into a lower bound on the NNLO LEC $c_F$, in terms of the
LO and NLO LECs, $F$, $\bar\ell_4$ and $\bar\ell_{12}$.  This upper
bound is not necessarily a monotonic function of $M_\pi$. Therefore,
unlike the lower bound of \eq{eq:l4-cstrt-nnlo} on $\bar\ell_4$, which
need only be satisfied at $M_\pi^\max$ for the $\xi$-expansion to
hold, the minimum of the bound on $c_F$ in the region $M_\pi\in
[0,M_\pi^\max]$ must be found and imposed as an upper bound on
$c_F$. Thus,
\be
\labell{eq:cF-bnd-nnlo}
c_F\le
\mathop{\min}\limits_{M_\pi\in [0,\,M_\pi^\max]}\left\{
\frac14
\left[\ln\left(\frac{\Omega_F}{M_\pi^2}\right)^2\right]^2
-\left(\frac{4\pi F}{M_\pi}\right)^2r_+^2\left[r_+^2-r_+-C\right]\right\}
\ ,\ee
with $r_+$ given above. This bound is very sensitive to the values of
the LECs, and is not very enlightening when LO and NLO LECs, such as
those given in \tab{tab:final-lec-results} are used, assuming no
correlations between them.  However, for a given fit, this bound may be quite
constraining. Thus, we impose this upper bound and the lower bound on
$\bar\ell_4$ given in \eq{eq:l4-cstrt-nnlo} when fitting lattice
results to NNLO $\xi$-expansion expressions.

The fixed-order bounds on LECs discussed above are mainly of technical
use here: they are enforced to avoid that the fitting routine gets lost in
exploring unphysical regions of parameter space. However, for theories
other than QCD which have $SU(2)$ $\chi$PT as a low-energy
description, one could imagine being in a situation where these bounds
suggest a failure of the effective theory in a region of pion masses
where it is not entirely clear what is meant by the requirement that
chiral corrections are ``small''.

For completeness we also provide here the analytical expression for the
physical $F_\pi$ solution of the NNLO expression for $F$
in \eq{eq:MpiFpi}. It is given by \cite{wiki:quartic-function}:
\be
F_\pi = F\left\{\frac14 + S + \frac12 \sqrt{-4 S^2-2p + \frac{q}{S}}\right\}
\ ,\ee
with
\bea
p &=& -\frac38 - C\\
q &=& \frac18 + \frac{C}2\ ,
\eea
and
\bea
S &=& \frac12\sqrt{-\frac23 p+\frac13(Q+\frac{\Delta_0}{Q})}\\
Q &=& \left[\frac{\Delta_1 + \sqrt{\Delta_1^2-4\Delta_0^3}}{2}\right]
\ ,\eea
where
\bea
\Delta_0 &=& C^2 - 12 D\\
\Delta_1 &=& -2 C^3-27 D-72 C D
\ .\eea

\end{appendices}

\bibliographystyle{elsarticle-num}
\bibliography{su2xpt}

\end{document}